\documentclass[acmsmall, screen, nonacm]{acmart}

\usepackage{booktabs}
\usepackage{multirow} 
\usepackage{pifont}
\usepackage{graphicx} %
\usepackage{makecell}
\usepackage{adjustbox}
\usepackage{forest}
\usepackage{subcaption}
\usepackage{wrapfig}
\usepackage{tablefootnote}
\usepackage{threeparttable}

\usepackage{colortbl}
\usepackage{color}
\usepackage{xcolor}
\usepackage{makecell}
\usepackage{enumitem}
\usepackage[normalem]{ulem}

\newcommand{\ulit}[1]{\uline{\textit{#1}}}

\definecolor{reviewOriginalGreen}{HTML}{008000}
\definecolor{reviewRevisionRed}{HTML}{CC0000}
\newcommand{\revSearch}[1]{{\color{reviewOriginalGreen}}}
\newcommand{\revReplace}[2]{{#2}}

\definecolor{lightTeal}{HTML}{add2ca}
\definecolor{softPurple}{HTML}{8c89ba}
\definecolor{paleGreen}{HTML}{e6f0db}
\definecolor{mutedOrange}{HTML}{dfa660}
\definecolor{limeGreen}{HTML}{97c04b}
\definecolor{lightBlue}{HTML}{dbeef3}
\definecolor{softRed}{HTML}{d77470}
\definecolor{dustyBlue}{HTML}{93b2d9}
\definecolor{tabletitle}{HTML}{E5DBF8} 

\definecolor{checkgreen}{HTML}{009900}
\definecolor{crossred}{HTML}{CC0000}

\newcommand{\cmark}{\textcolor{checkgreen}{\ding{51}}}
\newcommand{\xmark}{\textcolor{crossred}{\ding{55}}}
\definecolor{partialorange}{HTML}{E69F00}
\newcommand{\pmark}{\textcolor{partialorange}{\ding{109}}}

\definecolor{benchverified}{HTML}{00736D}
\definecolor{benchlite}{HTML}{D55E00}
\definecolor{benchpro}{HTML}{0072B2}
\definecolor{benchother}{HTML}{7B3294}
\newcommand{\resVerified}[1]{\textcolor{benchverified}{#1}}
\newcommand{\resLite}[1]{\textcolor{benchlite}{#1}}

\newcommand{\resOther}[1]{\textcolor{benchother}{#1}}

\AtBeginDocument{%
  }

\usepackage{microtype}
\usepackage{mdframed}
\newcounter{rq} 
\newcommand{\answerRQ}[1]{\refstepcounter{rq}
\begin{mdframed}[linecolor=gray,roundcorner=12pt,backgroundcolor=gray!15,linewidth=3pt,innerleftmargin=2pt,
leftmargin=0cm, rightmargin=0cm, topline=false, bottomline=false, rightline=false,
skipabove=3pt plus 1pt, skipbelow=0pt,
innertopmargin=4pt, innerbottommargin=4pt]
\textbf{RQ\arabic{rq} Summary:} #1
\end{mdframed}
}

\begin{document}

\title{Agentic Software Issue Resolution with Large Language Models: A Survey}
\author{Zhonghao Jiang}
\email{zhonghao.j@zju.edu.cn}
\affiliation{%
  \institution{College of Computer Science and Technology and the State Key Laboratory of Blockchain and Data Security, Zhejiang University}
  \city{Hangzhou}
  \country{China}
}
\author{David Lo}
\email{davidlo@smu.edu.sg}
\affiliation{%
  \institution{School of Computing and Information Systems, Singapore Management University}
  \country{Singapore}
}
\author{Zhongxin Liu}
\authornote{Corresponding author.}
\email{liu_zx@zju.edu.cn}
\affiliation{%
  \institution{College of Computer Science and Technology and the State Key Laboratory of Blockchain and Data Security, Zhejiang University}
  \city{Hangzhou}
  \country{China}
}

\renewcommand{\shortauthors}{Jiang et al.}

\begin{abstract}

Software issue resolution task aims to address real-world issues in software repositories based on natural language descriptions provided by users, representing a key aspect of software maintenance. 
With the rapid development of large language models (LLMs) in reasoning and generative capabilities, LLM-based approaches have made significant progress in automated software issue resolution.
However, real-world software issue resolution is inherently complex and requires long-horizon reasoning, iterative exploration, and feedback-driven decision-making that demand agentic capabilities beyond conventional single-step approaches.
Recently, LLM-based agentic systems have become a promising research direction for software issue resolution, since the related literature has experienced explosive growth.
Advancements in agentic software issue resolution can not only greatly enhance software maintenance efficiency and quality but also provide a realistic environment for validating agentic systems’ reasoning, planning, and execution capabilities, bridging AI and software engineering.
This work presents a systematic survey of 242 recent studies at the forefront of LLM-based agentic software issue resolution research. 
It outlines the general workflow of the task and establishes a taxonomy across three dimensions: benchmarks, techniques, and empirical studies. 
Furthermore, it highlights how reinforcement learning has become an increasingly important training paradigm for agentic systems in software engineering.
Finally, it summarizes key challenges and outlines promising directions for future research.
The artifacts' page accompanying this survey is at~\url{https://github.com/ZhonghaoJiang/Awesome-Issue-Solving}.

\end{abstract}

\maketitle

\vspace{-0.5em}
\section{Introduction}\label{sec:introduction}

In software engineering, software maintenance often accounts for about two-thirds of software lifecycle costs~\cite{lientz1978characteristics}.
Software issue resolution occupies a crucial position in software maintenance~\cite{kuramoto2024understanding}. 
This task aims to understand, locate, and resolve issues in real-world code repositories based on developers’ natural language descriptions of issues, covering diverse maintenance activities such as bug fixes, feature additions, and efficiency optimizations. 
Hereafter, we refer to this task as issue resolution for brevity.
Traditional issue resolution methods heavily rely on human expertise, making the process time-consuming, error-prone, and difficult to scale~\cite{rahman2024systematic}. 
Consequently, issue resolution has long been regarded as a critical bottleneck to efficient software evolution~\cite{berhe2023maintenance}. 

In recent years, LLMs have achieved significant success in multiple areas of software engineering~\cite{bouzenia2024repairagent} due to their rapid advancements in code understanding~\cite{shrivastava2023repofusion}, reasoning~\cite{deepseekr1}, and generation~\cite{jiang2024survey}, paving the way for automated issue resolution.
However, issue resolution in real-world software systems is inherently complex, typically involving long-horizon reasoning, iterative exploration, interaction with evolving codebases, and feedback-driven decision-making, which goes beyond the capabilities of conventional single-step automated approaches.
LLM-based agentic systems are autonomous, goal-driven AI architectures that can interpret objectives, plan multi-step tasks, and adapt their behavior based on environmental feedback~\cite{sapkota2025ai}.
\revReplace{R1.5}{They have become increasingly popular in recent issue resolution research and report the strongest results on public benchmarks~\cite{SWEAgent, openhands, agentless}.}
\revReplace{R1.5}{For example, related papers on agentic issue resolution systems rose from 35 in 2024 to 122 in 2025, achieving a 248.6\% increase.
Moreover, recent agentic systems built on the latest frontier model Claude Fable 5~\cite{anthropic2026claudefable5} can already resolve 80.3\% of issues on the challenging SWE-Bench Pro benchmark~\cite{SWEbenchPro}.}
On the other hand, because of the complexity and broad applicability, issue resolution has become a key task for evaluating LLMs and software engineering (SE) agentic systems~\cite{anthropic2026claudefable5}.
Advancing LLM-based agentic issue resolution not only holds the potential to significantly improve the efficiency and quality of software maintenance but also provides LLM-based agentic systems with a realistic environment to validate complex reasoning, planning, and execution capabilities, which serves as a crucial bridge for the bidirectional integration of AI and software engineering.
Given its significance and impact, a systematic survey of LLM-based agentic issue resolution is essential to provide a comprehensive overview of current research and to highlight future directions.

Although existing surveys~\cite{liu2024large, yang2025survey, tao2025retrieval, guo2025comprehensive} have organized a part of agentic issue resolution methods based on the taxonomy of automated program repair (APR) or code generation, they have not comprehensively covered the literature in the field of issue resolution for the following reasons.
\ding{182} Lack of taxonomies specifically tailored to issue resolution and workflow modeling.
Existing surveys~\cite{yang2025survey, tao2025retrieval, guo2025comprehensive} do not regard issue resolution as a holistic task but divide issue resolution-related research into existing areas, such as APR. 
However, unlike APR, issue resolution encompasses more diverse software maintenance activities, such as efficiency optimization~\cite{swe-perf} and feature addition~\cite{nocode-bench, FEABench}.
Even for bug-fixing issues, issue resolution does not assume the existence of tests that can trigger the bug.
These differences render existing taxonomies insufficient to systematically organize all research related to issue resolution.
Additionally, these surveys also ignore the examination of numerous issue resolution benchmarks and empirical studies.
\ding{183} Overlook the recent paradigm shift in issue resolution techniques.
\revReplace{R1.5}{
Recently, LLM-based issue resolution has gradually shifted from prompt-engineering-based scaffold design toward training dedicated models. 
For instance, since February 2025, the number of training studies has grown 11.4-fold, substantially outpacing the 3.1-fold growth of scaffold-design studies over the same period. 
Within this training-driven paradigm, reinforcement learning (RL) has emerged as the main approach, propelled by its remarkable success in LLMs~\cite{deepseekr1, rl_survey}.
24 of the 39 (61.5\%) representative models released from February 2025 to May 2026 use reinforcement learning.
}
Existing surveys~\cite{liu2024large, yang2025survey} did not capture this paradigm shift, overlooking the studies on model training strategies for issue resolution.
In summary, there is still no comprehensive survey on LLM-based agentic issue resolution.

To fill this gap, we present a comprehensive survey of LLM-based agentic issue resolution, aiming to provide researchers with a foundational reference for quickly understanding this emerging area.
\revReplace{R1.1}{Specifically, we mainly focus on software issue resolution, treating the end-to-end methods as the core task, the subtask-level methods (e.g., issue localization), the learning strategies (e.g., RL training) as part of this problem space, the benchmark construction as evaluation infrastructure, and the empirical studies as the supporting findings.
Based on this scope, we propose a taxonomy that spans three dimensions: benchmarks, techniques, and empirical studies, offering an analysis from task definition of issue resolution to solutions and reflections.}
In particular, under the technical dimension, we examine each phase of the issue resolution process, including localization, repair, and validation from two perspectives: solution design paradigms and learning strategies for domain-specific models. 
This highlights how the rise of reinforcement learning has brought a significant transformation to the field.
Furthermore, we identify key challenges and future opportunities in the current landscape, making this survey a valuable resource for researchers in both the software engineering and the natural language processing (NLP) communities to grasp the research trajectory, frontiers, limitations, and prospects of the issue resolution task.
\textit{To the best of our knowledge, this is the first survey that specifically focuses on LLM-based agentic issue resolution tasks.}

The remainder of this survey is organized as follows.
Section~\ref{sec:background} introduces the background of automatic issue resolution, LLM-based agentic systems, and related surveys.
Section~\ref{sec:method} \revReplace{R1.7/R1.11/R3.7/R3.12}{defines research questions to guide the analysis and} outlines the methodology for systematically conducting this survey.
Section~\ref{sec:benchmark} presents the construction and evolution of current issue resolution benchmarks.
Section~\ref{sec:technique} provides a detailed taxonomy of existing techniques from the perspectives of method design and learning strategies.
Section~\ref{sec:emprical_studies} categorizes and analyzes empirical studies related to the issue resolution task.
\revReplace{R3.11}{Section~\ref{sec:road_ahead} discusses potential research opportunities and future directions. 
Section~\ref{sec:threats} analyzes the threats to the validity of this survey. 
Finally, Section~\ref{sec:conclusion} concludes this survey.}

\vspace{-0.5em}
\section{Background}\label{sec:background}

\subsection{LLM-based Agentic Systems for Issue Resolution}\label{sec:2.2}

\revReplace{R1.1}{The automated, \textit{end-to-end} issue resolution task takes a natural-language issue description and a code repository as input and produces a repository-level patch resolving the issue.}
\revReplace{R3.4}{From the functional components of the surveyed systems, we abstract this task into five phases, namely repo preprocessing, localization, repair, patch validation, and patch selection, whose coverage is validated by the open coding procedure described in Section~\ref{sec:taxonomy_construction}.
These phases denote logical functions rather than a fixed execution order, so a system may execute them sequentially or interleave them dynamically at inference time.}
Appendix~\ref{ap:issue_resolution_overview} illustrates the resulting framework and describes each phase in detail.

LLM-based agentic systems for issue resolution are composed of LLMs and scaffolds~\cite{wang2025confucius}.
The LLM serves as the core reasoning and generation engine of the system, responsible for understanding natural-language issue descriptions, analyzing code semantics, and generating patches or intermediate reasoning steps.
In contrast, scaffolds provide external structured control and an operational framework by orchestrating task workflows, coordinating multi-step reasoning, and invoking tools for code retrieval and execution, thereby guiding the LLM through iterative refinement and enabling the system to resolve the issue in real-world software repositories.

\revReplace{R1.6/R3.5}{
Existing agentic systems can be divided into agentic pipelines and agents based on how their control flow is designed and executed~\cite{sapkota2025ai}. 
In pipelines, human designers predefine the control flow as staged steps or state-machine-like workflows, while the LLM performs the assigned subtask at each stage. 
In agents, the LLM determines the next action during execution based on environmental feedback, so the control flow emerges dynamically during execution.
In summary, pipelines emphasize determinism and controllability, whereas agents offer greater autonomy and flexibility.
}

\subsection{Related Surveys}\label{sec:related_surveys}

\revReplace{R1.10}{
We mainly compare surveys~\cite{liu2024large, yang2025survey, guo2025comprehensive, tao2025retrieval} conducted since the introduction of automated issue resolution~\cite{swebench} from several high-level issue-resolution-related dimensions, as shown in Table~\ref{tab:exisiting_surveys}.
Overall, existing surveys do not examine issue resolution as a distinct area.
Instead, they discuss it primarily in the context of related tasks, such as automated program repair (APR) or repository-level code generation, covering only the aspects that fall within their respective scopes.
For example, Liu et al.~\cite{liu2024large} focus on the design of agent modules for general software engineering tasks and describe issue resolution only as a coarse maintenance process, while overlooking the training strategies and issue-resolution-specified empirical findings.
Yang et al.~\cite{yang2025survey} organize issue resolution methods under APR design paradigms and discuss supervised fine-tuning (SFT) and reinforcement learning (RL), benchmark settings, and success-rate trends only for APR, while overlooking non-bug-fixing issues and the fact that issue resolution assumes no trigger tests and thus follows a different workflow.
Guo et al.~\cite{guo2025comprehensive} catalogue solutions, SFT and RL techniques, and task-based benchmarks for agentic software engineering, while modeling only a general agent workflow and overlooking benchmark construction and empirical findings.
Tao et al.~\cite{tao2025retrieval} concentrate on retrieval strategies, retriever and generator training, and the effectiveness of retrieval-augmented generation (RAG) for repository-level code generation, while overlooking the issue resolution workflow and the construction and evolution of its benchmarks.}

\revReplace{R1.10}{Different from these surveys, our work treats issue resolution as the central research topic.
Based on 242 collected papers, it models the five logical phases of the task, organizes techniques along both scaffold design and training strategies that cover data preparation, SFT, and RL, analyzes the construction and evolution of issue resolution benchmarks, and synthesizes empirical findings across all of these phases.
Furthermore, it systematically examines how the task has evolved from its proposal to the latest developments, revealing the paradigm-level shift in the field and elucidating the technical landscape both before and after this transition.
}

\begin{table}[t]
\centering
\caption{\revReplace{R1.10}{Comparison of related surveys. \cmark: systematic coverage; \pmark: partial coverage; \xmark: not covered.}}
\label{tab:exisiting_surveys}
\begin{adjustbox}{max width=\textwidth}
\begin{tabular}{lccccc}
\toprule
\textbf{Dimension}
& \textbf{Liu et al.~\cite{liu2024large}}
& \textbf{Yang et al.~\cite{yang2025survey}}
& \textbf{Tao et al.~\cite{tao2025retrieval}}
& \textbf{Guo et al.~\cite{guo2025comprehensive}}
& \textbf{Ours} \\
\midrule
\textbf{Primary Scope}
& \textbf{SE agents}
& \textbf{LLM-based APR}
& \textbf{Repository-level RACG}
& \textbf{Agentic SE}
& \textbf{Issue resolution} \\
\midrule
\textbf{Workflow Modeling}
& \makecell{\pmark~Maintenance Process}
& \xmark
& \xmark
& \makecell{\pmark~Agent Workflow}
& \makecell{\cmark~Five Logical Task Phases} \\
\midrule
\textbf{Scaffold Design}
& \makecell{\cmark~Agent Modules}
& \makecell{\cmark~APR Design Paradigms}
& \makecell{\cmark~Agent Autonomy Levels}
& \makecell{\cmark~Solution Paradigms}
& \makecell{\cmark~Pipeline and Agent Scaffolds} \\
\midrule
\textbf{Training Strategies}
& \xmark
& \makecell{\pmark~SFT and RL}
& \makecell{\pmark~Retriever and Generator Training}
& \makecell{\pmark~SFT and RL}
& \makecell{\cmark~Data Preparation, SFT, RL} \\
\midrule
\textbf{Benchmark Analysis}
& \makecell{\pmark~Benchmark Listing}
& \makecell{\pmark~Scope and Evaluation Settings}
& \makecell{\pmark~Benchmark Listing}
& \makecell{\pmark~Task-based Categories}
& \makecell{\cmark~Construction and Evolution} \\
\midrule
\textbf{Empirical Findings}
& \xmark
& \makecell{\pmark~Success-rate Trends}
& \makecell{\pmark~RAG Effectiveness}
& \xmark
& \makecell{\cmark~Findings across All Phases} \\
\midrule
\textbf{Collection}
& \makecell{124 papers,\\until Sept. 2024}
& \makecell{63 papers,\\Jan. 2022--Oct. 2025}
& \makecell{110 papers,\\Jan. 2023--Aug. 2025}
& \makecell{150+ papers,\\2023--2025}
& \makecell{242 papers,\\Oct. 2023--May 2026} \\
\bottomrule
\end{tabular}
\end{adjustbox}
\end{table}

\vspace{-0.5em}
\section{Survey Methodology}\label{sec:method}

This section details the systematic methodology employed for conducting the literature review. 
Following prior surveys~\cite{hou2024large,jiang2024survey,liu2024large}, we adopt the systematic approach proposed by Kitchenham et al.~\cite{kitchenham2009systematic} and Petersen et al.~\cite{petersen2015guidelines}. 

\subsection{Research Questions}\label{sec:rqs}

\revReplace{R1.7/R1.11/R3.7/R3.12}{To guide our systematic literature review and characterize this field, we formulate four research questions (RQs), as follows:
\begin{itemize}[left=0pt, topsep=0em, itemsep=0pt, parsep=0pt]
    \item \textit{\textbf{RQ1:} How is the issue resolution task evaluated?} This RQ investigates how the existing evaluation datasets are constructed and have evolved, as well as which metrics are used for evaluation.
    \item \textit{\textbf{RQ2:} How are the key components in existing agentic issue resolution systems developed?} This RQ examines existing scaffolds, focusing on the design of their core components and the training of the LLMs that power them.
    \item \textit{\textbf{RQ3:} What do empirical studies reveal about the trustworthiness of reported issue-resolution performance and the factors and processes underlying agent performance variation?} This RQ synthesizes empirical findings on the validity of existing evaluation results and the factors and mechanisms that cause agent performance to vary.
    \item \textit{\textbf{RQ4:} What key challenges remain in LLM-based agentic issue resolution, and what research opportunities do they suggest?} This RQ identifies unresolved challenges in agentic software issue resolution, and outlines promising directions for future research.
\end{itemize}
}

\subsection{Literature Search and Selection}

To answer the RQs presented above and perform a survey, we need to retain relevant studies published over a wide time range and conduct a comprehensive analysis on those relevant studies. 
To achieve this, we follow prior surveys and use a four-stage process comprising automated search, inclusion/exclusion screening, quality assessment, and backward/forward snowballing.
For space limitations, we include a detailed description of each step in Appendix~\ref{ap:paper_collect}.

\subsection{Taxonomy Construction}\label{sec:taxonomy_construction}

\looseness=-1
\revReplace{R2.5/R2.7/R2.8/R2.11/R3.8}{To provide systematic answers to RQ1--RQ3, we aim to construct taxonomies for benchmarks, techniques, and empirical studies.
These categories are derived from the primary research objectives of the included papers during the inclusion and exclusion stage.
Following prior work~\cite{seaman1999qualitative}, we construct all three taxonomies through the standard open coding procedure.
Specifically, we randomly sample 70\% of the papers of each study type to construct the initial taxonomy.
Two authors jointly read these papers multiple times, inspect their abstracts, introductions, related work, and conclusions to understand their research goals, assign short phrases summarizing their contributions, group similar phrases into categories in a bottom-up manner, and iteratively refine the categories while revisiting the papers.
If a paper is associated with more than one category, it is assigned to all relevant categories.
The remaining 30\% of the papers are then independently labeled by the two authors based on the initial taxonomies, and each paper is marked with the leaf categories of the corresponding taxonomy.
Papers that cannot be classified into the current taxonomy are temporarily placed in a pending category and later discussed.
We use Cohen's $\kappa$ to evaluate inter-rater agreement during independent labeling, reaching values of 0.8682, 0.8985, and 0.9107 for the benchmark, technique, and empirical-study taxonomies, respectively.
Disagreements in both phases, including pending cases, are resolved through discussion with a third author.
As a result, every paper maps to at least one leaf category of the corresponding taxonomy.}
\revReplace{R3.4}{In addition, this procedure validates the coverage of the five-phase workflow model in Section~\ref{sec:2.2}. Functional components that directly support issue resolution are mapped to the corresponding phases of the resolution workflow. Cross-cutting mechanisms, such as memory, context management, and self-improvement, are treated separately because they operate across multiple phases.}

\vspace{-0.5em}
\section{Benchmarks}\label{sec:benchmark}

To answer \textbf{RQ1}, we systematically introduce the evaluation for software engineering agentic systems on the issue resolution task and its subtasks, focusing on three dimensions, including benchmark construction, benchmark evolution, and evaluation metrics, as shown in Figure~\ref{fig:benchmark_overview}.
Its three dimensions reflect the life cycle of a benchmark, covering how it is constructed, how it evolves after release, and how it measures performance.
In addition, Section~\ref{sec:benchmark_tradeoff} examines benchmark validity and comparability to clarify what the reported evaluation scores represent.

\begin{figure*}[htbp]
  \centering 
  \begin{adjustbox}{width=\textwidth}
  \begin{forest}
for tree={
    rounded corners,
    child anchor=west,
    parent anchor=east,
    grow'=east,  
    text width=4cm,%
    draw=softPurple,
    anchor=west,
    node options={align=center},
    fit=rectangle,
    edge path={
      \noexpand\path[\forestoption{edge}]
        (.child anchor) -| +(-5pt,0) -- +(-5pt,0) |-
        (!u.parent anchor)\forestoption{edge label};
    },
    where n children=0{text width=12cm}{}
  },
  [Benchmarks
    [Benchmark Construction
        [Manual
        [{SWE-bench~\cite{swebench}, SWE-bench Lite~\cite{swebench}, SWE-bench Verified~\cite{swebenchverified}, SWE-bench Lite-S~\cite{agentless}, SWE-bench-java~\cite{swebench_java}, SWE-bench Multimodal~\cite{swebenchmutimodal}, SWE-Bench+~\cite{swebench_plus}, Visual SWE-bench~\cite{CodeV}, FEA-Bench~\cite{FEABench}, Multi-SWE-bench~\cite{mutiswebench}, LiveSWEBench~\cite{liveswebench2025}, SwingArena~\cite{xu2026swingarena}, SWE-PolyBench~\cite{swepoly}, OmniGIRL~\cite{omnigirl}, SWE-bench Multilingual~\cite{swebenchmutilingual}, GSO~\cite{gso}, NoCode-bench~\cite{nocode-bench}, SWE-Bench Pro~\cite{SWEbenchPro}, SWE-fficiency~\cite{ma2025swe}, SWE-EVO~\cite{thai2025swe}, SWE-Refactor~\cite{xu2026swe}, FeatureBench~\cite{zhou2026featurebench}, Rust-SWE-bench~\cite{xiang2026evaluating}, SWE-Bench Mobile~\cite{tian2026swe}, SWE-STEPS~\cite{shastry2026beyond}, SWE-Chain~\cite{lam2026swe}},
        text width=17cm],
        ],
        [Automatic
            [{SWEE-Bench~\cite{SWA}, SWA-Bench~\cite{SWA}, SWE-rebench~\cite{SWE-rebench}, SWE-Perf~\cite{swe-perf}, SWE-MERA~\cite{swe-mera}, SWE-bench-Live~\cite{swebench-live}, SWE-Bench++~\cite{wang2025swe}, SWE-CI~\cite{SWE-CI}},
            text width=17cm]
        ],
    ],
    [Benchmark Evolution
        [Enhancement
            [Clarity of Issue Description
                [{SWE-bench Verified~\cite{swebenchverified}, SWE-bench Lite-S~\cite{agentless}, SWE-Bench+~\cite{swebench_plus}, SPICE~\cite{oliva2025spice}}]
            ],
            [Coverage of Unit Tests
                [{SWE-bench Verified~\cite{swebenchverified}, SWE-Bench+~\cite{swebench_plus}, UTBoost~\cite{utboost}, SPICE~\cite{oliva2025spice}, SWE-ABS~\cite{yu2026swe}, SWE-Mutation~\cite{SWE-Mutation}}]
            ],
        ],
        [Extension
            [Multimodal
                [{SWE-bench Multimodal~\cite{swebenchmutimodal}, Visual SWE-bench~\cite{CodeV}, OmniGIRL~\cite{omnigirl}}]
            ],
            [Multilingual
                [{SWE-bench-java~\cite{swebench_java}, Multi-SWE-bench~\cite{mutiswebench}, SWE-PolyBench~\cite{swepoly}, OmniGIRL~\cite{omnigirl}, SWE-bench Multilingual~\cite{swebenchmutilingual}, SwingArena~\cite{xu2026swingarena}, Rust-SWE-bench~\cite{xiang2026evaluating}, SWE-Bench Mobile~\cite{tian2026swe}, SWE-Bench++~\cite{wang2025swe}}]
            ],
            [Timeliness
                [{LiveSWEBench~\cite{liveswebench2025}, SWE-bench-Live~\cite{swebench-live}, SWE-MERA~\cite{swe-mera}}]
            ],
            [Other Issue Types
                [{FEA-Bench~\cite{FEABench}, GSO~\cite{gso}, SWE-Perf~\cite{swe-perf}, NoCode-bench~\cite{nocode-bench}, SWE-fficiency~\cite{ma2025swe}, SWE-EVO~\cite{thai2025swe}, SWE-Refactor~\cite{xu2026swe}, FeatureBench~\cite{zhou2026featurebench}, SWE-CI~\cite{SWE-CI}, SWE-STEPS~\cite{shastry2026beyond}, SWE-Chain~\cite{lam2026swe}}]
            ],
            [Subtasks
                [{LocBench~\cite{locagent}, SWT-Bench~\cite{SWT_bench}, TestGenEval~\cite{testgeneval}, TDD-Bench~\cite{tddbench}, MULocBench~\cite{mulocbench}, ContextBench~\cite{li2026contextbench}, SWE Context Bench~\cite{zhu2026swe}}]
            ],
        ],
    ],
    [Evaluation Metrics
        [Execution-based
            [{Applied\%, Resolved\%, Pass@k, Reproduction Success Rate, Delta Change Coverage, Speedup, Opt@k}, text width=17cm]
        ],
        [Match-based
            [{File Matched Rate, Function Matched Rate, Top-k, MAP, MRR, Precision, Recall, F1}, text width=17cm]
        ],
        [Statistics-based
            [{\#Token, \$Cost, Time}, text width=17cm]
        ],
    ],
  ]
\end{forest}
\end{adjustbox}
\caption{Taxonomy of issue resolution benchmarks.} 
\label{fig:benchmark_overview}
\end{figure*}

\subsection{Benchmark Construction}\label{sec:benchmark_construction}

As shown in Table~\ref{tab:benchmarks} (see Appendix~\ref{ap: bench_stat}), based on the curation process of benchmark construction, we categorize benchmarks into \textit{manual construction} and \textit{automatic construction}.
\revReplace{R3.3}{We also provide statistics on the number of repositories, number of instances, programming languages, system kind, domain, issue type, release time, and data sources involved in all benchmarks.
This reveals clear cross-benchmark patterns.
Early benchmarks concentrate on bug fixing over Python libraries, frameworks, and developer tools from the data science, machine learning, and web development domains, reflecting their shared SWE-bench lineage, while recent benchmarks diversify toward more programming languages, more system kinds such as end-user applications and heterogeneous open-source projects, broader domains such as mobile and enterprise software, and richer issue types including feature addition, performance optimization, and refactoring.
Section~\ref{sec:benchmark_evolution} further details the purpose of each benchmark along these evolution directions, and Section~\ref{sec:benchmark_tradeoff} provides guidance on selecting appropriate benchmarks.}

\subsubsection{Manually Constructed Benchmarks}

\begin{figure}[h]
    \centering
    \includegraphics[width=0.7\linewidth]{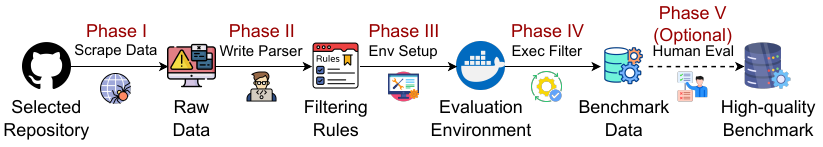}
    \caption{A typical pipeline of manual construction of issue resolution-related benchmarks.}
    \label{fig:benchmark_pipeline}
\end{figure}

The manual construction of issue resolution-related benchmarks typically involves five phases, as illustrated in Figure~\ref{fig:benchmark_pipeline}.
First, well-maintained repositories are carefully selected from code hosting platforms such as GitHub, and all pull request (PR) data is collected. 
Second, the raw data are processed with handcrafted parsing rules to filter out instances irrelevant to the target tasks. 
Third, the execution environments of the selected projects are manually configured by consulting the corresponding documentation to ensure their executability. 
Fourth, all test cases are executed before and after applying the patches associated with each PR, and instances that demonstrate consistent changes in test outcomes are preserved as valid benchmark data.
Finally, developers may optionally annotate factors such as the clarity of problem descriptions and the coverage of test cases, and use these criteria to filter out a high-quality subset of data for evaluation.

In the early stages of the issue resolution task, nearly all benchmarks relied on manual construction.
SWE-bench~\cite{swebench} is the first benchmark that implements the construction pipeline illustrated in Figure~\ref{fig:benchmark_pipeline}. 
It collects approximately 90,000 PRs from 12 carefully selected repositories, which, after rule-based filtering and execution-based validation, yield 2,294 instances for evaluation. 
Subsequent efforts generally take SWE-bench as a reference paradigm, enhancing and extending its methodology (as discussed in Section~\ref{sec:benchmark_evolution}) to incorporate multimodal~\cite{swebenchmutimodal, CodeV, omnigirl}, multilingual~\cite{swebench_java, swebenchmutilingual, mutiswebench}, and diverse code repositories~\cite{swebench-live, swepoly, liveswebench2025, SWEbenchPro}, thereby broadening the scope of issue resolution evaluation.

\subsubsection{Automatically Constructed Benchmarks}

The automatic construction of issue resolution benchmarks aims to rapidly collect large-scale issue resolution data to compensate for the limited timeliness of static benchmarks. 
In particular, existing methods primarily target environment setup, which is the most labor-intensive process. 
For example, SWEE/SWA-Bench~\cite{SWA} employs a SetupAgent that automates environment setup through three phases—Command Extraction, Iterative Testing and Improvement, and Validation—constructing installation commands, testing commands, and result parsers. 
SWE-bench-Live~\cite{swebench-live} adopts an end-to-end agentic workflow, RepoLaunch, which reads documentation, iteratively executes and debugs commands, and refines strategies based on feedback to complete environment setup.
Similarly, SWE-rebench~\cite{SWE-rebench} and SWE-CI~\cite{SWE-CI} build automated pipelines that continuously collect and validate tasks from GitHub repositories and continuous-integration histories, respectively.
Such automated construction methods can also be used to collect training datasets to train LLMs, which is further discussed in Section~\ref{sec:training_data}.C.

\subsection{Benchmark Evolution}\label{sec:benchmark_evolution}

Based on the evolution direction of existing benchmarks, we discuss the evolution of issue resolution benchmarks along two main routes: enhancement and extension.

\subsubsection{Benchmark Enhancement}
Benchmark enhancement aims at the process of filtering high-quality subsets or repairing existing benchmarks through manual or automated approaches, focusing on aspects such as the clarity of issue descriptions or the coverage of unit tests, in order to enable reliable evaluation.

\textbf{A. Clarity of Issue Description}

In benchmarks, issue descriptions are typically derived from natural language explanations that developers provide in real-world pull requests. 
However, because natural language is often ambiguous and vague, LLMs may fail to accurately capture the requirements, which in turn hinders their ability to solve the tasks.

Existing studies have investigated methods for identifying unclear issue descriptions. 
OpenAI~\cite{swebenchverified} employs 93 experienced software engineers to annotate the clarity of issue descriptions in 1,699 SWE-bench samples and filters instances based on inter-annotator agreement. 
Xia et al.~\cite{agentless} manually inspect issue descriptions to identify misleading or underspecified statements, filtering 252 high-quality tasks from SWE-bench Lite to create SWE-bench Lite-S. 
Aleithan et al.~\cite{swebench_plus} conduct an empirical study on SWE-bench and found that solutions to 32.67\% of tasks are explicitly provided in the issue descriptions. 
Building on these findings, Oliva et al.~\cite{oliva2025spice} propose SPICE, the first automated annotation method, which constructs rationale-informed prompts from OpenAI’s annotations and uses prompt engineering to automate clarity labeling.

\textbf{B. Coverage of Unit Tests}

In real-world software repositories, unit tests are often written for specific issues. 
As a result, the unit tests used to evaluate solution correctness may be overly specified or excessively broad, and in some cases, even unrelated to the underlying problem~\cite{oliva2025spice, swebenchverified}. 
This can lead to situations where correct solutions are rejected or incorrect solutions are accepted.
\revReplace{R1.8}{Plausible patches accepted by weak test suites can behave differently from ground-truth patches and thus inflate measured resolution rates~\cite{patchdiff} (see Section~\ref{sec:evaluation_oriented_studies}).}

Recent work has focused on automating the identification of issues in test case coverage. 
Specifically, OpenAI~\cite{swebenchverified} conducts manual annotation to identify overly specific test cases and, based on this effort, constructs SWE-bench Verified. 
Aleithan et al.~\cite{swebench_plus} study the performance of the state-of-the-art software engineering agent SWE-agent on SWE-bench and found that 31.08\% of the test cases were excessively broad.
Yu et al.~\cite{utboost} employ differential testing to detect discrepancies between ground-truth patches and candidate patches.
These discrepancies are then used as context to guide the LLM in automatically enhancing the tests, thereby reinforcing the unit tests in SWE-bench. 
Oliva et al.~\cite{oliva2025spice} leverage Aider to build an autonomous agent capable of exploring software repositories to assess the coverage of test cases automatically.
Beyond coverage, SWE-ABS~\cite{yu2026swe} adversarially strengthens test suites to expose inflated success rates, while SWE-Mutation~\cite{SWE-Mutation} applies mutation analysis to assess the reliability of the test suites.

\subsubsection{Benchmark Extension}

Benchmark extension aims to broaden the coverage of issue resolution benchmarks across multimodal inputs, multiple programming languages, time spans, issue types, and related subtasks, in order to achieve comprehensive and thorough evaluation.

\textbf{A. Multimodal}

Issue descriptions presented solely in textual~\cite{swebench,swebench_java,swebenchverified} form are insufficient to capture the diverse application scenarios in software engineering.
For instance, domains such as front-end development and game development often rely on visual elements, such as images, to express requirements. 
To broaden the coverage of issue resolution tasks, researchers have extended the current benchmark framework by incorporating multimodal problem descriptions as inputs.
Yang et al.~\cite{swebenchmutimodal} select four categories of tasks in JavaScript repositories, including diagramming, interactive mapping, syntax highlighting, and web frameworks, and construct SWE-bench Multimodal following the construction pipeline of SWE-bench~\cite{swebench}. 
Zhang et al.~\cite{CodeV} and Guo et al.~\cite{omnigirl} select tasks with image-based descriptions from SWE-bench and GitHub to construct Visual SWE-bench and OmniGIRL, respectively.

\textbf{B. Multilingual}

Early issue resolution benchmarks almost exclusively focus on tasks from Python repositories~\cite{swebench,swebenchverified,FEABench,SWA,swebench_plus}, which results in insufficient evaluation of large language models across the diverse software ecosystem. To address this limitation, researchers extend issue resolution tasks to multiple programming languages to enable broader evaluation.
Zan et al.~\cite{swebench_java} first migrate SWE-bench to the Java language, after which the Bytedance team further extends it to ten additional programming languages, constructing Multi-SWE-bench~\cite{mutiswebench}. 
Similarly, Rashid et al.~\cite{swepoly}, Guo et al.~\cite{omnigirl}, and Khandpur et al.~\cite{swebenchmutilingual} also construct multilingual issue resolution benchmarks named SWE-PolyBench~\cite{swepoly}, OmniGIRL~\cite{omnigirl}, and SWE-bench Multilingual~\cite{swebenchmutilingual}, with the specific programming languages involved summarized in Table~\ref{tab:benchmarks}.
Recent efforts further broaden language and platform coverage, including SwingArena~\cite{xu2026swingarena}, Rust-SWE-bench~\cite{xiang2026evaluating}, SWE-Bench Mobile~\cite{tian2026swe}, and SWE-Bench++~\cite{wang2025swe}.

\textbf{C. Timeliness}

Static issue resolution benchmarks~\cite{swebench,mutiswebench,swebenchmutimodal,swebenchmutilingual,swepoly,omnigirl,swebenchverified} cover only a limited number of repositories and remain unchanged after their release.
As more advanced LLMs are trained, the data in these benchmarks is likely to have already been exposed, leading to risks of overfitting and data contamination. 
\revReplace{R1.8}{For example, state-of-the-art models can infer buggy file paths from issue descriptions alone~\cite{liang2025Illusion} (see Section~\ref{sec:evaluation_oriented_studies}).}
To address this problem, researchers construct dynamic, continuously updated benchmarks.
LiveSWEBench~\cite{liveswebench2025} and SWE-MERA~\cite{swe-mera} use manually designed pipelines to automatically collect the latest issue data from target repositories. 
In contrast, SWE-bench-Live~\cite{swebench-live} designs a fully agentic workflow to mitigate performance saturation caused by benchmark obsolescence.

\textbf{D. Other Issue Types}

SWE-bench and its variants primarily focus on evaluating end-to-end bug-fixing tasks~\cite{swebench,swebenchmutilingual,swebenchmutimodal,mutiswebench,swebenchverified}. 
However, in real-world software repositories, issues also include non-bug-fixing tasks such as efficiency optimization. 
Therefore, exploring the performance of LLMs on non-bug-fixing tasks is crucial.
\revReplace{R3.18}{Researchers now extend benchmarks to four categories of non-bug-fixing tasks. For feature addition tasks~\cite{FEABench,nocode-bench,zhou2026featurebench} and performance optimization tasks~\cite{gso,swe-perf,ma2025swe}, benchmarks evaluate agents' ability to implement new functionality and improve program efficiency, respectively. For code refactoring tasks, SWE-Refactor~\cite{xu2026swe} evaluates repository-level tasks from real-world Java pull requests, requiring agents to preserve program behavior while improving code structure. Long-term software evolution tasks~\cite{shastry2026beyond,lam2026swe,thai2025swe} further evaluate agents across sequences of interdependent repository changes.}

\textbf{E. Subtasks}

As mentioned in Section~\ref{sec:2.2}, issue resolution is a multi-stage task.
However, most benchmarks focus solely on end-to-end accuracy evaluation and lack a fine-grained assessment of software engineering agentic systems on the subtasks of each stage. 
To address this challenge, researchers construct independent benchmarks for subtasks such as issue reproduction and issue localization to gain fine-grained insights into these subtasks.
Table~\ref{tab:sub_task_bench} (see Appendix~\ref{ap: bench_stat_sub}) presents the statistics of benchmarks that target the subtasks of issue resolution.
For issue localization, LocBench~\cite{locagent} collects four categories of issue data, including bug reports, feature requests, security issues, and performance issues, and applies traditional fault localization evaluation methods~\cite{li2026contextbench,zhu2026swe} to assess the fine-grained performance of LLMs in localizing different types of issues. 
MULocBench~\cite{mulocbench} extends the localization problem beyond source code to include other non-code files such as commits, comments, configurations, and documentation, offering greater diversity in terms of issue types, root causes, localization scopes, and file categories.
For issue reproduction, SWT-Bench~\cite{SWT_bench}, TestGenEval~\cite{testgeneval}, and TDD-Bench~\cite{tddbench} all derive instances from SWE-bench and, through different filtering criteria, evaluate the ability of LLMs to reproduce issues in SWE-bench.

\revReplace{R1.9}{Beyond enabling fine-grained assessment, performance on these subtasks is closely tied to end-to-end resolution outcomes.
Existing studies have demonstrated that dedicated localization methods help improve the end-to-end resolution rates when their localization results are fed into downstream patch generation pipelines~\cite{locagent, CoSIL, OrcaLoca}.
Additionally, generated reproduction tests are widely used to filter and rerank candidate patches, improving the precision of submitted patches~\cite{agentless, SWT_bench, e-otter}.
Nevertheless, stronger subtask performance is helpful but not sufficient for end-to-end success, since correct localization does not guarantee a correct patch, and weak reproduction tests may accept plausible but behaviorally incorrect patches~\cite{patchdiff}.
Subtask benchmarks should therefore be regarded as diagnostic instruments that decompose end-to-end scores and expose stage-level bottlenecks, rather than substitutes for end-to-end evaluation.}

\subsection{Evaluation Metrics}\label{sec:eval_metrics}

Evaluation metrics play an important role in both end-to-end issue resolution evaluation and fine-grained stage-specific evaluation, as they quantify the issue resolution capabilities of software engineering agentic systems.
\revReplace{R2.6}{Following the taxonomy construction procedure in Section~\ref{sec:taxonomy_construction}, two authors extract and normalize the metrics used in surveyed papers, including their names, definitions, and target stages.
We retain recurring and widely used metrics for cross-study comparison.
Specifically, we identify all metrics used in the surveyed papers, retain those reported in more than 5\% of the papers, and group the retained metrics into execution-based, match-based, and statistics-based metrics.}
We further discuss these three types of metrics in Appendix~\ref{ap:metrics}.

\looseness=-1
\revReplace{R2.6}{Across the surveyed papers, we observe that evaluation relies almost exclusively on automatic assessment, and that the choice of metrics is highly correlated with the evaluation target.
Specifically, end-to-end evaluation is entirely execution-based, and 76.4\% of the surveyed papers report Resolved\%, indicating that it serves as the most common metric for the issue resolution task.
Beyond Resolved\%, metric selection is further differentiated along two dimensions.
First, it depends on the task stage. 
The localization stage is evaluated exclusively with match-based metrics.
For example, Top-k, MAP, and MRR for techniques that output explicit localization results, and File/Function Matched Rate derived from the final patch for agentic systems that do not expose such intermediate results.
The reproduction stage uses Reproduction Success Rate and Delta Change Coverage to measure whether a generated test reproduces the issue and how well it covers the ground-truth fix.
Second, it depends on the issue type.
While bug fixing and feature addition tasks are assessed with the correctness-oriented Resolved\%, performance issues additionally require Speedup and Opt@k, which extend execution-based evaluation from functional correctness to runtime efficiency.
Statistics-based metrics are task-agnostic and serve only as complements to effectiveness metrics, reflecting that current evaluation practice prioritizes effectiveness over efficiency.}
\revReplace{R3.13}{Manual patch assessment is common in classical APR evaluations because passing the available tests does not always indicate semantic correctness~\cite{chatrepair,bouzenia2024repairagent}.
Accordingly, prior APR studies~\cite{chatrepair} manually inspect each test-passing patch against the developer patch or intended behavior to determine whether it is actually correct, and typically report the number of correct patches together with the correct-to-plausible ratio~\cite{chatrepair,bouzenia2024repairagent}.
Issue resolution benchmarks face the same oracle limitation, since issue-specific tests may not identify some behavioral differences~\cite{patchdiff}. However, the diversity of issue types further complicates manual patch assessment.
For example, feature-addition issues may involve about 180 lines of edits on average~\cite{nocode-bench}, making it more difficult to manually determine whether such a generated patch fully implements the intended functionality without introducing unintended behavior than bug-fixing tasks~\cite{chatrepair}.
Additionally, the scale of hundreds of instances makes per-patch manual assessment impractical for routine benchmark evaluation. 
Manual assessment therefore appears mainly in complementary roles, such as benchmark curation~\cite{swebenchverified} and empirical comparisons between generated and developer patches~\cite{swebench_plus,chen2025evaluating}.}

\subsection{Benchmark Validity and Comparability}\label{sec:benchmark_tradeoff}

\revReplace{R1.7}{
To clarify what an evaluation score represents, we examine benchmark validity along three dimensions: realism versus control, contamination and timeliness, and comparability across benchmark families.
Specifically, benchmarks built from raw PRs are closer to real development settings, but this realism costs control, since issue descriptions may leak their own solutions~\cite{swebenchverified}, and overly broad test suites may let plausible but incorrect patches pass~\cite{swebench_plus,patchdiff}.
Filtered benchmarks (e.g., SWE-bench Verified~\cite{swebenchverified}) alleviate this problem through human annotation, but they discard the harder, more ambiguous instances that made the benchmark realistic.
Thus, we recommend using raw benchmarks comprising real-world, minimally filtered issues for robustness evaluation, and using filtered benchmarks for reliable comparison across models and methods.

Moreover, contamination poses another threat to benchmark validity, because benchmark instances are mainly collected from public repositories, where both the issue and its ground-truth fix remain openly accessible.}
\revReplace{R1.7/R3.13}{Specifically, such contamination arises through two main aspects.
The first operates at evaluation time.
An agent granted internet access or the full commit history of the repository could directly retrieve the known fix instead of reasoning about the issue.
To overcome this, existing benchmarks typically execute agents in isolated environments that provide only a pre-fix snapshot of the repository, without access to the internet or the subsequent commit history~\cite{swebench}.
The second is training-data contamination, where the fixed instances of a benchmark gradually enter the training corpora of newer models~\cite{liang2025Illusion}. 
For example, prior studies~\cite{liang2025Illusion,prathifkumar2025does} show that models can identify buggy file paths from issue descriptions without a repository environment.
Such contamination cannot be excluded by isolation and inevitably grows as the benchmark ages, which leads to a trade-off between static and dynamic benchmarks.
Static benchmarks support reproducible and controlled comparison, but their fixed instances face growing training-data contamination risk over time.
Dynamic benchmarks~\cite{swebench-live,liveswebench2025,swe-mera} alleviate this risk by continuously collecting issues created after training cutoffs, but they are hard to sustain, since repeated manual construction and validation are prohibitively costly and automated construction struggles to guarantee instance quality.
Thus, we recommend static benchmarks for rigorous method comparison under a stable protocol, and dynamic benchmarks for assessing the genuine capability of the latest models where contamination is the primary concern.

These differences also undermine comparability across benchmark families.
Influenced by factors such as programming language and repository difficulty, resolution rates on different families of benchmarks do not support strict absolute comparison~\cite{martinez2025dissecting}.
However, the relative rankings of different models remain a useful reference, since robustness across languages and repositories is an essential dimension of models' capability~\cite{chen2025evaluating, liu2025empirical}.
}
\revReplace{R1.8}{Section~\ref{sec:evaluation_oriented_studies} reviews in detail the empirical studies that provide systematic evidence for these evaluation concerns.}

\answerRQ{\revReplace{R1.11/R3.3}{
Manual curation with execution-based validation remains the common practice of benchmark construction, while benchmarks keep evolving toward multilingual, multimodal, dynamic, and subtask-level evaluation, with execution-based Resolved\% as the main metric.
Since no single benchmark is universally valid, benchmarks should first be scoped by the target language, system kind, domain, and issue type, and then selected by evaluation purpose.
For example, filtered benchmarks for reliable comparison, dynamic benchmarks for the latest models, and subtask benchmarks for fine-grained diagnosis.
}}

\vspace{-0.5em}
\section{Techniques}\label{sec:technique}

\begin{figure*}[htbp]
  \centering 
  \begin{adjustbox}{width=\textwidth}
  \begin{forest}
for tree={
    rounded corners,
    child anchor=west,
    parent anchor=east,
    grow'=east,  
    text width=3cm,%
    draw=softPurple,
    anchor=west,
    node options={align=center},
    edge path={
      \noexpand\path[\forestoption{edge}]
        (.child anchor) -| +(-5pt,0) -- +(-5pt,0) |-
        (!u.parent anchor)\forestoption{edge label};
    },
    where n children=0{text width=16cm}{}
  },
  [Techniques
    [Scaffold/Method Design
        [End-to-End Scaffold
            [Agent-Based
                [{MAGIS~\cite{MAGIS}, AutoCodeRover~\cite{AutoCodeRover}, SWE-agent~\cite{SWEAgent}, CodeR~\cite{CodeR}, MASAI~\cite{MASAI}, Alibaba LingmaAgent~\cite{repounderstander}, OpenHands CodeAct~\cite{openhands}, CodeXGraph~\cite{CodeXGraph}, SpecRover~\cite{SpecRover}, MarsCode Agent~\cite{marscode}, HyperAgent~\cite{hyperagent}, SuperCoder~\cite{supercoder}, SWE-Search~\cite{SWE-Search}, Infant Agent~\cite{infant}, Learn-by-interact~\cite{Learn-by-interact}, DARS~\cite{DARS}, InfantAgent-Next~\cite{InfantAgent-Next}, OpenHands-Versa~\cite{OpenHands-Versa}, EXPEREPAIR~\cite{experepair}, Agent KB~\cite{AgentKB}, Prometheus~\cite{Prometheus}, SWE-Exp~\cite{SWE-EXP}, SWE-Debate~\cite{SWE-Debate}, TRAE~\cite{trae2025,tian2026agentbased}, SE-Agent~\cite{SE-Agent}, Lita~\cite{lita}, Lingxi~\cite{yang2025lingxi}, InfCode~\cite{li2025infcode}, Confucius Code Agent~\cite{wang2025confucius}, Agyn~\cite{benkovich2026agyn}, Debug2Fix~\cite{garg2026debug2fix}, SGAgent~\cite{zhang2026sgagent}, SWE-Adept~\cite{he2026swe}, iSWE Agent~\cite{ganhotra2026resolving}, REAgent~\cite{kuang2026reagent}, Agent-CoEvo~\cite{li2026beyond}}],
            ],
            [Pipeline-Based
                [{Agentless~\cite{agentless}, RepoGraph~\cite{RepoGraph}, SWESynInfer~\cite{SWE-GPT}, SWE-Fixer~\cite{SWE-Fixer}, Agentless-Mini~\cite{SWE-RL}, CodeV~\cite{CodeV}, CodeMonkeys~\cite{CodeMonkeys}, PatchPilot~\cite{PatchPilot}, KGCompass~\cite{KGCompass}, Jiang et al.~\cite{LCLM}, CGM~\cite{CGM}, GUIRepair~\cite{GUIRepair}, SemAgent~\cite{SEMAgent}, Nemotron-Cortexa~\cite{Nemotron-Cortexa}, SynFix~\cite{SynFix}, SIADAFIX~\cite{cao2025siadafix}, TDFlow~\cite{han-etal-2026-tdflow}, Think-Search-Patch~\cite{xiong2025think}, SVRepair~\cite{tang2026svrepair}, RepoRepair~\cite{pan2026reporepair}, RAIM~\cite{liu2026architecture}, ARISE~\cite{seddik2026arise}}],
            ],
        ],
        [Scaffold Improvement
            [Memory
                [{EXPEREPAIR~\cite{experepair}, Agent KB~\cite{AgentKB}, SWE-Exp~\cite{SWE-EXP}, TOM-SWE~\cite{zhou2025tom}, EET~\cite{guo2026eet}, Confucius Code Agent~\cite{wang2025confucius}, MemGovern~\cite{wang2026memgovern}, FailureMem~\cite{ma2026failuremem}}]
            ],
            [Context Management
                [{AgentDiet~\cite{xiao2026reducing}, ConRAD~\cite{li2026outcome}, SWE-Pruner~\cite{wang2026swe}, Compressing Code Context~\cite{jia2026compressing}, CODESTRUCT~\cite{kim2026codestruct}, SWE-Edit~\cite{SWE-Edit}, CodeScout~\cite{suri2026codescout}}]
            ],
            [Self-Evolution
                [{SICA~\cite{robeyns2025a}, SE-Agent~\cite{SE-Agent}, SAGE~\cite{hayashi2025self}, Live-SWE-agent~\cite{xia2025live}, MemCoder~\cite{deng2026your}, HGM~\cite{HGM}, DGM~\cite{DGM}}]
            ]
        ],
        [Single-Phased Methods
            [Issue Localization
                [{DEVLoRe~\cite{feng2025integrating}, BLAZE~\cite{BLAZE}, OrcaLoca~\cite{OrcaLoca}, BugCerberus~\cite{BugCerberus}, LocAgent~\cite{locagent}, CoSIL~\cite{CoSIL}, SweRank~\cite{SweRank}, CoRet~\cite{coret}, SACL~\cite{SACL}, Meta-RAG~\cite{meta-rag}, RepoSearcher~\cite{RepoSearcher}, CoRNStack~\cite{CoRNStack}, RepoMem~\cite{wang2025improving}, HiLoRM~\cite{zhang2025hierarchical}, SweRank+~\cite{reddy2025swerank+}, RepoNavigator~\cite{zhang2025one}, GraphLocator~\cite{liu2025graphlocator}, RGFL~\cite{sepidband2026rgfl}, FuseSearch~\cite{xu2026learning}, RPG-Encoder~\cite{luo2026closing}, CodeScout~\cite{sutawika2026codescout}, LogicLoc~\cite{xu2026neurosymbolic}, BLAgent~\cite{mamun2026blagent}}]
            ],
            [Issue Reproduction
                [{Libro~\cite{Libro}, AEGIS~\cite{AEGIS}, BRT~\cite{BRT}, EvoCoder~\cite{lin2024llms}, Otter~\cite{Otter}, Issue2Test~\cite{issue2test}, AssertFlip~\cite{AssertFlip}, e-Otter++~\cite{e-otter}, BLAST~\cite{BLAST}, SWE-Tester~\cite{soni2026swe}, Echo~\cite{fei2026echo}, iCoRe~\cite{wang2026icore}, e-Otter++ for Java~\cite{ahmed2026reproduction}}]
            ],
            [Patch Selection (Reward Model)
                [{Agentic Rubrics~\cite{raghavendra2026agentic}, SWE-Replay~\cite{ding2026swe}, Critic Rubrics~\cite{wang2026rubric}, SWE-PRM~\cite{SWE-PRM}, GRM~\cite{huang2026beyond}, SWE-RM~\cite{shum2026swerm}, RTV~\cite{kim2026scaling}}]
            ]
        ],
    ],
    [Learning Strategies
        [Data Preparation
            [Real-world Data
                [{SWE-bench(train)~\cite{swebench}, SWE-Gym~\cite{SWE-Gym}, SWE-bench-extra~\cite{swe-bench-extra}, SWE-Fixer~\cite{SWE-Fixer}, Multi-SWE-RL~\cite{mutiswebench}, SWE-rebench~\cite{SWE-rebench}, SWE-Dev(a)~\cite{swe-dev_sjtu}, Skywork-SWE~\cite{Skywork-SWE}, SWE-Universe~\cite{SWE-Universe}, Scale-SWE-Data~\cite{zhao2026immersion}, SWE-rebench V2~\cite{badertdinov2026swe}, SWE-Next~\cite{liang2026swe}, STITCH~\cite{team2026yet}}]
            ],
            [Synthetic Data
                [{R2E~\cite{R2E}, SWE-Synth~\cite{SWE-Synth}, R2E-Gym~\cite{r2e_gym}, SWE-Flow~\cite{zhang2025synthesizing},SWE-smith~\cite{SWE-smith}, SWE-Dev(b)~\cite{swe-dev_thu}, SWE-Mirror~\cite{swe-mirror}, Hybrid-Gym~\cite{xie2026hybrid}}]
            ],
            [Automated Environment Infrastructure
                [{SWE-Factory~\cite{guo2025swe}, EvoConfig~\cite{guo2026evoconfig}, SWE-MiniSandbox~\cite{yuan2026swe}, SWE-Hub~\cite{zeng2026swe}, SWE-Universe~\cite{SWE-Universe}, MEnvAgent~\cite{guo2026menvagent}, daVinci-Env~\cite{fu2026davinci}}]
            ]
        ],
        [Training
            [SFT
                [{SWE-GPT~\cite{SWE-GPT}, ReSAT~\cite{ReSAT}, SWE-Fixer~\cite{SWE-Fixer}, SoRFT~\cite{SoRFT}, SWE-Reasoner~\cite{swe-resoner}, Co-PatcheR~\cite{Co-PatcheR}, Satori-SWE~\cite{Satori-SWE}, MCTS-Refine~\cite{MCTS-Refined}, RepoForge~\cite{RepoForge}, SWE-Gym-32B~\cite{SWE-Gym}, R2E-Gym-32B~\cite{r2e_gym}, SWE-agent-LM-32B~\cite{SWE-smith}, SWE-Dev-32B~\cite{swe-dev_thu}, Skywork-SWE~\cite{Skywork-SWE}, SWE-Mirror-LM-32B~\cite{swe-mirror}, Devstral-Small~\cite{devstal_small}, Kimi-Dev~\cite{kimi-dev}, CWM~\cite{CWM}, EntroPO~\cite{yu2025building}, SWE-Swiss~\cite{SWESwiss}, FrogBoss~\cite{sonwane2025bugpilot}, SWE-Play~\cite{zhu2025training}, SWE-Compressor~\cite{liu2025context}, SWE-Lego~\cite{tao2026swe}, daVinci-Dev~\cite{zeng2026davinci}, SERA~\cite{shen2026sera}, SWE-Spot~\cite{peng2026swe}, SWE-World~\cite{SWE-World}, SWE-Master~\cite{song2026swe}, SWE-Protégé~\cite{kon2026swe}, SWE-Fuse~\cite{wen2026swe}, SWE-HERO~\cite{ludwig2026swe}, SWE-AGILE~\cite{lian2026swe}, SWE-TRACE~\cite{SWE-TRACE}, BoostAPR~\cite{li2026boostapr}, HHD~\cite{wang2026hindsight}}]
            ],
            [RL
                [{SWE-RL~\cite{SWE-RL}, SoRFT~\cite{SoRFT}, SWE-Reasoner~\cite{swe-resoner}, SEAlign~\cite{SEAlign}, Satori-SWE~\cite{Satori-SWE}, Agent-RLVR~\cite{Agent-RLVR}, RepoForge~\cite{RepoForge}, Golubev et al.~\cite{golubev2025training}, SWE-Dev-32B~\cite{swe-dev_thu}, Devstral-Small~\cite{devstal_small}, Kimi-Dev~\cite{kimi-dev}, CWM~\cite{CWM}, EntroPO~\cite{yu2025building}, DeepSWE~\cite{DeepSWE}, SWE-Swiss~\cite{SWESwiss}, SSR~\cite{wei2025toward}, SWE-RM~\cite{shum2026swerm}, SWE-Master~\cite{song2026swe}, SWE-Protégé~\cite{kon2026swe}, SWE-Fuse~\cite{wen2026swe}, SWE-AGILE~\cite{lian2026swe}, SWE-TRACE~\cite{SWE-TRACE}, BoostAPR~\cite{li2026boostapr}, SWE-World~\cite{SWE-World}}]
            ],
        ],
    ],
  ]
\end{forest}
\end{adjustbox}
\caption{Taxonomy of issue resolution techniques.} 
\label{fig:techniques_overview}
\end{figure*}

\revReplace{R3.12}{To answer \textbf{RQ2}, we} present the evolution of issue resolution techniques before and after the paradigm shift along two dimensions, i.e., scaffold/method design and learning strategy, as shown in Figure~\ref{fig:techniques_overview}.
\revReplace{R2.7/R2.11}{The two dimensions are distinguished by whether a technique designs the scaffold around a backbone LLM or trains the underlying model itself.}

\subsection{Scaffold/Method Design}\label{sec:scaffold_design}

Because the context window of LLMs is limited~\cite{deepseekr1,openai2024gpt4technicalreport}, it is infeasible to feed all code from a repository into an LLM at once to localize focal methods or generate patches.
Therefore, researchers design scaffolds for LLMs to solve the issue resolution task or its subtasks.
Before February 2025, researchers favored designing complex scaffolds for LLMs to improve performance on issue resolution.
According to the scope of the target task, we preliminarily categorize existing scaffolds/methods into end-to-end scaffolds and single-phase methods.
\revReplace{R3.18}{In addition, some existing works improve existing scaffolds with scaffold-agnostic mechanisms, which we categorize as scaffold improvement.}

\subsubsection{End-to-End Scaffold}\label{sec:e2e_scaffold}

\revReplace{R1.6/R3.5}{Following the criterion defined in Section~\ref{sec:2.2}, we categorize end-to-end scaffolds into agent-based and pipeline-based according to whether the control flow is determined by the LLM at inference time or predefined by human designers.}
Agent-based scaffolds and pipeline-based scaffolds both generally follow the common work phases of Section~\ref{sec:2.2}, which Appendix~\ref{ap:issue_resolution_overview} describes in detail.
Agent-based scaffolds~\cite{benkovich2026agyn, garg2026debug2fix, zhang2026sgagent, he2026swe, ganhotra2026resolving, kuang2026reagent, li2026beyond, li2025infcode, lita}, such as SWE-agent~\cite{SWEAgent} and OpenHands~\cite{openhands}, equip one or more agents with tools to autonomously navigate repositories and generate patches, offering high flexibility and autonomy. 
However, constrained by the planning, decision-making, and instruction-following capabilities of the underlying LLM~\cite{CoSIL, SEAlign}, agent behavior may deviate from the intended pipeline, potentially leading to task failure or system crashes.
Pipeline-based scaffolds~\cite{cao2025siadafix, han-etal-2026-tdflow, xiong2025think, tang2026svrepair, pan2026reporepair, liu2026architecture, seddik2026arise}, such as Agentless~\cite{agentless} and PatchPilot~\cite{PatchPilot}, instead execute human-designed, fixed phases for localization, repair, validation, and patch selection sequentially.
Each phase is explicitly controlled, resulting in more stable and reliable execution~\cite{PatchPilot} but lower autonomy.
\revReplace{R1.6/R3.5}{For scaffolds that combine both designs, such as pipelines whose individual phases internally contain bounded feedback loops, we classify them by their global control flow.}

Table~\ref{tab:e2e_scaffolds} (see Appendix~\ref{ap:e2e_scaffold_overview}) presents the chronological development of existing end-to-end scaffolds, highlighting their designs across the phases of the pipeline described in Appendix~\ref{ap:issue_resolution_overview}, as well as their unique characteristics \revReplace{R3.6}{and performance-cost trade-off.
Since the performance score and cost are self-reported, we draw quantitative comparisons only between methods evaluated on the same benchmark with the same backbone.
Under such matched settings, pipeline-based methods deliver comparable effectiveness at lower and more predictable cost.
For example, Agentless~\cite{agentless} resolves more instances than the agent-based MASAI~\cite{MASAI} at roughly one-third of the cost under the same setting.
This cost difference arises because a pipeline bounds the number of LLM calls in each phase, whereas an agent trajectory accumulates context over multiple rounds of tool interactions, making its cost higher and more variable.
Beyond such matched pairs, the highest reported resolved rates come from agent-based scaffolds~\cite{experepair, yang2025lingxi, wang2025confucius}.
However, these systems also adopt the strongest and most recent backbones, so scaffold type and backbone capability are confounded in the reported numbers, and we therefore do not attribute the effectiveness gap to the scaffold design alone.
In practice, pipeline-based methods are preferable under tight budgets or with weaker backbones, where the predefined workflow compensates for limited planning ability, whereas agent-based methods are preferable when frontier backbones are available and maximizing the resolution rate justifies the higher and less predictable per-issue cost.}
In the following, we categorize and analyze existing methods with respect to the key design dimensions of each phase in the issue resolution pipeline.
We do not distinguish the ``Repair'' phase in detail, as all scaffolds share the same action at this step: editing the codebase.
They differ only in engineering implementation, which is beyond the scope of this survey.

\textbf{A. Repository Representation (Repo Preprocessing Phase)}

Constrained by the semantic gap between natural language requirements and source code, LLMs face challenges in capturing the complex dependencies and semantic information within repositories.
To better understand repository structures and code semantics, some approaches generate structured representations of the codebase during the repo preprocessing phase, thereby bridging the gap between natural language and code.
According to the relations they preserve, existing representations can be divided into two types: \textit{tree-based representations}, which organize repository files, classes, and functions into containment hierarchies, and \textit{code graphs}, which additionally encode relations such as function calls, inheritance, imports, references, and data flows (detailed in Appendix~\ref{ap:repo_representation}).

The two types trade construction cost against the relations they expose.
Trees are cheap and require no execution, but preserve only containment relations, leaving cross-file dependencies such as calls and imports invisible.
Code graphs recover these dependencies at the cost of additional construction and retrieval mechanisms, and their coverage is bounded by static analysis, so dynamic behavior remains outside the graph.
Code graphs have themselves evolved along two axes, from single-relation call graphs toward multi-source graphs that also absorb documentation~\cite{marscode, Prometheus}, and from graphs consumed as flat context toward graphs exposed as queryable interfaces~\cite{CodeXGraph, Prometheus}.
Trees therefore remain the default in pipeline-based scaffolds, where the representation must fit the context window at negligible cost, whereas graphs are adopted by systems that can afford a dedicated retrieval mechanism to exploit them.

\textbf{B. Localization (Localization Phase)}

Accurate localization serves as a critical basis for effective issue resolution~\cite{meng2024empirical}.
Its objective is to identify a set of potentially faulty code snippets within a specified scope of repository files.
According to the evidence they use to link an issue to code, existing localization methods can be divided into five types: \textit{BM25} by lexical overlap, \textit{spectrum-based methods} by execution coverage, \textit{embedding-based methods} by semantic similarity, \textit{navigation-based methods} by LLM-driven repository exploration, and \textit{graph-based methods} by structural relations over the repository graphs of Section~\ref{sec:e2e_scaffold}.A.2 (detailed in Appendix~\ref{ap:localization}).

The five types primarily differ in the cost of the evidence they rely on and the failure modes that this evidence introduces.
BM25 is the cheapest signal, requiring neither training, execution, nor model calls, but it is vulnerable to lexical mismatch.
Embedding retrieval alleviates this mismatch, yet chunking disrupts cross-file context, and general-purpose encoders poorly capture the asymmetry between issue descriptions and code.
Spectrum-based localization adds execution evidence and complements text-based methods, but it depends on a trigger test, which often lacks in issue resolution task, and therefore remains mainly auxiliary.
Navigation trades search coverage for cost predictability. Open-ended exploration can recover from early mistakes but may fail to converge, whereas staged cascades cannot revisit pruned files.
Graph-based methods capture structural dependencies, but their recall is ultimately bounded by omissions in the underlying static-analysis graph.

\textbf{C. Reproduction Test Generation (Patch Validation Phase)}

Reproduction tests help localize an issue, filter candidate patches, and check whether a patch satisfies the reported behavior, but they are often absent from issue descriptions.
Scaffolds therefore either prompt an LLM directly to generate tests~\cite{SpecRover, agentless, SWE-RL, CodeMonkeys, PatchPilot, GUIRepair, SEMAgent, Nemotron-Cortexa} or let an agent inspect the repository before generation~\cite{SWEAgent, openhands, MASAI, CodeR, marscode, DARS, OpenHands-Versa, experepair, AgentKB, Prometheus, trae2025, KGCompass, SynFix}.
For example, Agentless~\cite{agentless} prompts the LLM for a self-contained script reproducing the reported behavior and keeps it only if it fails on the unpatched repository, so that a candidate patch is accepted only when the same script later passes.
This pre-patch check makes the mechanism usable without ground truth, but it verifies only that the script fails, not that it fails for the reported reason.
Agentic generation instead inspects fixtures and helpers before writing the test, raising the chance of a repository-consistent test at the cost of additional interaction rounds.
Dedicated reproduction methods are discussed in Section~\ref{sec:single_phased_scaffold}.B.

\textbf{D. Regression Test Selection (Patch Validation Phase)}

Regression test selection checks whether candidate patches break existing behavior by selecting tests related to the modified code.
Current scaffolds rely mainly on LLM-based selection~\cite{openhands, SpecRover, OpenHands-Versa, AgentKB, trae2025, agentless, SWE-RL, PatchPilot, SynFix}, which may discard correct patches when irrelevant tests are selected or miss regressions when relevant tests are omitted.
For example, Agentless~\cite{agentless} records which tests pass before patching, asks the LLM which of them relate to the edited code, and rejects any candidate patch that turns a selected passing test into a failing one.
The mechanism is unsound in both directions because the selection step is a model judgment rather than a dependency analysis.
Although program-analysis techniques exist~\cite{kauhanen2021regression}, they have not been integrated into these scaffolds, and available Python techniques remain unsafe~\cite{kauhanen2021regression, bouzenia2024dypybench}.
This category has thus seen the least design evolution of the whole validation phase, as scaffolds have converged on the same LLM-based selection rather than importing the analysis-based selection developed by the testing community.
\revReplace{R1.8}{Trajectory-level failure analyses reinforce the importance of the patch validation phase, showing that many unresolved instances stem from frequent runtime errors surfaced during test execution and from stage-specific mistakes~\cite{chen2025unveiling, liu2025empirical} (see Section~\ref{sec:technique_oriented_studies}).}

\textbf{E. Patch Rerank (Patch Selection Phase)}

\looseness=-1
After test-based filtering, scaffolds commonly rerank plausible patches through self-consistency-based majority voting~\cite{marscode, trae2025, agentless, RepoGraph, SWE-RL, CodeV, CodeMonkeys, KGCompass, GUIRepair, Nemotron-Cortexa} or LLM-as-a-Judge assessment~\cite{MASAI, SpecRover, DARS, experepair, trae2025, CodeMonkeys, PatchPilot, SEMAgent}.
For example, Agentless~\cite{agentless} normalizes the sampled patches so that formatting differences do not create spurious variants, and submits a representative of the largest equivalence class.
Voting needs no additional model capability but is blind to the content of the patch and rewards whatever the backbone is biased toward, whereas LLM-as-a-Judge inspects the semantics of each candidate against the issue but inherits the judging model's own limits.
These strategies can be combined: TRAE applies code review before voting~\cite{trae2025,tian2026agentbased}, while KGCompass performs rule-guided review before selection~\cite{KGCompass}.
Such hybrids mark the evolution of the category from a single scoring rule toward staged selection that filters semantically before counting.
\revReplace{R1.8}{However, existing empirical studies (see Section~\ref{sec:evaluation_oriented_studies}) show that patches passing all validation tests can still deviate behaviorally from ground-truth patches due to limited test coverage~\cite{patchdiff}, and may fall short of developer patches in code quality and requirement compliance~\cite{chen2025evaluating}.
Thus, test-based rerank selects the most promising candidate rather than guaranteeing correctness.}
\revReplace{R3.18}{Methods dedicated to patch selection will be discussed in Section~\ref{sec:single_phased_scaffold}.C.}

\subsubsection{Scaffold Improvement}\label{sec:scaffold_improvement}

\revReplace{R3.18}{
Besides building new scaffolds, complementary works develop scaffold-agnostic mechanisms that improve existing scaffolds.
According to the target of improvement, we categorize these mechanisms into three types: memory, which accumulates reusable knowledge across issues, context management, which steers the content fed into the context window, and self-evolution, which lets the agentic system refine its own design.

\textbf{A. Memory}

Since most scaffolds resolve each issue in isolation and discard the trajectory afterwards, they ignore the knowledge and lessons that can be learned in previous exploration.
Memory mechanisms equip a scaffold with a long-term store that carries such knowledge and lessons across issues.
According to the source of the stored knowledge, existing memory mechanisms work in two ways: distilling experience from the system's own trajectories, and importing knowledge from sources beyond the system.
Specifically, trajectory-based methods summarize previous resolution processes into structured entries and retrieve relevant ones to guide new resolutions~\cite{experepair, AgentKB, SWE-EXP, ma2026failuremem, guo2026eet}.
For example, EXPEREPAIR~\cite{experepair} stores concrete demonstrations as episodic memory and summarized repair insights as semantic memory, and retrieves both to guide a new issue, while SWE-Exp~\cite{SWE-EXP} extracts transferable experience from both successful and failed trajectories.
External-source methods instead structure knowledge held by humans, such as developer experience on GitHub~\cite{wang2026memgovern}, user intents~\cite{zhou2025tom}, and notes accumulated across sessions~\cite{wang2025confucius}.
For example, MemGovern~\cite{wang2026memgovern} transforms unstructured GitHub discussions into governed experience cards and matches them to agents through logic-driven retrieval.
The trade-off separating the two is provenance: self-distilled memory is automatic and repository-specific but can entrench the system's own mistakes, whereas external memory carries human-validated knowledge but requires curation and may not match the current codebase.

\textbf{B. Context Management}

Since agent trajectories accumulate context over multiple rounds of tool interactions, as discussed in Section~\ref{sec:e2e_scaffold}, the accumulated context inflates cost and dilutes the attention of the LLM~\cite{liu2024lost}.
Context management mechanisms control what enters the context window of the LLM during resolution.
According to the direction of adjustment, existing methods work in two ways: compressing low-value content and enriching informative content.
Specifically, compression methods shrink the context either by removing content that no longer contributes to resolution~\cite{xiao2026reducing, wang2026swe, jia2026compressing} or by reorganizing the interaction interface so that the same information occupies less context~\cite{kim2026codestruct, SWE-Edit}.
For example, AgentDiet~\cite{xiao2026reducing} identifies useless, redundant, and expired information in the trajectory and removes it during execution, and CODESTRUCT~\cite{kim2026codestruct} lets agents read and edit abstract syntax tree entities instead of matching raw text.
Enrichment methods instead improve the quality of the supplied context~\cite{suri2026codescout, li2026outcome}.
For example, CodeScout~\cite{suri2026codescout} pre-explores the codebase to rewrite underspecified issue descriptions into actionable problem statements.
The two directions trade against each other, since compression lowers cost but risks discarding evidence the agent later needs, whereas enrichment improves grounding at the price of additional pre-processing.

\textbf{C. Self-Evolution}

\looseness=-1
Since handcrafted scaffolds are fixed at development time, adapting them to new backbones or task distributions requires costly manual redesign~\cite{xia2025live}.
Self-evolution mechanisms enable the agentic system to optimize its own scaffold or resolution strategy based on execution feedback.
According to the object of evolution, existing methods work at two levels: evolving the scaffold implementation and evolving the resolution strategy.
Specifically, implementation-level methods let the agent modify its own code and retain variants that are validated by benchmark performance~\cite{robeyns2025a, DGM, HGM, xia2025live}.
For example, the Darwin G\"{o}del Machine~\cite{DGM} maintains an archive of agent variants and grows it through validated self-modifications, and Live-SWE-agent~\cite{xia2025live} evolves a basic scaffold with only bash tools on the fly while solving real issues.
Strategy-level methods instead keep the scaffold fixed and refine how it resolves issues, such as revising and recombining prior trajectories~\cite{SE-Agent}, abstracting plans from grounded experience~\cite{hayashi2025self}, and accumulating project knowledge from commit history~\cite{deng2026your}.
For example, SE-Agent~\cite{SE-Agent} revises, recombines, and refines prior trajectories to explore the solution space more effectively.
Implementation-level evolution can reach designs a human would not write but needs a trusted validation signal, since the archive is only as meaningful as the benchmark that admits variants into it, whereas strategy-level evolution is cheaper and safer but cannot escape the limits of the fixed scaffold.\par
}

\subsubsection{Single-Phased Methods}\label{sec:single_phased_scaffold}

Current single-phased methods focus on three subtasks: issue localization, issue reproduction, and patch selection.
Issue localization requires the agentic system to identify the code locations that need modification, serving as the foundation for subsequent editing.
Issue reproduction aims to generate a test case for the target issue that fails in the buggy repository but passes in the fixed one, thereby assisting in both localization and verification of issue resolution.
\revReplace{R3.18}{Patch selection aims to identify, among multiple candidate patches, the one most likely to resolve the issue, determining the quality of the final submission.}
\revReplace{R1.9}{Since these subtasks are critical steps in issue resolution, improvements on them have been shown to translate into higher end-to-end resolution rates~\cite{locagent, CoSIL, OrcaLoca, SWT_bench}.}

\textbf{A. Issue Localization}

\looseness=-1
Different from the localization component of the scaffold discussed in Section~\ref{sec:e2e_scaffold}.B, we discuss methods dedicated to issue localization, whose results can be leveraged to enhance the repair phase~\cite{feng2025integrating, liu2025graphlocator, mamun2026blagent, sepidband2026rgfl, xu2026learning, xu2026neurosymbolic, sutawika2026codescout, zhang2025one, zhang2025hierarchical, reddy2025swerank+, luo2026closing}.
These methods split by whether localization capability is placed in the inference-time procedure or in the model parameters.

\ulit{A.1 RAG-based Methods.}
Retrieval-augmented generation localizes issues through vector, graph, or navigation-based repository retrieval.
Vector RAG ranks code segments by embedding similarity and is discussed with trained embedding models in Section~\ref{sec:single_phased_scaffold}.A.2.
Graph RAG performs multi-hop structural retrieval.
For example, LocAgent~\cite{locagent} parses the repository into a heterogeneous graph of files, classes, and functions and equips an agent with traversal tools, so that it can hop from an entity named in the issue to its callers before ranking suspicious locations.
Navigation-based RAG instead lets agents inspect repositories dynamically.
It improve search through action scheduling and context pruning~\cite{OrcaLoca}, multi-agent codebase compression~\cite{meta-rag}, or related guided navigation designs~\cite{wang2025improving, locagent}.
Hybrid approaches combine these signals, as SACL does with vector retrieval and agent navigation~\cite{SACL}.
The shared advantage is that these methods need no training and transfer to any repository, so they follow a frontier backbone as it improves.
The shared cost is that every issue pays for repeated retrieval and reasoning at inference time.

\ulit{A.2 Fine-Tuning-based Methods.}
Fine-tuning-based localization trains either generation models to predict locations or embedding models to retrieve them.
For generation models, it learn defect patterns from segmented code or intermediate program representations~\cite{BLAZE, BugCerberus}.
For embedding models, it improve retrieval through multilingual contrastive data~\cite{CoRNStack}, real issue--modification pairs~\cite{SweRank}, or repository structure and call dependencies~\cite{coret}.
For example, SweRank~\cite{SweRank} contrastively trains a retriever on issue-modification pairs mined from commit history and pairs it with a trained reranker, turning localization into a retrieve-then-rerank pass.
Compared to RAG-based methods, this moves cost from inference to training, giving cheap and reusable localization, but the learned signal reflects the repositories and issue styles seen in training and must be refreshed as projects drift.

\textbf{B. Issue Reproduction}

Automatically reproducing issues from issue descriptions enables timely localization and fixing, thereby improving development efficiency. 
Existing approaches have focused on narrow bug categories including Android issues~\cite{huang2024crashtranslator} and configuration-induced faults~\cite{fu2024missconf}, but these constitute only a small fraction of real-world defects.

Given the strong understanding and generation capabilities of LLMs, some works use basic promptingt to generate reproduction test.
For example, AssertFlip~\cite{AssertFlip} first generates tests that pass on the buggy version and then flips assertions to obtain failing reproduction tests.
Building on basic prompting, several approaches introduce systematic multi-turn refinement.
For example, AEGIS~\cite{AEGIS} employs a finite-state-machine-based optimization module to orchestrate multi-turn interactions and script adjustments for controlled refinement.
Another line of work enhances LLM-based generation with repository inspection and rule-based reasoning.
Otter~\cite{Otter} augments LLM outputs with rule-based analysis and a self-reflection planning stage to support test-driven development.
More recent methods further improve reproduction through graph-enhanced retrieval and execution feedback~\cite{fei2026echo, wang2026icore}, continuous learning for defective-code reproduction~\cite{lin2024llms}, dedicated training for issue reproduction~\cite{soni2026swe}, and extension to additional languages~\cite{ahmed2026reproduction}.

\textbf{C. Patch Selection}

\revReplace{R3.18}{
Dedicated patch selection methods build verifiers that assess candidate patches or their resolution trajectories and select the most promising one for submission.
They follow two designs: scoring-based methods independently rate candidates with repository-derived rubrics, outcome-trained critics, or process signals~\cite{raghavendra2026agentic, huang2026beyond, wang2026rubric, shum2026swerm, SWE-PRM}, whereas comparison-based methods rank candidates through rollout tournaments or branch from promising archived steps~\cite{kim2026scaling, ding2026swe}.
SWE-RM exemplifies execution-free fine-grained scoring~\cite{shum2026swerm}, while RTV compares structured rollout summaries in recursive small-group tournaments~\cite{kim2026scaling}, so that the verifier only ever performs local comparisons instead of scoring the whole pool at once.
Scoring is cheap and parallel but requires the verifier to be calibrated on an absolute scale, whereas comparison avoids calibration by asking only which of two candidates is better, at the cost of more verifier calls.
Unlike the in-scaffold rerank of Section~\ref{sec:e2e_scaffold}.E, these verifiers are trained artifacts, which is what allows them to be reused as reward models.
Scoring verifiers can also supply rewards for reinforcement learning~\cite{shum2026swerm, huang2026beyond}, as discussed in Section~\ref{sec:training}.B.
}

\subsection{Learning Strategies}\label{sec:learning_strategies}
\revReplace{R1.5}{After the release of DeepSeek-R1~\cite{deepseekr1}, agentic reinforcement learning has become an increasingly important strategy for training issue-resolution agentic systems.
As shown in Table~\ref{tab:training_method}, reinforcement learning is used by none of the four representative models released before February 2025, but by 24 of the 39 released thereafter, accounting for 61.5\%.
Moreover, the highest reported SWE-bench Verified performance in Table~\ref{tab:training_method} is achieved by an RL-based model.
These results indicate a recent shift toward reinforcement-learning-based training in issue resolution.}

\subsubsection{Data Preparation}\label{sec:training_data}

The process of preparing real-world training datasets for training issue resolution domain-specific models is similar to the benchmark construction process described in Section~\ref{sec:benchmark_construction}.
Given the scarcity of annotated data and the gradual depletion of real issue resolution samples, researchers have also explored synthetic data generation methods based on real repositories to expand the scale of training data.
Table~\ref{tab:training_data} (see Appendix~\ref{ap:train_data}) summarizes the existing real-world and synthetic datasets.
\revReplace{R3.18}{Since both directions depend on executable environments, recent work further builds automated infrastructure for environment construction.}

\textbf{A. Real-world Data}

Real-world datasets are constructed from open-source repositories and issue–commit pairs.
SWE-bench (Train)~\cite{swebench} is the first large-scale dataset to organize 19k real issue-fix pairs, providing training samples that reflect real-world distributions.
However, the lack of executable environments prevents models trained on these datasets from obtaining stable reward signals, limiting their use to supervised fine-tuning.
With the growing success of RL, recent models such as DeepSeek-R1~\cite{deepseekr1} show that rule-based rewards can be effective, making executable environments a key requirement for new datasets.
Execution-backed datasets are built either by curating an environment per instance~\cite{SWE-Gym, swe-dev_sjtu} or by automated pipelines that continuously harvest and validate new tasks~\cite{SWE-rebench, Skywork-SWE, mutiswebench, badertdinov2026swe, liang2026swe, zhao2026immersion, team2026yet}.

\textbf{B. Synthetic Data}

Since curating real instances and their executable environments still demands hundreds of hours of manual effort and often suffers low success rates~\cite{SWE-smith, swe-mirror}, synthetic data instead generates instances and environments from existing repositories.
According to where the defect of a synthesized instance originates, existing methods work in two ways: deriving tasks from real code changes and injecting faults into prepared environments.
Specifically, PR-derived methods recover a task from a change that developers actually made, reverse-translating the commit into a problem statement and generating the tests that witness it, which removes the reliance on human-written issue descriptions and unit tests while staying within the distribution of real development~\cite{R2E, r2e_gym, swe-dev_thu, zhang2025synthesizing}.
For example, SWE-Flow~\cite{zhang2025synthesizing} recovers an incremental test-driven development schedule from the unit tests of an existing repository.
Bug-injection methods instead start from a repository whose environment is already configured and introduce the fault themselves, either mutating working code~\cite{SWE-smith, SWE-Synth} or transferring bug-fix logic abstracted from other projects~\cite{swe-mirror}.
For example, SWE-Synth~\cite{SWE-Synth} has LLM agents inject bugs into a working repository and then simulate the human debugging process over them to obtain verifiable instances.
Injection is the cheaper of the two because one configured environment amortizes over many generated instances, but its defects are defined by the mutation applied rather than by a behavior a user reported, so it drifts further from real issue distributions.
Therefore, recent work mixes synthetic with real tasks rather than replacing them~\cite{xie2026hybrid}.

\textbf{C. Automated Environment Infrastructure}

\revReplace{R3.18}{

Because both real-world collection and data synthesis rely on executable environments, whose construction dominates the manual effort and overhead of dataset production~\cite{SWE-smith, swe-mirror}, automated environment infrastructure is proposed to construct and manage executable environments without manual configuration.
Existing works can be divided into two aspects: automating environment setup and environment execution cost reduction.
Specifically, automated setup methods employ agentic pipelines that generate environment configurations, iteratively diagnose and repair build failures, and validate the results through executed tests~\cite{guo2025swe, guo2026evoconfig, guo2026menvagent, zeng2026swe, fu2026davinci, SWE-Universe}.
For example, SWE-Factory~\cite{guo2025swe} builds environments with a multi-agent system and validates instances through exit-code-based log parsing and fail-to-pass checks.
Cost reduction methods aim to reduce the runtime cost of using massive environments.
For example, SWE-MiniSandbox~\cite{yuan2026swe} replaces per-task containers with kernel-level isolation and environment pre-caching.
The two aspects address the two costs of environments respectively, namely the one-off cost of making a task runnable and the recurring cost of running it once per rollout, and the emergence of the latter reflects that reinforcement learning executes environments continuously rather than once.
}

\subsubsection{Training}\label{sec:training}

The training of domain-specific models for the issue resolution task has evolved rapidly alongside the expansion of SWE-bench-style datasets and the emergence of open scaffolds such as OpenHands~\cite{openhands}, Agentless~\cite{agentless}, and SWE-agent~\cite{SWEAgent}.
Table~\ref{tab:training_method} (see Appendix~\ref{ap:train_method}) summarizes representative models, their training paradigms, and their performance on SWE-bench Verified.

\textbf{A. Supervised Fine-tuning}

Supervised Fine-tuning (SFT) is a fundamental approach for training domain-specific models by injecting task knowledge through input–output pairs.
In issue resolution, common SFT strategies include process-oriented methods and teacher-model distillation.

\revReplace{R3.20}{Process-oriented approaches decompose issue resolution into pipeline stages such as localization and repair, and supervise each stage separately.}
Existing methods model these intermediate processes by simulating developer interactions and tool use~\cite{SWE-GPT}, enriching stage-level supervision with repository structure or specialized retrieval and editing modules~\cite{ReSAT, SWE-Fixer, Co-PatcheR}, or training deeper reasoning trajectories with rejection-based refinement~\cite{swe-resoner, MCTS-Refined}.
Thus, they shift supervision from static issue--patch pairs toward the structure and reasoning of the resolution process.

\looseness=-1
\revReplace{R3.20}{Teacher-model distillation methods run a powerful model (e.g., Claude-4.5~\cite{Claude-4.5}, GPT-4o~\cite{openai2024gpt4technicalreport}) within a scaffold to sample complete resolution trajectories, retain the trajectories whose final patches pass validation, and fine-tune a smaller open model on them, so that the student model imitates the teacher's multi-step reasoning and tool usage.}
Models released with new datasets commonly use this strategy to validate dataset effectiveness~\cite{SWE-Gym, r2e_gym, SWE-smith, Skywork-SWE, swe-mirror}.
Distillation also supplies cold-start policies before RL~\cite{li2026boostapr, yu2025building, SWE-World, SWE-TRACE}, while other open models rely predominantly on SFT over curated trajectories~\cite{sonwane2025bugpilot, zhu2025training, liu2025context}.
The two strategies differ in where supervision comes from and what it costs.
Process-oriented supervision is derivable from commit history alone and therefore scales with text-only corpora, but its golden intermediate outputs are reconstructed from the final fix rather than observed, so they describe a path the developer never took.
Distillation instead observes genuine trajectories, yet it is bounded by the teacher, requires executable environments to filter trajectories by validation outcome, and incurs the teacher's inference cost.
Because filtering already presumes an execution oracle, distillation naturally became the cold start for the reinforcement learning that follows.

\textbf{B. Reinforcement Learning}

\looseness=-1
Reinforcement learning (RL) has demonstrated a distinctive ability to cultivate long-horizon decision-making behaviors that static supervised methods fail to capture~\cite{deepseekr1, DeepSWE}.
Recent open-source efforts have started exploring this complexity through RL-based training frameworks.
\revReplace{R3.20}{During RL training, the model rolls out resolution trajectories by interacting with a repository environment, receives a reward that scores each trajectory, and updates its parameters with policy optimization algorithms such as PPO~\cite{schulman2017proximal} to make high-reward behaviors more likely.
Unlike SFT, which imitates fixed demonstrations, RL lets the model learn from feedback on its own explored trajectories.}
According to the type of reward model used, RL training methods can be categorized into process reward models (PRM) and outcome reward models (ORM).

\looseness=-1
PRM-based reinforcement learning rewards intermediate states or actions to teach effective resolution procedures rather than only successful outcomes.
For example, SoRFT~\cite{SoRFT} rewards file, function, and line localization and code editing separately against the developer's fix, so that credit is assigned to the stage that produced the error rather than to the trajectory as a whole.
PRMs provide fine-grained guidance but require reliable process-level signals.
This is the binding constraint on the category, since intermediate correctness has no execution oracle and must itself be approximated by rules or a learned critic, which is why PRM adoption remains far behind ORM despite its finer credit assignment.

ORM-based reinforcement learning instead uses verifiable final outcomes.
Existing methods derive rewards from commit histories or candidate checks~\cite{SWE-RL, swe-resoner}, use unit-test outcomes directly~\cite{Agent-RLVR, swe-dev_thu, SWESwiss, kimi-dev, devstal_small}, or combine outcome verification with structured exploration and policy optimization~\cite{SEAlign, Satori-SWE, golubev2025training}.
In the canonical workflow, the patch produced by a rollout is executed against the fail-to-pass and pass-to-pass tests, and the binary outcome becomes the reward for the whole trajectory, so supervision requires no human annotation but arrives only at the end of a long horizon.
ORMs scale because outcomes are automatically verifiable, but their supervision is only as reliable as the underlying oracle.
\revReplace{R1.8}{Nevertheless, the reliability of outcome rewards is affected by the test suites that produce them.
Since weak test suites may accept plausible patches that are behaviorally incorrect~\cite{patchdiff, chen2025evaluating}, purely outcome-driven rewards risk reinforcing such patches during training.
Efficiency analyses further reveal token snowballing and costly failure patterns in current systems~\cite{SWE-Effi}, suggesting that reward designs should also account for the cost of the resolution process.
We further discuss in Section~\ref{sec:evaluation_oriented_studies}.}

\answerRQ{\revReplace{R1.11}{
Existing techniques advance along two complementary dimensions: scaffold design and model training.
Among scaffolds, pipeline-based designs suit tight budgets and weaker backbones, agent-based designs suit frontier backbones, and scaffold-agnostic mechanisms for memory, context management, and self-evolution improve both without redesign.
Training has shifted decisively toward agentic reinforcement learning, but its ORM remains bounded by test suite quality.
Thus, PRM may provide a promising direction.
}}

\vspace{-0.5em}
\section{Empirical Studies}\label{sec:emprical_studies}

\begin{figure*}[htbp]
  \centering
  \begin{adjustbox}{width=\textwidth}
  \begin{forest}
for tree={
    rounded corners,
    child anchor=west,
    parent anchor=east,
    grow'=east,
    text width=4cm,%
    draw=softPurple,
    anchor=west,
    node options={align=center},
    fit=rectangle,
    edge path={
      \noexpand\path[\forestoption{edge}]
        (.child anchor) -| +(-5pt,0) -- +(-5pt,0) |-
        (!u.parent anchor)\forestoption{edge label};
    },
    where n children=0{text width=12cm}{}
  },
  [Empirical Studies
    [Evaluation-oriented Studies
        [Evaluation Validity
            [{Yadavally et al.~\cite{yadavally2025critics}, PatchDiff~\cite{patchdiff}, Liang et al.~\cite{liang2025Illusion}, Garg et al.~\cite{garg2025saving}, Prathifkumar et al.~\cite{prathifkumar2025does}, Martinez et al.~\cite{martinez2025dissecting}}]
        ],
        [Quality and Requirement Compliance
            [{Chen et al.~\cite{chen2025evaluating}, Sajadi et al.~\cite{sajadi2025Secure}, Yu et al.~\cite{yu2026does}}]
        ],
        [Efficiency and Resource Consumption
            [{SWE-Effi~\cite{SWE-Effi}, Gao et al.~\cite{gao2026more}, Tripathy et al.~\cite{tripathy2026swenergy}}]
        ],
        [Robustness and Process-aware Evaluation
            [{Vijayvargiya et al.~\cite{vijayvargiya2025interactive}, SWE-Bench-CL~\cite{swebench-cl}, Liu et al.~\cite{liu2026process}, TRAJEVAL~\cite{kim2026trajeval}, Gloaguen et al.~\cite{gloaguen2026coding}, AgentLens~\cite{sahoo2026agentlens}, RepoMirage~\cite{li2026repomirage}}]
        ]
    ],
    [Technique-oriented Studies
        [Performance Variation and Influencing Factors
            [{DEI~\cite{DEI}, Meng et al.~\cite{meng2024empirical,meng2026llmbased}, Zhao et al.~\cite{zhao2026beyond}, ORACLE-SWE~\cite{li2026oracle}, Zhang et al.~\cite{zhang2026agent}, Cheng et al.~\cite{cheng2026dynamic}}]
        ],
        [Agent Behavior and Resolution Process
            [{Bouzenia et al.~\cite{bouzenia2025understanding}, Majgaonkar et al.~\cite{majgaonkar2025understanding}, Vijayvargiya et al.~\cite{vijayvargiya2025interactive}, Liu et al.~\cite{liu2026plan}}]
        ],
        [Failure Modes and Root Causes
            [{Chen et al.~\cite{chen2025unveiling}, PAGENT~\cite{xue2025pagent}, Liu et al.~\cite{liu2025empirical}, Cuadron et al.~\cite{cuadron2026saber}}]
        ]
    ]
  ]
\end{forest}
\end{adjustbox}
\caption{Taxonomy of empirical studies of issue resolution.}
\label{fig:empirical_structure}
\end{figure*}

\revReplace{R3.7/R3.12}{To answer \textbf{RQ3}, this section synthesizes what empirical studies establish about the reliability of reported resolution rates and about the causes of the performance differences observed among agentic systems.}
Figure~\ref{fig:empirical_structure} presents the resulting taxonomy of empirical studies constructed through the open coding procedure in Section~\ref{sec:taxonomy_construction}. \revReplace{R2.8/R2.11/R3.7}{Its top level separates evaluation-oriented studies, which scrutinize what a reported resolution rate does and does not certify (Section~\ref{sec:evaluation_oriented_studies}), from technique-oriented studies, which trace performance differences to the information, control, and process characteristics of individual agentic systems (Section~\ref{sec:technique_oriented_studies}). We then analyze the findings of each group and identify what is still missing before these systems can be used in real projects.}

\subsection{Evaluation-oriented Studies}\label{sec:evaluation_oriented_studies}

\revReplace{R3.7}{Evaluation-oriented studies show that a reported resolution rate is a useful signal under a fixed protocol, but it is not sufficient evidence of practical issue-resolution capability. They expose validity threats, assess properties hidden by outcome metrics, and test whether conclusions persist beyond standard benchmark settings.}

\textit{Evaluation Validity.}
\revReplace{R3.7}{To determine whether benchmarks, tests, and evaluation protocols faithfully reflect the capability of an agentic system, evaluation-validity studies examine three distinct threats to reported resolution rates. 
First, weak tests can accept patches that differ behaviorally from ground-truth patches, as shown by PatchDiff~\cite{patchdiff}. 
Second, model memory and contamination can inflate resolution rates on SWE-bench and SWE-bench Verified~\cite{liang2025Illusion,prathifkumar2025does}. 
Third, benchmark prompts and leaderboard settings can obscure the gap between formal tasks and realistic developer requests~\cite{garg2025saving,martinez2025dissecting}.
Execution-free evaluation mechanisms provide a complementary signal for judging repository-level changes without relying on a prescribed test outcome~\cite{yadavally2025critics}. Together, these studies show that reported resolution rates support controlled comparison, but do not alone establish trustworthy real-world capability.}

\textit{Quality and Requirement Compliance.}
\revReplace{R3.7}{To assess whether test-passing patches meet the broader requirements of real projects, quality and requirement-compliance studies examine evidence beyond functional correctness. 
Comparisons with developer patches reveal differences in code quality and requirement compliance~\cite{chen2025evaluating}, security analyses expose vulnerabilities introduced by generated fixes~\cite{sajadi2025Secure}, and design-constraint evaluation identifies violations that conventional tests may miss~\cite{yu2026does}. 
Behavioral correctness, maintainability, security, and project constraints are therefore complementary requirements rather than interchangeable metrics that a high resolution rate can compensate for.}

\looseness=-1
\textit{Efficiency and Resource Consumption.}
\revReplace{R3.7}{To assess whether an agentic system is practically useful under resource constraints, efficiency studies relate successful resolution to the resources required to obtain a successful patch.
SWE-Effi identifies token snowballing and costly failure patterns~\cite{SWE-Effi}, while turn-control strategies, energy consumption, runtime, and memory use reveal further trade-offs among agentic systems~\cite{gao2026more,tripathy2026swenergy}.
These findings distinguish raw effectiveness from deployable effectiveness, since a modest gain in resolution rate may not justify substantially greater resource consumption.}

\textit{Robustness and Process-aware Evaluation.}
\revReplace{R3.7}{To assess whether agentic systems remain reliable when conditions depart from standard benchmarks, robustness and process-aware studies examine interactive requirements, continual learning, trajectory quality, and perturbed repository contexts. 
Interactive tasks show that agentic systems struggle with ambiguous requirements but can improve by seeking clarification~\cite{vijayvargiya2025interactive}. 
SWE-Bench-CL examines continual learning, whereas process analyses and trajectory-level evaluation identify properties that outcome metrics conceal~\cite{swebench-cl,liu2026process,kim2026trajeval,sahoo2026agentlens}. 
Repository perturbations reveal whether agentic systems reason over context rather than exploit spurious cues~\cite{li2026repomirage}. Most evidence remains centered on fixed SWE-bench-style protocols, which enable controlled comparison but bound the generalizability of their conclusions. 
Well-timed actions and reliable trajectories are therefore necessary for trusting a reported success~\cite{gloaguen2026coding}.}

\subsection{Technique-oriented Studies}\label{sec:technique_oriented_studies}

\revReplace{R3.7}{Technique-oriented studies explain performance variation through the information available to an agentic system, the control mechanisms that govern its actions, and the reliability of its resolution process.}

\textit{Performance Variation and Influencing Factors.}
\revReplace{R3.7}{To explain why agentic systems achieve different outcomes on the same task, performance-variation studies examine the information and control available to them. 
Different agentic systems solve complementary subsets of tasks~\cite{DEI}, and performance differences across leading agentic systems depend on the instances they can or cannot resolve~\cite{meng2024empirical,meng2026llmbased}. 
Localization quality, repository context, search budget, and oracle signals such as tests, locations, and APIs determine how much useful information reaches the agentic system~\cite{zhao2026beyond,li2026oracle}. 
Cogeneration experiments further show that coupling bug reproduction tests with fixes influences how an agentic system behaves while preserving the generation rate of plausible fixes, indicating that workflow structure is another performance factor~\cite{cheng2026dynamic}. 
Rules, guidance, and guardrails further shape whether the agentic system can use that information reliably~\cite{zhang2026agent}. 
Thus, performance cannot be attributed to the backbone model alone. It reflects the joint design of information access and control.}

\textit{Agent Behavior and Resolution Process.}
\revReplace{R3.7}{To identify process-level determinants of success and failure, behavior studies analyze how agentic systems interpret requirements, plan, and act during resolution. Performance variation is also process-level rather than solely model-level. 
Agentic systems must recognize ambiguity and seek clarification when requirements are underspecified~\cite{vijayvargiya2025interactive}.
Successful and failed trajectories differ in action sequences, iteration patterns, and semantic consistency~\cite{bouzenia2025understanding,majgaonkar2025understanding}. 
Agentic systems do not always execute their own plans faithfully~\cite{liu2026plan}.
These findings shift the explanation of failure from an undifferentiated lack of model ability to observable weaknesses in interaction and trajectory discipline.}

\textit{Failure Modes and Root Causes.}
\revReplace{R3.7}{To locate recurrent breakdowns that the design of an agentic system can target, failure-mode studies diagnose where and why resolution processes fail. 
Such failures are recurrent and diagnosable rather than random residual errors. 
Large-scale trajectory and test-log analyses identify frequent runtime errors and process-level mistakes~\cite{chen2025unveiling}. 
Failed-patch categories can support targeted repair~\cite{xue2025pagent}, and stage-based taxonomies expose where unresolved instances break down~\cite{liu2025empirical}. 
Safeguards for error-prone mutating actions provide a concrete example of using such diagnoses to improve the reliability of agentic systems~\cite{cuadron2026saber}. 
These studies motivate phase-level logging, diagnostic evaluation, and targeted safeguards instead of relying exclusively on end-to-end resolution rates.}

\answerRQ{\revReplace{R1.8/R1.11/R3.7}{
Reported resolution rates support controlled comparison but do not certify practical capability, since weak tests~\cite{patchdiff}, contamination~\cite{liang2025Illusion,prathifkumar2025does}, unrealistic task framing~\cite{garg2025saving,martinez2025dissecting}, and unreliable trajectories~\cite{sahoo2026agentlens,gloaguen2026coding} can make a resolved instance overstate real ability.
Performance variation is driven more by information access, control design, and recurrent process-level failures than by the backbone model or the aggregate rank of an agentic system.
In practice, assessment requires the resolution rate to be reported alongside patch quality, security, robustness, and resource evidence, supported by phase-level and trajectory-level diagnostics.
These conclusions calibrate the benchmark claims in Section~\ref{sec:benchmark} and the technique claims in Section~\ref{sec:technique}, while motivating the multi-perspective and fine-grained evaluation agenda in Section~\ref{sec:road_ahead}.
}}

\vspace{-0.5em}
\section{Research Opportunities and Road Ahead}\label{sec:road_ahead}

\looseness=-1
Despite substantial progress by large language model-based agentic systems on the issue resolution task, achieving fully autonomous and reliable automated software maintenance remains highly challenging.
\revReplace{R3.9}{ 
To answer \textbf{RQ4}, this section outlines the current challenges extracted from previous sections and surveyed papers and identifies research opportunities and future directions on the path toward automated issue resolution.
Specifically, during the open coding procedure in Section~\ref{sec:taxonomy_construction}, we additionally record the limitations and future work explicitly stated in the coded papers, cross-check them against the cross-study insights in Section~\ref{sec:emprical_studies}, and identify unique challenges according to previous parts of this survey.}
\revReplace{R3.9/R3.10}{We present the challenges in the organizational order of this survey, moving from evaluation to techniques and then to emerging directions, so that each challenge follows the sections whose evidence it draws on.
This order is presentational and does not itself encode priority.
To indicate where the field places its own emphasis, we rank the challenges by community attention.
Specifically, we report for each challenge below the number of surveyed papers that explicitly raise it as a limitation or future work during the open coding described above.
By this measure, the community concentrates most on training domain-specific models (65 of 242 papers, 26.9\%), repository knowledge representation (60, 24.8\%), and validation via tests (49, 20.2\%).
Although the two evaluation-related challenges are discussed less frequently than the others (in 37 and 25 papers, respectively), the empirical evidence presented in Section~\ref{sec:emprical_studies} indicates that unreliable evaluation makes it difficult to assess progress toward addressing the remaining challenges.
We therefore regard this mismatch between limited attention and foundational importance as an important finding in its own right.}

\subsection{Evaluation of Agentic System}

\textit{Multi-Perspective Evaluation.}
\revReplace{R3.9}{This challenge is distilled from the metric analysis in Section~\ref{sec:eval_metrics} and the evaluation-oriented studies in Section~\ref{sec:evaluation_oriented_studies}, which show that resolution rate alone omits deployment-critical properties such as robustness, security, code quality and maintainability, design/architectural-constraint compliance, and performance~\cite{chen2025evaluating, sajadi2025Secure, vijayvargiya2025interactive, yu2026does, liu2025empirical}.
In total, 37 of the 242 surveyed papers (15.3\%) explicitly raise this challenge.}
\revReplace{R2.9}{A pragmatic route is to improve existing issue benchmarks by associating them with property-specific checkers, such as perturbed issue descriptions for robustness, static security scanning of generated patches, static code-quality and maintainability metrics (e.g., complexity, duplication, code smells) benchmarked against the pre-patch baseline, automated checks of project-specific design/architectural constraints, and performance profiling of patched code, rather than building new benchmarks from scratch. Such benchmarks speak primarily to industrial adopters deciding whether a system is fit for deployment, whereas designing the underlying metrics and testbeds remains a task for academic researchers.}

\noindent\textit{Fine-Grained Evaluation.}
\revReplace{R3.9}{This challenge follows from the end-to-end benchmark design in Section~\ref{sec:benchmark} and the trajectory and failure analyses in Section~\ref{sec:technique_oriented_studies}, which show that aggregate resolution rates conceal stage-specific failures~\cite{chen2025unveiling, bouzenia2025understanding, liu2025empirical}.
In total, 25 of the 242 surveyed papers (10.3\%) explicitly raise this challenge.}
\revReplace{R2.9}{Concretely, such frameworks should measure phase-level capabilities along the workflow in Section~\ref{sec:technique}, including localization accuracy at the file and function levels, the fault-revealing quality of reproduction tests, and patch correctness beyond passing the available test suite.
Beyond final task success, trajectory-level process quality and efficiency, including successful tool execution, redundant actions or interactions, and token consumption, deserve comparable attention.
Such evaluation can be made more rigorous and informative when benchmarks provide phase-level reference information, such as ground-truth fault locations, and when agent scaffolds record execution trajectories using a standardized format.
These diagnostic signals matter most to researchers and scaffold developers iterating on system designs.}

\subsection{Repository Knowledge Representation}

\looseness=-1
\revReplace{R3.9}{The repo-preprocessing designs in Section~\ref{sec:e2e_scaffold} predominantly represent code as natural language, sometimes augmented with trees or entity-relation graphs~\cite{agentless, PatchPilot, SWE-Fixer}.
In total, 60 of the 242 surveyed papers (24.8\%) explicitly raise this challenge.}
Future representations should more directly encode program semantics and execution behavior while integrating repository artifacts beyond code, including documentation, images, and demonstration videos.
Research should determine which representations improve localization and architectural understanding, how they remain synchronized after edits, and whether their benefits justify construction and retrieval costs.
\revReplace{R2.9}{Progress on both fronts rests largely with the repository representation research community, whose advances can then be adopted by builders of agentic systems.}

\subsection{Validation Via Tests}

\revReplace{R3.9}{This challenge arises from the patch-validation methods in Section~\ref{sec:e2e_scaffold}: reproduction methods still solve only 49\% of SWT-bench Verified instances~\cite{AssertFlip, SWT_bench}.
In total, 49 of the 242 surveyed papers (20.2\%) explicitly raise this challenge.}
\revReplace{R1.2}{Current LLM-based systems mainly rely on heuristic or prompt-based regression-test retrieval, which remains incomplete or redundant~\cite{SWEAgent, PatchPilot}.}
\revReplace{R1.2}{TestLoc~\cite{chen2025old} shows that selecting relevant regression tests can improve reproduction test generation for new issues and strengthen patch validation, but reliably identifying such tests remains challenging.}
Future work should combine repository-aware reproduction-test generation with program-analysis-based regression-test selection~\cite{kauhanen2021regression}, while addressing the unsafe omission of affected tests in current techniques~\cite{kauhanen2021regression, bouzenia2024dypybench}.
\revReplace{R2.9}{Advancing both problems calls for expertise in software testing, and mature solutions would immediately benefit the patch validation stage of nearly every system surveyed in this work.}

\subsection{Training Domain-Specific Models for Issue Resolution}

\revReplace{R3.9}{This challenge follows from the learning strategies in Section~\ref{sec:learning_strategies} and the performance evidence in Section~\ref{sec:technique_oriented_studies}: backbone choice can outweigh scaffold differences, yet leading commercial models remain inaccessible in privacy-sensitive settings and open models still lag behind~\cite{swebenchverified, Claude-4.5}.
In total, 65 of the 242 surveyed papers (26.9\%) explicitly raise this challenge, the highest among the challenges we identify.}
\revReplace{R1.2}{HE-SNR~\cite{wang2026he} shows that token-level entropy predicts downstream SWE-bench performance more reliably than aggregate perplexity during mid-training. Future work should use such diagnostics to select data, identify ineffective checkpoints, and preserve reasoning capability during alignment.}
\revReplace{R3.14}{Existing fine-tuned models are typically trained on trajectories collected from a single scaffold~\cite{r2e_gym, SWEAgent}, so the learned policy silently binds to that scaffold's tool set and observation format and transfers poorly to others. We therefore advocate treating the scaffold interface as a randomization dimension during training~\cite{tobin2017domain}, in analogy to domain randomization in robotics, exposing the model to varied tool sets and observation formats so that it acquires scaffold-agnostic resolution capability rather than scaffold-specific habits.}

Current agentic RL predominantly uses outcome rewards~\cite{rl_survey, DeepSWE, SWESwiss, kimi-dev}, which may reward successful patches produced through unreliable processes.
\revReplace{R3.14}{Moreover, prevailing training pipelines keep only successful trajectories for supervision and discard failures~\cite{SWE-Gym,r2e_gym}, although failed trajectories encode recurring dead-end patterns that can train critics to recognize and abandon unpromising explorations early.}
\revReplace{R2.9}{The phase-level signals discussed above, such as localization accuracy and reproduction test verdicts, are natural sources of process rewards. This avenue is best pursued by model providers and research groups with sufficient training resources, and the resulting open models chiefly serve organizations for which commercial APIs are not an option.}

\subsection{Domain-Specific Issue Resolution}

\looseness=-1
\revReplace{R3.15}{This direction is motivated by both benchmark- and instance-level evidence presented in Sections~\ref{sec:benchmark} and~\ref{sec:technique_oriented_studies}.
Specifically, different systems solve complementary subsets of tasks, while their performance degrades in settings beyond Python libraries~\cite{DEI, meng2024empirical, swebenchmutimodal, omnigirl}.
Moreover, 49 of the 242 surveyed papers (20.2\%) explicitly call for further research in this direction.
These indicate substantial room for domain-specific issue resolution along two axes.
Along the bug-family axis, each family exhibits distinct symptoms and admits distinct verification oracles, motivating an upfront classifier that routes each issue to a pipeline specialized for that family.
For example, cosmetic and interaction bugs demand visual observation and rendering-aware oracles, whereas calculation bugs are better served by input-space exploration and numerical assertions.
Along the system-type axis, each type of system requires a different execution harness, calling for agents that integrate the domain-specific toolkit of each ecosystem as first-class tools.
For example, resolving issues in web applications relies on browser automation, while mobile apps require device emulators~\cite{huang2024crashtranslator}.
Progress hinges on benchmarks stratified by bug family and system type, which make specialization gains measurable.
Constructing such benchmarks falls to academic researchers, while domain toolkits are best contributed by practitioners within each ecosystem.}

\subsection{SE for Agentic System}

\revReplace{R3.9}{This direction is motivated by the recurrent, diagnosable failure modes identified in Section~\ref{sec:technique_oriented_studies}~\cite{chen2025unveiling, xue2025pagent}.
In total, 26 of the 242 surveyed papers (10.7\%) explicitly raise this direction.}
Software Engineering for Agentic Systems (SE4AS) should extend testing, debugging, versioning, and continuous integration to agent lifecycles, with tools that expose internal states and feed diagnostic evidence back into training~\cite{TRAIL2025}.
\revReplace{R3.14}{We see particular promise in porting classical debugging techniques to agent trajectories. Counterfactual replay~\cite{ding2026swe}, which intervenes on a single step and re-executes the remainder of a trajectory, can isolate the minimal decisions responsible for a failure in the spirit of delta debugging. Statistics over large collections of successful and failed trajectories can rank suspicious action patterns, mirroring spectrum-based fault localization, and process invariants mined from normal runs, such as never editing a file before reading it, can be monitored at runtime to abort abnormal executions early~\cite{nanda2026wink}.}
\revReplace{R2.9}{Establishing this field is a joint undertaking, in which academia supplies methodologies and tooling while industrial teams operating agents in production ground them with real workloads and failure data.}

\answerRQ{\revReplace{R1.11/R3.15}{
Due to the limitations identified above, issue resolution may need to move beyond single-score evaluation and assess agentic systems in a fine-grained manner from perspectives such as security, efficiency, robustness, and code quality.
It may also require more effective and multimodal knowledge representations for software repositories, more reliable reproduction test generation and regression test selection, process-oriented reinforcement learning for open-source models, domain-specific customization across bug families and system types, and dedicated software engineering methods for developing, debugging, and maintaining agentic systems.
}}

\vspace{-0.5em}
\section{Threats to Validity}\label{sec:threats}

\revReplace{R3.11}{
\noindent\textbf{Internal Validity.}
The main internal threat lies in the manual paper screening and classification process, where subjective judgment may exclude relevant studies or misclassify papers into study types and taxonomy categories.
To mitigate this threat, two authors independently apply the inclusion and exclusion criteria using a conservative union rule that retains a paper if either author votes to include it.
They also independently assign study types and apply the type-adapted QAC.
For taxonomy construction, two authors jointly derive each initial taxonomy from 70\% of the corresponding papers and independently label the remaining 30\%.
Cohen's $\kappa$ is reported for the independent stages, and remaining disagreements are discussed with a third author.
We further anchor the study type assignment to the inclusion criteria each paper satisfies, \revReplace{R1.2/R2.3}{conduct searches over four databases, using full-text search where it is supported and title-based metadata search on DBLP,} together with backward and forward snowballing to reduce the risk of missing relevant work, and release the paper lists of every filtering stage on the artifacts page for public scrutiny.

\noindent\textbf{External Validity.}
Our conclusions may not generalize beyond the collection cutoff of May 31, 2026, as issue resolution evolves rapidly.
We mitigate this threat by treating our findings as a snapshot of the field as of the cutoff date.
Additionally, 50.8\% of the collected papers are arXiv preprints that have not undergone peer review, which may affect taxonomies.
We mitigate this threat by applying the same type-adapted quality assessment to all papers and \revReplace{R1.4}{by reconstructing the taxonomy from the peer-reviewed subset alone, which yields an unchanged top-level taxonomy.}
}

\vspace{-0.5em}
\section{Conclusion}\label{sec:conclusion}

In this survey, we present a comprehensive and systematic literature review of LLM-based agentic systems for issue resolution from three perspectives: benchmarks, techniques, and empirical studies, and distill the research opportunities they suggest.
\revReplace{R1.11}{Across these perspectives, three higher-level conclusions emerge.
First, effective design is conditional rather than universal.
Benchmarks should be selected by evaluation purpose, pipeline-based scaffolds suit tight budgets and weaker backbones, agent-based scaffolds suit frontier backbones, and agentic reinforcement learning has become the dominant training paradigm.
Second, the persistent limitation is the gap between measured and actual capability, since weak test suites, data contamination, and unreliable trajectories inflate reported resolution rates and bound the outcome rewards used for training.
Moreover, even for issues counted as resolved, prevailing execution-based evaluation overlooks the software engineering quality of generated patches, such as their maintainability and conformance to project-specific design and architectural constraints.
Third, the field is likely to evolve from maximizing a single resolution rate toward process-aware development and evaluation, including fine-grained multi-perspective assessment, process-level supervision, richer repository knowledge representations, domain-specific customization, and software engineering methods for agentic systems themselves.}
Together, these contributions position this survey as a timely and foundational reference that consolidates fragmented progress, clarifies emerging trends, and supports the development of next-generation research on automated issue resolution.

\bibliographystyle{ACM-Reference-Format}
\bibliography{reference}

@article{jiang2024survey,
  title={A survey on large language models for code generation},
  author={Jiang, Juyong and Wang, Fan and Shen, Jiasi and Kim, Sungju and Kim, Sunghun},
  journal={ACM Transactions on Software Engineering and Methodology},
  volume={35},
  number={2},
  pages={1--72},
  year={2026}
}

@article{kitchenham2009systematic,
  title={Systematic literature reviews in software engineering--a systematic literature review},
  author={Kitchenham, Barbara and Brereton, O Pearl and Budgen, David and Turner, Mark and Bailey, John and Linkman, Stephen},
  journal={Information and software technology},
  volume={51},
  number={1},
  pages={7--15},
  year={2009},
}

@article{hou2024large,
  title={Large language models for software engineering: A systematic literature review},
  author={Hou, Xinyi and Zhao, Yanjie and Liu, Yue and Yang, Zhou and Wang, Kailong and Li, Li and Luo, Xiapu and Lo, David and Grundy, John and Wang, Haoyu},
  journal={ACM Transactions on Software Engineering and Methodology},
  volume={33},
  number={8},
  pages={1--79},
  year={2024},
}

@article{liu2024large,
  title={Large language model-based agents for software engineering: A survey},
  author={Liu, Junwei and Wang, Kaixin and Chen, Yixuan and Peng, Xin and Chen, Zhenpeng and Zhang, Lingming and Lou, Yiling},
  journal={arXiv preprint arXiv:2409.02977},
  year={2024}
}

@inproceedings{
    swebench,
    title={{SWE}-bench: Can Language Models Resolve Real-world Github Issues?},
    author={Carlos E Jimenez and John Yang and Alexander Wettig and Shunyu Yao and Kexin Pei and Ofir Press and Karthik R Narasimhan},
    booktitle={The Twelfth International Conference on Learning Representations},
    year={2024},
}

@inproceedings{
    swebenchmutimodal,
    title={{SWE}-bench Multimodal: Do {AI} Systems Generalize to Visual Software Domains?},
    author={John Yang and Carlos E Jimenez and Alex L Zhang and Kilian Lieret and Joyce Yang and Xindi Wu and Ori Press and Niklas Muennighoff and Gabriel Synnaeve and Karthik R Narasimhan and Diyi Yang and Sida Wang and Ofir Press},
    booktitle={The Thirteenth International Conference on Learning Representations},
    year={2025},
}

@misc{swebenchverified,
  title={Introducing {SWE}-bench Verified},
  author={Chowdhury, Neil and Aung, James and Shern, Chan Jun and Jaffe, Oliver and Sherburn, Dane and Starace, Giulio and Mays, Evan and Dias, Rachel and Aljubeh, Marwan and Glaese, Mia and Jimenez, Carlos E. and Yang, John and Ho, Leyton and Patwardhan, Tejal and Liu, Kevin and Madry, Aleksander},
  year={2024},
  url={https://openai.com/index/introducing-swe-bench-verified/},
}

@inproceedings{locagent,
  title={Locagent: Graph-guided llm agents for code localization},
  author={Chen, Zhaoling and Tang, Robert and Deng, Gangda and Wu, Fang and Wu, Jialong and Jiang, Zhiwei and Prasanna, Viktor and Cohan, Arman and Wang, Xingyao},
  booktitle={Proceedings of the 63rd Annual Meeting of the Association for Computational Linguistics (Volume 1: Long Papers)},
  pages={8697--8727},
  year={2025}
}

@article{agentless,
  title={Demystifying llm-based software engineering agents},
  author={Xia, Chunqiu Steven and Deng, Yinlin and Dunn, Soren and Zhang, Lingming},
  journal={Proceedings of the ACM on Software Engineering},
  volume={2},
  number={FSE},
  pages={801--824},
  year={2025},
}

@article{CGM,
  title={Code graph model (cgm): A graph-integrated large language model for repository-level software engineering tasks},
  author={Tao, Hongyuan and Zhang, Ying and Tang, Zhenhao and Peng, Hongen and Zhu, Xukun and Liu, Bingchang and Yang, Yingguang and Zhang, Ziyin and Xu, Zhaogui and Zhang, Haipeng and others},
  journal={Advances in Neural Information Processing Systems},
  volume={38},
  pages={15869--15909},
  year={2026}
}

@article{MAGIS,
  title={Magis: Llm-based multi-agent framework for github issue resolution},
  author={Tao, Wei and Zhou, Yucheng and Wang, Yanlin and Zhang, Wenqiang and Zhang, Hongyu and Cheng, Yu},
  journal={Advances in Neural Information Processing Systems},
  volume={37},
  pages={51963--51993},
  year={2024}
}

@inproceedings{AutoCodeRover,
  title={Autocoderover: Autonomous program improvement},
  author={Zhang, Yuntong and Ruan, Haifeng and Fan, Zhiyu and Roychoudhury, Abhik},
  booktitle={Proceedings of the 33rd ACM SIGSOFT International Symposium on Software Testing and Analysis},
  pages={1592--1604},
  year={2024}
}

@article{SWEAgent,
  title={Swe-agent: Agent-computer interfaces enable automated software engineering},
  author={Yang, John and Jimenez, Carlos E and Wettig, Alexander and Lieret, Kilian and Yao, Shunyu and Narasimhan, Karthik and Press, Ofir},
  journal={Advances in Neural Information Processing Systems},
  volume={37},
  pages={50528--50652},
  year={2024}
}

@article{CodeR,
  title={Coder: Issue resolving with multi-agent and task graphs},
  author={Chen, Dong and Lin, Shaoxin and Zeng, Muhan and Zan, Daoguang and Wang, Jian-Gang and Cheshkov, Anton and Sun, Jun and Yu, Hao and Dong, Guoliang and Aliev, Artem and others},
  journal={arXiv preprint arXiv:2406.01304},
  year={2024}
}

@inproceedings{repounderstander,
  title={Alibaba lingmaagent: Improving automated issue resolution via comprehensive repository exploration},
  author={Ma, Yingwei and Yang, Qingping and Cao, Rongyu and Li, Binhua and Huang, Fei and Li, Yongbin},
  booktitle={Proceedings of the 33rd ACM International Conference on the Foundations of Software Engineering},
  pages={238--249},
  year={2025}
}

@inproceedings{
  MASAI,
  title={{MASAI}: Modular Architecture for Software-engineering {AI} Agents},
  author={Nalin Wadhwa and Atharv Sonwane and Daman Arora and Abhav Mehrotra and Saiteja Utpala and Ramakrishna B Bairi and Aditya Kanade and Nagarajan Natarajan},
  booktitle={NeurIPS 2024 Workshop on Open-World Agents},
  year={2024},
}

@article{SWT_bench,
  title={SWT-bench: Testing and validating real-world bug-fixes with code agents},
  author={M{\"u}ndler, Niels and M{\"u}ller, Mark and He, Jingxuan and Vechev, Martin},
  journal={Advances in Neural Information Processing Systems},
  volume={37},
  pages={81857--81887},
  year={2024}
}

@article{openhands,
  title={Openhands: An open platform for ai software developers as generalist agents},
  author={Wang, Xingyao and Li, Boxuan and Song, Yufan and Xu, Frank F and Tang, Xiangru and Zhuge, Mingchen and Pan, Jiayi and Song, Yueqi and Li, Bowen and Singh, Jaskirat and others},
  journal={arXiv preprint arXiv:2407.16741},
  year={2024}
}

@article{BLAZE,
  title={BLAZE: Cross-language and cross-project bug localization via dynamic chunking and hard example learning},
  author={Chakraborty, Partha and Alfadel, Mahmoud and Nagappan, Meiyappan},
  journal={IEEE Transactions on Software Engineering},
  year={2025},
}

@inproceedings{SpecRover,
  title={SpecRover: Code Intent Extraction via LLMs},
  author={Ruan, Haifeng and Zhang, Yuntong and Roychoudhury, Abhik},
  booktitle={2025 IEEE/ACM 47th International Conference on Software Engineering},
  pages={963--974},
  year={2025},
}

@inproceedings{CodeXGraph,
  title={Codexgraph: Bridging large language models and code repositories via code graph databases},
  author={Liu, Xiangyan and Lan, Bo and Hu, Zhiyuan and Liu, Yang and Zhang, Zhicheng and Wang, Fei and Shieh, Michael Qizhe and Zhou, Wenmeng},
  booktitle={Proceedings of the 2025 Conference of the Nations of the Americas Chapter of the Association for Computational Linguistics: Human Language Technologies (Volume 1: Long Papers)},
  pages={142--160},
  year={2025}
}

@inproceedings{DEI,
  title={Diversity empowers intelligence: Integrating expertise of software engineering agents},
  author={Zhang, Kexun and Yao, Weiran and Liu, Zuxin and Feng, Yihao and Liu, Zhiwei and Ramapura Narasimha Murthy, Rithesh and Lan, Tian and Li, Lei and Lou, Renze and Xu, Jiacheng and others},
  booktitle={The Thirteenth International Conference on Learning Representations},
  year={2025}
}

@article{swebench_java,
  title={Swe-bench-java: A github issue resolving benchmark for java},
  author={Zan, Daoguang and Huang, Zhirong and Yu, Ailun and Lin, Shaoxin and Shi, Yifan and Liu, Wei and Chen, Dong and Qi, Zongshuai and Yu, Hao and Yu, Lei and others},
  journal={arXiv preprint arXiv:2408.14354},
  year={2024}
}

@article{marscode,
  title={Marscode agent: Ai-native automated bug fixing},
  author={Liu, Yizhou and Gao, Pengfei and Wang, Xinchen and Liu, Jie and Shi, Yexuan and Zhang, Zhao and Peng, Chao},
  journal={arXiv preprint arXiv:2409.00899},
  year={2024}
}

@article{supercoder,
  title={SuperCoder2.0: Technical Report on Exploring the feasibility of LLMs as Autonomous Programmer},
  author={Gautam, Anmol and Kumar, Kishore and Jha, Adarsh and NS, Mukunda and Bhola, Ishaan},
  journal={arXiv preprint arXiv:2409.11190},
  year={2024}
}

@article{hyperagent,
  title={Hyperagent: Generalist software engineering agents to solve coding tasks at scale},
  author={Phan, Huy Nhat and Nguyen, Tien N and Nguyen, Phong X and Bui, Nghi DQ},
  journal={arXiv preprint arXiv:2409.16299},
  year={2024}
}

@article{swebench_plus,
  title={Swe-bench+: Enhanced coding benchmark for llms},
  author={Aleithan, Reem and Xue, Haoran and Mohajer, Mohammad Mahdi and Nnorom, Elijah and Uddin, Gias and Wang, Song},
  journal={arXiv preprint arXiv:2410.06992},
  year={2024}
}

@inproceedings{RepoGraph,
  title={Repograph: Enhancing ai software engineering with repository-level code graph},
  author={Ouyang, Siru and Yu, Wenhao and Ma, Kaixin and Xiao, Zilin and Zhang, Zhihan and Jia, Mengzhao and Han, Jiawei and Zhang, Hongming and Yu, Dong},
  booktitle={The Thirteenth International Conference on Learning Representations},
  year={2025}
}

@inproceedings{SWE-Search,
  title={Swe-search: Enhancing software agents with monte carlo tree search and iterative refinement},
  author={Antoniades, Antonis and {\"O}rwall, Albert and Zhang, Kexun and Xie, Yuxi and Goyal, Anirudh and Wang, William},
  booktitle={The Thirteenth International Conference on Learning Representations},
  year={2025}
}

@article{SWE-GPT,
  title={Swe-gpt: A process-centric language model for automated software improvement},
  author={Ma, Yingwei and Cao, Rongyu and Cao, Yongchang and Zhang, Yue and Chen, Jue and Liu, Yibo and Liu, Yuchen and Li, Binhua and Huang, Fei and Li, Yongbin},
  journal={Proceedings of the ACM on Software Engineering},
  volume={2},
  number={ISSTA},
  pages={2362--2383},
  year={2025},
}

@article{infant,
  title={Infant agent: A tool-integrated, logic-driven agent with cost-effective api usage},
  author={Lei, Bin and Li, Yuchen and Zeng, Yiming and Ren, Tao and Luo, Yi and Shi, Tianyu and Gao, Zitian and Hu, Zeyu and Kang, Weitai and Chen, Qiuwu},
  journal={arXiv preprint arXiv:2411.01114},
  year={2024}
}

@inproceedings{CodeV,
  title={Codev: issue resolving with visual data},
  author={Zhang, Linhao and Zan, Daoguang and Yang, Quanshun and Huang, Zhirong and Chen, Dong and Shen, Bo and Liu, Tianyu and Gong, Yongshun and Pengjie, Huang and Lu, Xudong and others},
  booktitle={Findings of the Association for Computational Linguistics: ACL 2025},
  pages={7350--7361},
  year={2025}
}

@article{ReSAT,
  title={Repository Structure-Aware Training Makes SLMs Better Issue Resolver},
  author={Ma, Zexiong and An, Shengnan and Lin, Zeqi and Zou, Yanzhen and Xie, Bing},
  journal={arXiv preprint arXiv:2412.19031},
  year={2024}
}

@inproceedings{
  SWE-Gym,
  title={Training Software Engineering Agents and Verifiers with {SWE}-Gym},
  author={Jiayi Pan and Xingyao Wang and Graham Neubig and Navdeep Jaitly and Heng Ji and Alane Suhr and Yizhe Zhang},
  booktitle={Forty-second International Conference on Machine Learning},
  year={2025},
}

@inproceedings{SWE-Fixer,
  title={Swe-fixer: Training open-source llms for effective and efficient github issue resolution},
  author={Xie, Chengxing and Li, Bowen and Gao, Chang and Du, He and Lam, Wai and Zou, Difan and Chen, Kai},
  booktitle={Findings of the Association for Computational Linguistics: ACL 2025},
  pages={1123--1139},
  year={2025}
}

@inproceedings{Learn-by-interact,
  title={Learn-by-interact: A data-centric framework for self-adaptive agents in realistic environments},
  author={Su, Hongjin and Sun, Ruoxi and Yoon, Jinsung and Yin, Pengcheng and Yu, Tao and Arik, Sercan},
  booktitle={The Thirteenth International Conference on Learning Representations},
  year={2025}
}

@article{CodeMonkeys,
  title={Codemonkeys: Scaling test-time compute for software engineering},
  author={Ehrlich, Ryan and Brown, Bradley and Juravsky, Jordan and Clark, Ronald and R{\'e}, Christopher and Mirhoseini, Azalia},
  journal={arXiv preprint arXiv:2501.14723},
  year={2025}
}

@inproceedings{
  OrcaLoca,
  title={OrcaLoca: An {LLM} Agent Framework for Software Issue Localization},
  author={Zhongming Yu and Hejia Zhang and Yujie Zhao and Hanxian Huang and Matrix Yao and Ke Ding and Jishen Zhao},
  booktitle={Forty-second International Conference on Machine Learning},
  year={2025},
}

@article{PatchPilot,
  title={Patchpilot: A stable and cost-efficient agentic patching framework},
  author={Li, Hongwei and Tang, Yuheng and Wang, Shiqi and Guo, Wenbo},
  journal={arXiv e-prints},
  year={2025}
}

@inproceedings{
  Otter,
  title={Otter: Generating Tests from Issues to Validate {SWE} Patches},
  author={Toufique Ahmed and Jatin Ganhotra and Rangeet Pan and Avraham Shinnar and Saurabh Sinha and Martin Hirzel},
  booktitle={Forty-second International Conference on Machine Learning},
  year={2025},
}

@article{BugCerberus,
  title={Bridging bug localization and issue fixing: A hierarchical localization framework leveraging large language models},
  author={Chang, Jianming and Zhou, Xin and Lulu, Lulu and Lo, David and Li, Bixin},
  journal={IEEE Transactions on Software Engineering},
  year={2026},
}

@article{SWE-RL,
  title={Swe-rl: Advancing llm reasoning via reinforcement learning on open software evolution},
  author={Wei, Yuxiang and Duchenne, Olivier and Copet, Jade and Carbonneaux, Quentin and Zhang, Lingming and Fried, Daniel and Synnaeve, Gabriel and Singh, Rishabh and Wang, Sida},
  journal={Advances in Neural Information Processing Systems},
  volume={38},
  pages={78500--78525},
  year={2026}
}

@inproceedings{SoRFT,
  title={Sorft: Issue resolving with subtask-oriented reinforced fine-tuning},
  author={Ma, Zexiong and Peng, Chao and Gao, Pengfei and Meng, Xiangxin and Zou, Yanzhen and Xie, Bing},
  booktitle={Proceedings of the 63rd Annual Meeting of the Association for Computational Linguistics (Volume 1: Long Papers)},
  pages={11427--11441},
  year={2025}
}

@inproceedings{DARS,
  title={Dars: Dynamic action re-sampling to enhance coding agent performance by adaptive tree traversal},
  author={Aggarwal, Vaibhav and Kamal, Ojasv and Japesh, Abhinav and Jin, Zhijing and Sch{\"o}lkopf, Bernhard},
  booktitle={Proceedings of the 63rd Annual Meeting of the Association for Computational Linguistics (Volume 1: Long Papers)},
  pages={19808--19855},
  year={2025}
}

@inproceedings{SEAlign,
  author    = {Kechi Zhang and Huangzhao Zhang and Ge Li and
               Jinliang You and Jia Li and Yunfei Zhao and Zhi Jin},
  title     = {SEAlign: Alignment Training for Software Engineering Agent},
  booktitle = {Proceedings of the 48th IEEE/ACM International Conference on Software Engineering},
  year      = {2026},
}

@article{KGCompass,
  title={Enhancing repository-level software repair via repository-aware knowledge graphs},
  author={Yang, Boyang and Ren, Jiadong and Jin, Shunfu and Liu, Yang and Liu, Feng and Le, Bach and Tian, Haoye},
  journal={arXiv preprint arXiv:2503.21710},
  year={2025}
}

@article{CoSIL,
  title={Issue localization via llm-driven iterative code graph searching},
  author={Jiang, Zhonghao and Ren, Xiaoxue and Yan, Meng and Jiang, Wei and Li, Yong and Liu, Zhongxin},
  journal={arXiv preprint arXiv:2503.22424},
  year={2025}
}

@article{mutiswebench,
  title={Multi-swe-bench: A multilingual benchmark for issue resolving},
  author={Zan, Daoguang and Huang, Zhirong and Liu, Wei and Chen, Hanwu and Xin, Shulin and Zhang, Linhao and Liu, Qi and Aoyan, Li and Chen, Lu and Zhong, Xiaojian and others},
  journal={Advances in Neural Information Processing Systems},
  volume={38},
  year={2026}
}

@inproceedings{
  r2e_gym,
  title={R2E-Gym: Procedural Environments and Hybrid Verifiers for Scaling Open-Weights {SWE} Agents},
  author={Naman Jain and Jaskirat Singh and Manish Shetty and Tianjun Zhang and Liang Zheng and Koushik Sen and Ion Stoica},
  booktitle={NeurIPS 2025 Fourth Workshop on Deep Learning for Code},
  year={2025},
}

@article{SWE-Synth,
  title={SWE-Synth: Synthesizing Verifiable Bug-Fix Data to Enable Large Language Models in Resolving Real-World Bugs},
  author={Pham, Minh VT and Phan, Huy N and Phan, Hoang N and Chi, Cuong Le and Nguyen, Tien N and Bui, Nghi DQ},
  journal={arXiv preprint arXiv:2504.14757},
  year={2025}
}

@article{SWE-smith,
  title={Swe-smith: Scaling data for software engineering agents},
  author={Yang, John and Lieret, Kilian and Jimenez, Carlos and Wettig, Alexander and Khandpur, Kabir and Zhang, Yanzhe and Hui, Binyuan and Press, Ofir and Schmidt, Ludwig and Yang, Diyi},
  journal={Advances in Neural Information Processing Systems},
  volume={38},
  year={2026}
}

@inproceedings{
  SweRank,
  title={{SWER}ank: Software Issue Localization with Code Ranking},
  author={Revanth Gangi Reddy and Tarun Suresh and JaeHyeok Doo and Ye Liu and Xuan-Phi Nguyen and Yingbo Zhou and Semih Yavuz and Caiming Xiong and Heng Ji and Shafiq Joty},
  booktitle={The Fourteenth International Conference on Learning Representations},
  year={2026},
}

@article{LCLM,
  title={Putting It All into Context: Simplifying Agents with LCLMs},
  author={Jiang, Mingjian and Ruan, Yangjun and Lastras, Luis and Kapanipathi, Pavan and Hashimoto, Tatsunori},
  journal={arXiv preprint arXiv:2505.08120},
  year={2025}
}

@article{InfantAgent-Next,
  title={InfantAgent-Next: A Multimodal Generalist Agent for Automated Computer Interaction},
  author={Lei, Bin and Kang, Weitai and Zhang, Zijian and Chen, Winson and Xie, Xi and Zuo, Shan and Xie, Mimi and Payani, Ali and Hong, Mingyi and Yan, Yan and others},
  journal={Advances in Neural Information Processing Systems},
  volume={38},
  pages={34494--34520},
  year={2026}
}

@article{Co-PatcheR,
  title={Co-PatcheR: Collaborative Software Patching with Component-specific Small Reasoning Models},
  author={Tang, Yuheng and Li, Hongwei and Zhu, Kaijie and Yang, Michael and Ding, Yangruibo and Guo, Wenbo},
  journal={Advances in Neural Information Processing Systems},
  volume={38},
  pages={42041--42069},
  year={2026}
}

@article{SWE-rebench,
  title={Swe-rebench: An automated pipeline for task collection and decontaminated evaluation of software engineering agents},
  author={Badertdinov, Ibragim and Golubev, Alexander and Nekrashevich, Maksim and Shevtsov, Anton and Karasik, Simon and Andriushchenko, Andrei and Trofimova, Maria and Litvintseva, Daria and Yangel, Boris},
  journal={Advances in Neural Information Processing Systems},
  volume={38},
  year={2026}
}

@article{swebench-live,
  title={Swe-bench goes live!},
  author={Zhang, Linghao and He, Shilin and Zhang, Chaoyun and Kang, Yu and Li, Bowen and Xie, Chengxing and Wang, Junhao and Wang, Maoquan and Huang, Yufan and Fu, Shengyu and others},
  journal={Advances in Neural Information Processing Systems},
  volume={38},
  year={2026}
}

@inproceedings{AEGIS,
  author       = {Xinchen Wang and
                  Pengfei Gao and
                  Xiangxin Meng and
                  Chao Peng and
                  Ruida Hu and
                  Yun Lin and
                  Cuiyun Gao},
  title        = {{AEGIS:} An Agent-based Framework for Bug Reproduction from Issue
                  Descriptions},
  booktitle    = {Companion Proceedings of the 33rd ACM International Conference on the Foundations of Software Engineering},
  pages        = {331--342},
  year         = {2025}
}

@article{openai2024gpt4technicalreport,
  title={Gpt-4 technical report},
  author={Achiam, Josh and Adler, Steven and Agarwal, Sandhini and Ahmad, Lama and Akkaya, Ilge and Aleman, Florencia Leoni and Almeida, Diogo and Altenschmidt, Janko and Altman, Sam and Anadkat, Shyamal and others},
  journal={arXiv preprint arXiv:2303.08774},
  year={2023}
}

@article{deepseekr1,
  title={Deepseek-r1: Incentivizing reasoning capability in llms via reinforcement learning},
  author={Guo, Daya and Yang, Dejian and Zhang, Haowei and Song, Junxiao and Zhang, Ruoyu and Xu, Runxin and Zhu, Qihao and Ma, Shirong and Wang, Peiyi and Bi, Xiao and others},
  journal={arXiv preprint arXiv:2501.12948},
  year={2025}
}

@article{bouzenia2024repairagent,
  title={Repairagent: An autonomous, llm-based agent for program repair},
  author={Bouzenia, Islem and Devanbu, Premkumar and Pradel, Michael},
  journal={arXiv preprint arXiv:2403.17134},
  year={2024}
}

@inproceedings{chatrepair,
  title={Automated program repair via conversation: Fixing 162 out of 337 bugs for \$0.42 each using ChatGPT},
  author={Xia, Chunqiu Steven and Zhang, Lingming},
  booktitle={Proceedings of the 33rd ACM SIGSOFT International Symposium on Software Testing and Analysis},
  pages={819--831},
  year={2024}
}

@article{trae2025,
  title={Trae Agent: An LLM-based Agent for Software Engineering with Test-time Scaling},
  author={Gao, Pengfei and Tian, Zhao and Meng, Xiangxin and Wang, Xinchen and Hu, Ruida and Xiao, Yuanan and Liu, Yizhou and Zhang, Zhao and Chen, Junjie and Gao, Cuiyun and others},
  journal={arXiv preprint arXiv:2507.23370},
  year={2025}
}

@inproceedings{
  SWA,
  title={Automated Benchmark Generation for Repository-Level Coding Tasks},
  author={Konstantinos Vergopoulos and Mark Niklas Mueller and Martin Vechev},
  booktitle={Forty-second International Conference on Machine Learning},
  year={2025},
}

@inproceedings{FEABench,
  title={Fea-bench: A benchmark for evaluating repository-level code generation for feature implementation},
  author={Li, Wei and Zhang, Xin and Guo, Zhongxin and Mao, Shaoguang and Luo, Wen and Peng, Guangyue and Huang, Yangyu and Wang, Houfeng and Li, Scarlett},
  booktitle={Proceedings of the 63rd Annual Meeting of the Association for Computational Linguistics (Volume 1: Long Papers)},
  pages={17160--17176},
  year={2025}
}

@misc{liveswebench2025,
  author       = {{LiveSWEBench Team}},
  title        = {LiveSWEBench: A Contamination‑Free Benchmark for AI Software Engineers},
  howpublished = {Official Website},
  year         = 2025,
  url          = {https://liveswebench.ai/},
  note         = {Accessed 2025-06-14}
}

@article{swepoly,
  title={SWE-PolyBench: A multi-language benchmark for repository level evaluation of coding agents},
  author={Rashid, Muhammad Shihab and Bock, Christian and Zhuang, Yuan and Buchholz, Alexander and Esler, Tim and Valentin, Simon and Franceschi, Luca and Wistuba, Martin and Sivaprasad, Prabhu Teja and Kim, Woo Jung and others},
  journal={arXiv preprint arXiv:2504.08703},
  year={2025}
}

@article{omnigirl,
  title={Omnigirl: A multilingual and multimodal benchmark for github issue resolution},
  author={Guo, Lianghong and Tao, Wei and Jiang, Runhan and Wang, Yanlin and Chen, Jiachi and Liu, Xilin and Ma, Yuchi and Mao, Mingzhi and Zhang, Hongyu and Zheng, Zibin},
  journal={Proceedings of the ACM on Software Engineering},
  volume={2},
  number={ISSTA},
  pages={24--46},
  year={2025},
}

@misc{swebenchmutilingual,
  author       = {Kabir Khandpur},
  title        = {SWE‑bench Multilingual: Evaluating LLMs on Software Engineering Across Multiple Programming Languages},
  howpublished = {Kabir Khandpur’s Blog},
  month        = may,
  day          = 6,
  year         = 2025,
  url          = {https://kabirk.com/multilingual},
  note         = {Accessed 2025-06-15}
}

@inproceedings{testgeneval,
  title={Testgeneval: A real world unit test generation and test completion benchmark},
  author={Jain, Kush and Synnaeve, Gabriel and Rozi{\`e}re, Baptiste},
  booktitle={The Thirteenth International Conference on Learning Representations},
  year={2025}
}

@article{tddbench,
  title={TDD-Bench Verified: Can LLMs Generate Tests for Issues Before They Get Resolved?},
  author={Ahmed, Toufique and Hirzel, Martin and Pan, Rangeet and Shinnar, Avraham and Sinha, Saurabh},
  journal={arXiv preprint arXiv:2412.02883},
  year={2024}
}

@inproceedings{utboost,
  title={Utboost: Rigorous evaluation of coding agents on swe-bench},
  author={Yu, Boxi and Zhu, Yuxuan and He, Pinjia and Kang, Daniel},
  booktitle={Proceedings of the 63rd Annual Meeting of the Association for Computational Linguistics (Volume 1: Long Papers)},
  pages={3762--3774},
  year={2025}
}

@inproceedings{swe-mera,
  title={SWE-MERA: A Dynamic Benchmark for Agenticly Evaluating Large Language Models on Software Engineering Tasks},
  author={Pavel, Adamenko and Mikhail, Ivanov and Valeev, Aidar and Levichev, Rodion and Zadorozhny, Pavel and Lopatin, Ivan and Babaev, Dmitrii and Fenogenova, Alena and Malykh, Valentin},
  booktitle={Proceedings of the 2025 Conference on Empirical Methods in Natural Language Processing: System Demonstrations},
  pages={440--452},
  year={2025}
}

@inproceedings{
  swe-perf,
  title={{SWE}-Perf: Can Language Models Optimize Code Performance on Real-World Repositories?},
  author={Xinyi He and Qian Liu and Mingzhe Du and Lin Yan and ZhiJie Fan and Yiming Huang and Zejian Yuan and Zejun MA},
  booktitle={NeurIPS 2025 Fourth Workshop on Deep Learning for Code},
  year={2025},
}

@article{swebench-cl,
  title={SWE-Bench-CL: Continual Learning for Coding Agents},
  author={Joshi, Thomas and Chowdhury, Shayan and Uysal, Fatih},
  journal={arXiv preprint arXiv:2507.00014},
  year={2025}
}

@article{nocode-bench,
  title={NoCode-bench: A Benchmark for Evaluating Natural Language-Driven Feature Addition},
  author={Deng, Le and Jiang, Zhonghao and Cao, Jialun and Pradel, Michael and Liu, Zhongxin},
  journal={arXiv preprint arXiv:2507.18130},
  year={2025}
}

@INPROCEEDINGS{oliva2025spice,
  author={Oliva, Gustavo A. and Rajbahadur, Gopi Krishnan and Bhatia, Aaditya and Zhang, Haoxiang and Chen, Yihao and Chen, Zhilong and Leung, Arthur and Lin, Dayi and Chen, Boyuan and Hassan, Ahmed E.},
  booktitle={2025 40th IEEE/ACM International Conference on Automated Software Engineering}, 
  title={SPICE: An Automated SWE-Bench Labeling Pipeline for Issue Clarity, Test Coverage, and Effort Estimation}, 
  year={2025},
  pages={2325--2337},
}

@article{gso,
  title={Gso: Challenging software optimization tasks for evaluating swe-agents},
  author={Shetty, Manish and Jain, Naman and Liu, Jinjian and Kethanaboyina, Vijay and Sen, Koushik and Stoica, Ion},
  journal={Advances in Neural Information Processing Systems},
  volume={38},
  year={2026}
}

@inproceedings{
  OpenHands-Versa,
  title={Coding Agents with Multimodal Browsing are Generalist Problem Solvers},
  author={Aditya Bharat Soni and Boxuan Li and Xingyao Wang and Valerie Chen and Graham Neubig},
  booktitle={ICML 2025 Workshop on Computer Use Agents},
  year={2025},
}

@article{experepair,
  author  = {Fangwen Mu and Junjie Wang and Lin Shi and
             Song Wang and Shoubin Li and Qing Wang},
  title   = {ExpeRepair: Dual-Memory Enhanced LLM-Based Repository-Level Program Repair},
  journal = {Proceedings of the ACM on Software Engineering},
  volume  = {3},
  number  = {FSE},
  pages   = {FSE174:1--FSE174:23},
  year    = {2026},
}

@inproceedings{GUIRepair,
  title={Seeing is Fixing: Cross-Modal Reasoning with Multimodal LLMs for Visual Software Issue Repair},
  author={Huang, Kai and Zhang, Jian and Xie, Xiaofei and Chen, Chunyang},
  booktitle={2025 40th IEEE/ACM International Conference on Automated Software Engineering},
  pages={1156--1168},
  year={2025},
}

@article{SEMAgent,
  title={SemAgent: A Semantics Aware Program Repair Agent},
  author={Pabba, Anvith and Mathai, Alex and Chakraborty, Anindya and Ray, Baishakhi},
  journal={arXiv preprint arXiv:2506.16650},
  year={2025}
}

@article{AgentKB,
  title={Agent kb: Leveraging cross-domain experience for agentic problem solving},
  author={Tang, Xiangru and Qin, Tianrui and Peng, Tianhao and Zhou, Ziyang and Shao, Daniel and Du, Tingting and Wei, Xinming and Xia, Peng and Wu, Fang and Zhu, He and others},
  journal={arXiv preprint arXiv:2507.06229},
  year={2025}
}

@article{Prometheus,
  title={Prometheus: Unified Knowledge Graphs for Issue Resolution in Multilingual Codebases},
  author={Chen, Zimin and Pan, Yue and Lu, Siyu and Xu, Jiayi and Goues, Claire Le and Monperrus, Martin and Ye, He},
  journal={arXiv preprint arXiv:2507.19942},
  year={2025}
}

@article{SWE-EXP,
  title={SWE-Exp: Experience-Driven Software Issue Resolution},
  author={Chen, Silin and Lin, Shaoxin and Gu, Xiaodong and Shi, Yuling and Lian, Heng and Yun, Longfei and Chen, Dong and Sun, Weiguo and Cao, Lin and Wang, Qianxiang},
  journal={arXiv preprint arXiv:2507.23361},
  year={2025}
}

@inproceedings{SWE-Debate,
  author    = {Han Li and Yuling Shi and Shaoxin Lin and
               Xiaodong Gu and Heng Lian and Xin Wang and
               Yantao Jia and Tao Huang and Qianxiang Wang},
  title     = {SWE-Debate: Competitive Multi-Agent Debate for Software Issue Resolution},
  booktitle = {Proceedings of the 48th IEEE/ACM International Conference on Software Engineering},
  year      = {2026},
}

@article{SE-Agent,
  title={SE-agent: Self-evolution trajectory optimization in multi-step reasoning with LLM-based agents},
  author={Guo, Yifu and Lin, Jiaye and Wang, Huacan and Han, Yuzhen and Hu, Sen and Ni, Ziyi and Wang, Licheng and Chen, Mingguang},
  journal={Advances in Neural Information Processing Systems},
  volume={38},
  pages={116314--116341},
  year={2026}
}

@inproceedings{
  Nemotron-Cortexa,
  title={Nemotron-{CORTEXA}: Enhancing {LLM} Agents for Software Engineering Tasks via Improved Localization and Solution Diversity},
  author={Atefeh Sohrabizadeh and Jialin Song and Mingjie Liu and Rajarshi Roy and Chankyu Lee and Jonathan Raiman and Bryan Catanzaro},
  booktitle={Forty-second International Conference on Machine Learning},
  year={2025},
}

@inproceedings{SynFix,
  title={SynFix: Dependency-aware program repair via RelationGraph analysis},
  author={Tang, Xunzhu and Gao, Jiechao and Xu, Jin and Sun, Tiezhu and Song, Yewei and Ezzini, Saad and Ou{\'e}draogo, Wendk{\^u}uni C and Klein, Jacques and Bissyand{\'e}, Tegawend{\'e} F},
  booktitle={Findings of the Association for Computational Linguistics: ACL 2025},
  pages={4878--4894},
  year={2025}
}

@article{bm25,
  title={The probabilistic relevance framework: BM25 and beyond},
  author={Robertson, Stephen and Zaragoza, Hugo and others},
  journal={Foundations and Trends{\textregistered} in Information Retrieval},
  volume={3},
  number={4},
  pages={333--389},
  year={2009},
}

@inproceedings{jones2002visualization,
  title={Visualization of test information to assist fault localization},
  author={Jones, James A and Harrold, Mary Jean and Stasko, John},
  booktitle={Proceedings of the 24th international conference on Software engineering},
  pages={467--477},
  year={2002}
}

@inproceedings{abreu2007accuracy,
  title={On the accuracy of spectrum-based fault localization},
  author={Abreu, Rui and Zoeteweij, Peter and Van Gemund, Arjan JC},
  booktitle={Testing: Academic and industrial conference practice and research techniques-MUTATION},
  pages={89--98},
  year={2007},
}

@inproceedings{coret,
  title={CoRet: Improved Retriever for Code Editing},
  author={Fehr, Fabio James and Franceschi, Luca and Zappella, Giovanni and others},
  booktitle={Proceedings of the 63rd Annual Meeting of the Association for Computational Linguistics (Volume 2: Short Papers)},
  pages={775--789},
  year={2025}
}

@inproceedings{SACL,
    title = "{SACL}: Understanding and Combating Textual Bias in Code Retrieval with Semantic-Augmented Reranking and Localization",
    author = "Gupta, Dhruv  and
      Lakshmy, Gayathri Ganesh  and
      Xie, Yiqing",
    booktitle = "Findings of the Association for Computational Linguistics: EMNLP 2025",
    year = "2025",
    pages = "25052--25065",
}

@article{meta-rag,
  title={Meta-RAG on Large Codebases Using Code Summarization},
  author={Tawosi, Vali and Alamir, Salwa and Liu, Xiaomo and Veloso, Manuela},
  journal={arXiv preprint arXiv:2508.02611},
  year={2025}
}

@article{RepoSearcher,
  title={Tool-integrated Reinforcement Learning for Repo Deep Search},
  author={Ma, Zexiong and Peng, Chao and Zeng, Qunhong and Gao, Pengfei and Zou, Yanzhen and Xie, Bing},
  journal={arXiv preprint arXiv:2508.03012},
  year={2025}
}

@article{BRT,
  title={Agentic bug reproduction for effective automated program repair at google},
  author={Cheng, Runxiang and Tufano, Michele and Cito, J{\"u}rgen and Cambronero, Jos{\'e} and Rondon, Pat and Wei, Renyao and Sun, Aaron and Chandra, Satish},
  journal={arXiv preprint arXiv:2502.01821},
  year={2025}
}

@inproceedings{issue2test,
  author    = {Noor Nashid and Islem Bouzenia and Michael Pradel and Ali Mesbah},
  title     = {Issue2Test: Generating Reproducing Test Cases from Issue Reports},
  booktitle = {Proceedings of the 48th IEEE/ACM International Conference on Software Engineering},
  year      = {2026},
}

@inproceedings{AssertFlip,
  author    = {Lara Khatib and Noble Saji Mathews and Meiyappan Nagappan},
  title     = {AssertFlip: Reproducing Bugs via Inversion of LLM-Generated Passing Tests},
  booktitle = {Proceedings of the 48th IEEE/ACM International Conference on Software Engineering},
  year      = {2026},
}

@inproceedings{e-otter,
  author    = {Toufique Ahmed and Jatin Ganhotra and
               Avraham Shinnar and Martin Hirzel},
  title     = {Heterogeneous Prompting and Execution Feedback for SWE Issue Test Generation and Selection},
  booktitle = {Proceedings of the 48th IEEE/ACM International Conference on Software Engineering},
  year      = {2026},
}

@INPROCEEDINGS{BLAST,
  author={Kitsios, Konstantinos and Castelluccio, Marco and Bacchelli, Alberto},
  booktitle={2025 40th IEEE/ACM International Conference on Automated Software Engineering}, 
  title={Automated Generation of Issue-Reproducing Tests by Combining LLMs and Search-Based Testing},
  year={2025},
  pages={1982--1994},
}

@inproceedings{Libro,
  title={Large Language Models are Few-shot Testers: Exploring LLM-based General Bug Reproduction},
  author={Kang, Sungmin and Yoon, Juyeon and Yoo, Shin},
  booktitle={45th IEEE/ACM International Conference on Software Engineering},
  pages={2312--2323},
  year={2023},
}

@inproceedings{swe-dev_thu,
  title={Swe-dev: Building software engineering agents with training and inference scaling},
  author={Wang, Haoran and Hou, Zhenyu and Wei, Yao and Tang, Jie and Dong, Yuxiao},
  booktitle={Findings of the Association for Computational Linguistics: ACL 2025},
  pages={3742--3761},
  year={2025}
}

@inproceedings{
  swe-dev_sjtu,
  title={{SWE}-Dev: Evaluating and Training Autonomous Feature-Driven Software Development},
  author={Yaxin Du and Yuzhu Cai and Yifan Zhou and Cheng Wang and Yu Qian and Xianghe Pang and Qian Liu and Yue Hu and Siheng Chen},
  booktitle={NeurIPS 2025 Fourth Workshop on Deep Learning for Code},
  year={2025}
}

@article{swe-mirror,
  title={SWE-Mirror: Scaling Issue-Resolving Datasets by Mirroring Issues Across Repositories},
  author={Wang, Junhao and Zan, Daoguang and Xin, Shulin and Liu, Siyao and Wu, Yurong and Shen, Kai},
  journal={arXiv preprint arXiv:2509.08724},
  year={2025}
}

@inproceedings{swe-resoner,
  title={Thinking longer, not larger: Enhancing software engineering agents via scaling test-time compute},
  author={Ma, Yingwei and Li, Yongbin and Dong, Yihong and Jiang, Xue and Li, Yanhao and Liu, Yue and Cao, Rongyu and Chen, Jue and Huang, Fei and Li, Binhua},
  booktitle={2025 40th IEEE/ACM International Conference on Automated Software Engineering},
  pages={3730--3741},
  year={2025},
}

@article{Satori-SWE,
  title={Satori-SWE: Evolutionary Test-Time Scaling for Sample-Efficient Software Engineering},
  author={Zeng, Guangtao and Shen, Maohao and Chen, Delin and Qi, Zhenting and Das, Subhro and Gutfreund, Dan and Cox, David and Wornell, Gregory and Lu, Wei and Hong, Zhang-Wei and others},
  journal={arXiv preprint arXiv:2505.23604},
  year={2025}
}

@article{Agent-RLVR,
  title={Agent-RLVR: Training Software Engineering Agents via Guidance and Environment Rewards},
  author={Da, Jeff and Wang, Clinton and Deng, Xiang and Ma, Yuntao and Barhate, Nikhil and Hendryx, Sean},
  journal={arXiv preprint arXiv:2506.11425},
  year={2025}
}

@INPROCEEDINGS{MCTS-Refined,
  author={Wang, Yibo and Peng, Zhihao and Wang, Ying and Wei, Zhao and Yu, Hai and Zhu, Zhiliang},
  booktitle={2025 40th IEEE/ACM International Conference on Automated Software Engineering}, 
  title={MCTS-Refined CoT: High-Quality Fine-Tuning Data for LLM-Based Repository Issue Resolution}, 
  year={2025},
  pages={1844--1855},
}

@article{Skywork-SWE,
  title={Skywork-SWE: Unveiling Data Scaling Laws for Software Engineering in LLMs},
  author={Zeng, Liang and Li, Yongcong and Xiao, Yuzhen and Li, Changshi and Liu, Chris Yuhao and Yan, Rui and Wei, Tianwen and He, Jujie and Song, Xuchen and Liu, Yang and others},
  journal={arXiv preprint arXiv:2506.19290},
  year={2025}
}

@article{RepoForge,
  title={RepoForge: Training a SOTA Fast-thinking SWE Agent with an End-to-End Data Curation Pipeline Synergizing SFT and RL at Scale},
  author={Chen, Zhilong and Zhao, Chengzong and Chen, Boyuan and Lin, Dayi and Chen, Yihao and Leung, Arthur and Rajbahadur, Gopi Krishnan and Oliva, Gustavo A and Zhang, Haoxiang and Bhatia, Aaditya and others},
  journal={arXiv preprint arXiv:2508.01550},
  year={2025}
}

@article{golubev2025training,
  title={Training Long-Context, Multi-Turn Software Engineering Agents with Reinforcement Learning},
  author={Golubev, Alexander and Trofimova, Maria and Polezhaev, Sergei and Badertdinov, Ibragim and Nekrashevich, Maksim and Shevtsov, Anton and Karasik, Simon and Abramov, Sergey and Andriushchenko, Andrei and Fisin, Filipp and others},
  journal={arXiv preprint arXiv:2508.03501},
  year={2025}
}

@inproceedings{
  SWE-PRM,
  title={When Agents go Astray: Course-Correcting {SWE} Agents with {PRM}s},
  author={Shubham Gandhi and Jason Tsay and Jatin Ganhotra and Kiran Kate and Yara Rizk},
  booktitle={Workshop on Scaling Environments for Agents},
  year={2025},
}

@online{Claude-4.5,
  author       = {Anthropic},
  title        = {Introducing Claude Sonnet 4.5},
  year         = {2025},
  month        = {September},
  day          = {29},
  url          = {https://www.anthropic.com/news/claude-sonnet-4-5},
  note         = {Accessed: 2025-10-09},
}

@misc{swe-bench-extra,
  title={Scaling data collection for training software engineering agents},
  author={Badertdinov, Ibragim and Trofimova, Maria and Anapolskiy, Yury and Abramov, Sergey and Zainullina, Karina and Golubev, Alexander and Polezhaev, Sergey and Litvintseva, Daria and Karasik, Simon and Fisin, Filipp and Skvortsov, Sergey and Nekrashevich, Maxim and Shevtsov, Anton and Yangel, Boris},
  year={2024},
  howpublished={Nebius Blog},
}

@inproceedings{R2E,
  title={R2e: Turning any github repository into a programming agent environment},
  author={Jain, Naman and Shetty, Manish and Zhang, Tianjun and Han, King and Sen, Koushik and Stoica, Ion},
  booktitle={Forty-first International Conference on Machine Learning},
  year={2024}
}

@article{rl_survey,
  title={The landscape of agentic reinforcement learning for llms: A survey},
  author={Zhang, Guibin and Geng, Hejia and Yu, Xiaohang and Yin, Zhenfei and Zhang, Zaibin and Tan, Zelin and Zhou, Heng and Li, Zhongzhi and Xue, Xiangyuan and Li, Yijiang and others},
  journal={arXiv preprint arXiv:2509.02547},
  year={2025}
}

@article{devstal_small,
  title={Devstral: Fine-tuning Language Models for Coding Agent Applications},
  author={Rastogi, Abhinav and Yang, Adam and Jiang, Albert Q and Liu, Alexander H and Sablayrolles, Alexandre and H{\'e}liou, Am{\'e}lie and Martin, Am{\'e}lie and Agarwal, Anmol and Ehrenberg, Andy and Lo, Andy and others},
  journal={arXiv preprint arXiv:2509.25193},
  year={2025}
}

@inproceedings{
  kimi-dev,
  title={Kimi-Dev: Agentless Training as Skill Prior for {SWE}-agents},
  author={Zonghan Yang and Shengjie Wang and Kelin Fu and Wenyang He and Weimin Xiong and Yibo Liu and Yibo Miao and Bofei Gao and Yejie Wang and YINGWEI MA and Yanhao Li and Yue Liu and Zhenxing Hu and kaitai zhang and Shuyi Wang and Huarong Chen and Flood Sung and Yang Liu and Yang Gao and Zhilin Yang and Tianyu Liu},
  booktitle={The Fourteenth International Conference on Learning Representations},
  year={2026},
}

@article{CWM,
  title={CWM: An Open-Weights LLM for Research on Code Generation with World Models},
  author={Carbonneaux, Quentin and Cohen, Gal and Gehring, Jonas and Kahn, Jacob and Kossen, Jannik and Kreuk, Felix and McMilin, Emily and Meyer, Michel and Wei, Yuxiang and Zhang, David and others},
  journal={arXiv preprint arXiv:2510.02387},
  year={2025}
}

@misc{DeepSWE,
  title={DeepSWE: Training a State-of-the-Art Coding Agent from Scratch by Scaling RL},
  author={Michael Luo and Naman Jain and Jaskirat Singh and Sijun Tan and Ameen Patel and Qingyang Wu and Alpay Ariyak and Colin Cai and Tarun Venkat, Shang Zhu and Ben Athiwaratkun and Manan Roongta and Ce Zhang and Li Erran Li and Raluca Ada Popa and Koushik Sen and Ion Stoica},
  howpublished={\url{https://pretty-radio-b75.notion.site/DeepSWE-Training-a-Fully-Open-sourced-State-of-the-Art-Coding-Agent-by-Scaling-RL-22281902c1468193aabbe9a8c59bbe33}},
  note={Notion Blog},
  year={2025}
}

@misc{SWESwiss,
  title={SWE-Swiss: A Multi-Task Fine-Tuning and RL Recipe for High-Performance Issue Resolution},
  author={He, Zhenyu and Yang, Qingping and Sheng, Wei and Zhong, Xiaojian and Zhang, Kechi and An, Chenxin and Shi, Wenlei and Cai, Tianle and He, Di and Chen, Jiaze and Xu, Jingjing},
  howpublished={\url{https://www.notion.so/SWE-Swiss-A-Multi-Task-Fine-Tuning-and-RL-Recipe-for-High-Performance-Issue-Resolution-21e174dedd4880ea829ed4c861c44f88}},
  note={Notion Blog},
  year={2025}
}

@inproceedings{chen2025evaluating,
  title={Evaluating software development agents: Patch patterns, code quality, and issue complexity in real-world github scenarios},
  author={Chen, Zhi and Jiang, Lingxiao},
  booktitle={2025 IEEE International Conference on Software Analysis, Evolution and Reengineering},
  pages={657--668},
  year={2025},
}

@article{meng2024empirical,
  title={An empirical study on llm-based agents for automated bug fixing},
  author={Meng, Xiangxin and Ma, Zexiong and Gao, Pengfei and Peng, Chao},
  journal={arXiv preprint arXiv:2411.10213},
  year={2024}
}

@article{yadavally2025critics,
  title={Large language model critics for execution-free evaluation of code changes},
  author={Yadavally, Aashish and Nguyen, Hoan and Callot, Laurent and Guinet, Gauthier},
  journal={arXiv preprint arXiv:2501.16655},
  year={2025}
}

@inproceedings{vijayvargiya2025interactive,
  title={Ambig-swe: Interactive agents to overcome underspecificity in software engineering},
  author={Vijayvargiya, Sanidhya and Zhou, Xuhui and Yerukola, Akhila and Sap, Maarten and Neubig, Graham},
  booktitle={The Fourteenth International Conference on Learning Representations},
  year={2026}
}

@inproceedings{chen2025unveiling,
  title={Beyond Final Code: A Process-Oriented Error Analysis of Software Development Agents in Real-World GitHub Scenarios},
  author={Chen, Zhi and Ma, Wei and Jiang, Lingxiao},
  booktitle={Proceedings of the 48th IEEE/ACM International Conference on Software Engineering},
  year={2025}
}

@inproceedings{patchdiff,
  author    = {You Wang and Michael Pradel and Zhongxin Liu},
  title     = {Are ``Solved Issues'' in SWE-bench Really Solved Correctly? An Empirical Study},
  booktitle = {Proceedings of the 48th IEEE/ACM International Conference on Software Engineering},
  year      = {2026},
}

@article{liang2025Illusion,
  title={The SWE-Bench Illusion: When State-of-the-Art LLMs Remember Instead of Reason},
  author={Liang, Shanchao and Garg, Spandan and Moghaddam, Roshanak Zilouchian},
  journal={arXiv preprint arXiv:2506.12286},
  year={2025}
}

@inproceedings{martinez2025dissecting,
  author    = {Matias Martinez and Xavier Franch},
  title     = {What's in a Benchmark? The Case of SWE-Bench in Automated Program Repair},
  booktitle = {Proceedings of the 48th IEEE/ACM International Conference on Software Engineering: Software Engineering in Practice},
  year      = {2026},
  pages      = {1--12},
}

@article{xue2025pagent,
  title={PAGENT: Learning to Patch Software Engineering Agents},
  author={Xue, Haoran and Uddin, Gias and Wang, Song},
  journal={arXiv preprint arXiv:2506.17772},
  year={2025}
}

@inproceedings{bouzenia2025understanding,
  title={Understanding Software Engineering Agents: A Study of Thought-Action-Result Trajectories},
  author={Bouzenia, Islem and Pradel, Michael},
  booktitle={2025 40th IEEE/ACM International Conference on Automated Software Engineering},
  pages={2846--2857},
  year={2025}
}

@article{sajadi2025Secure,
  title={Are AI-Generated Fixes Secure? Analyzing LLM and Agent Patches on SWE-bench},
  author={Sajadi, Amirali and Damevski, Kostadin and Chatterjee, Preetha},
  journal={arXiv preprint arXiv:2507.02976},
  year={2025}
}

@article{SWE-Effi,
  title={SWE-Effi: Re-Evaluating Software AI Agent System Effectiveness Under Resource Constraints},
  author={Fan, Zhiyu and Vasilevski, Kirill and Lin, Dayi and Chen, Boyuan and Chen, Yihao and Zhong, Zhiqing and Zhang, Jie M and He, Pinjia and Hassan, Ahmed E},
  journal={arXiv preprint arXiv:2509.09853},
  year={2025}
}

@article{TRAIL2025,
  title={TRAIL: Trace Reasoning and Agentic Issue Localization},
  author={Deshpande, Darshan and Gangal, Varun and Mehta, Hersh and Krishnan, Jitin and Kannappan, Anand and Qian, Rebecca},
  journal={arXiv preprint arXiv:2505.08638},
  year={2025}
}

@inproceedings{CoRNStack,
  title={CoRNStack: High-Quality Contrastive Data for Better Code Retrieval and Reranking},
  author={Suresh, Tarun and Reddy, Revanth Gangi and Xu, Yifei and Nussbaum, Zach and Mulyar, Andriy and Duderstadt, Brandon and Ji, Heng},
  booktitle={The Thirteenth International Conference on Learning Representations},
  year={2025}
}

@article{liu2025empirical,
  title={An Empirical Study on Failures in Automated Issue Solving},
  author={Liu, Simiao and Liu, Fang and Li, Liehao and Tan, Xin and Zhu, Yinghao and Lian, Xiaoli and Zhang, Li},
  journal={arXiv preprint arXiv:2509.13941},
  year={2025}
}

@inproceedings{
  wang2025improving,
  title={Improving Code Localization with Repository Memory},
  author={Boshi Wang and Weijian Xu and Yunsheng Li and Xuemei Gao and Yujia Xie and Huan Sun and Dongdong Chen},
  booktitle={The Fourteenth International Conference on Learning Representations},
  year={2026},
}

@article{SWEbenchPro,
  title={SWE-Bench Pro: Can AI Agents Solve Long-Horizon Software Engineering Tasks?},
  author={Deng, Xiang and Da, Jeff and Pan, Edwin and He, Yannis Yiming and Ide, Charles and Garg, Kanak and Lauffer, Niklas and Park, Andrew and Pasari, Nitin and Rane, Chetan and others},
  journal={arXiv preprint arXiv:2509.16941},
  year={2025}
}

@inproceedings{garg2025saving,
  author    = {Spandan Garg and Benjamin Steenhoek and Yufan Huang},
  title     = {Saving SWE-Bench: A Benchmark Mutation Approach for Realistic Agent Evaluation},
  booktitle = {Proceedings of the IEEE/ACM International Conference on AI Engineering -- Software Engineering for AI},
  year      = {2026},
}

@article{mulocbench,
  title={A Benchmark for Localizing Code and Non-Code Issues in Software Projects},
  author={Zhang, Zejun and Wang, Jian and Yang, Qingyun and Pan, Yifan and Tang, Yi and Li, Yi and Xing, Zhenchang and Zhang, Tian and Li, Xuandong and Zhang, Guoan},
  journal={arXiv preprint arXiv:2509.25242},
  year={2025}
}

@article{lita,
  title={Lita: Light Agent Uncovers the Agentic Coding Capabilities of LLMs},
  author={Dai, Hankun and Wang, Maoquan and Qi, Mengnan and Zhang, Yikai and Jin, Zijian and Yao, Yongqiang and Huang, Yufan and Fu, Shengyu and Nallipogu, Elsie},
  journal={arXiv preprint arXiv:2509.25873},
  year={2025}
}

@article{guo2025comprehensive,
  title={A Comprehensive Survey on Benchmarks and Solutions in Software Engineering of LLM-Empowered Agentic System},
  author={Guo, Jiale and Huang, Suizhi and Li, Mei and Huang, Dong and Chen, Xingsheng and Zhang, Regina and Guo, Zhijiang and Yu, Han and Yiu, Siu-Ming and Jensen, Christian and others},
  journal={arXiv preprint arXiv:2510.09721},
  year={2025}
}

@article{tao2025retrieval,
  title={Retrieval-Augmented Code Generation: A Survey with Focus on Repository-Level Approaches},
  author={Tao, Yicheng and Qin, Yao and Liu, Yepang},
  journal={arXiv preprint arXiv:2510.04905},
  year={2025}
}

@article{shrivastava2023repofusion,
  title={Repofusion: Training code models to understand your repository},
  author={Shrivastava, Disha and Kocetkov, Denis and De Vries, Harm and Bahdanau, Dzmitry and Scholak, Torsten},
  journal={arXiv preprint arXiv:2306.10998},
  year={2023}
}

@article{lientz1978characteristics,
  title={Characteristics of application software maintenance},
  author={Lientz, Bennet P and Swanson, E. Burton and Tompkins, Gail E},
  journal={Communications of the ACM},
  volume={21},
  number={6},
  pages={466--471},
  year={1978},
}

@article{rahman2024systematic,
  title={A Systematic Literature Review on Software Maintenance Offshoring Decisions},
  author={Rahman, Hanif Ur and da Silva, Alberto Rodrigues and Alzayed, Asaad and Raza, Mushtaq},
  journal={Information and Software Technology},
  volume={172},
  pages={107475},
  year={2024},
}

@article{berhe2023maintenance,
  title={Maintenance cost of software ecosystem updates},
  author={Berhe, Solomon and Maynard, Marc and Khomh, Foutse},
  journal={Procedia Computer Science},
  volume={220},
  pages={608--615},
  year={2023},
}

@article{kuramoto2024understanding,
  title={Understanding the characteristics and the role of visual issue reports},
  author={Kuramoto, Hiroki and Wang, Dong and Kondo, Masanari and Kashiwa, Yutaro and Kamei, Yasutaka and Ubayashi, Naoyasu},
  journal={Empirical Software Engineering},
  volume={29},
  number={4},
  pages={89},
  year={2024},
}

@article{chen2025old,
  author={Chen, Yang and Ahmed, Toufique and Jabbarvand, Reyhaneh and Hirzel, Martin},
  title     = {Can Old Tests Do New Tricks for Resolving SWE Issues?},
  journal   = {Proceedings of the ACM on Software Engineering},
  volume    = {3},
  number    = {FSE},
  pages     = {141:1--141:22},
  year      = {2026},
}

@inproceedings{kauhanen2021regression,
  title={Regression test selection tool for python in continuous integration process},
  author={Kauhanen, Eero and Nurminen, Jukka K and Mikkonen, Tommi and Pashkovskiy, Matvei},
  booktitle={2021 IEEE International Conference on Software Analysis, Evolution and Reengineering},
  pages={618--621},
  year={2021},
}

@inproceedings{huang2024crashtranslator,
  title={Crashtranslator: Automatically reproducing mobile application crashes directly from stack trace},
  author={Huang, Yuchao and Wang, Junjie and Liu, Zhe and Wang, Yawen and Wang, Song and Chen, Chunyang and Hu, Yuanzhe and Wang, Qing},
  booktitle={Proceedings of the 46th ieee/acm international conference on software engineering},
  pages={1--13},
  year={2024}
}

@inproceedings{fu2024missconf,
  title={Missconf: Llm-enhanced reproduction of configuration-triggered bugs},
  author={Fu, Ying and Wang, Teng and Li, Shanshan and Ding, Jinyan and Zhou, Shulin and Jia, Zhouyang and Li, Wang and Jiang, Yu and Liao, Xiangke},
  booktitle={Proceedings of the 2024 IEEE/ACM 46th International Conference on Software Engineering: Companion Proceedings},
  pages={484--495},
  year={2024}
}

@article{bouzenia2024dypybench,
  title={DyPyBench: A benchmark of executable python software},
  author={Bouzenia, Islem and Krishan, Bajaj Piyush and Pradel, Michael},
  journal={Proceedings of the ACM on Software Engineering},
  volume={1},
  number={FSE},
  pages={338--358},
  year={2024},
}

@article{yang2025survey,
  title={A Survey of LLM-based Automated Program Repair: Taxonomies, Design Paradigms, and Applications},
  author={Yang, Boyang and Cai, Zijian and Liu, Fengling and Le, Bach and Zhang, Lingming and Bissyand{\'e}, Tegawend{\'e} F and Liu, Yang and Tian, Haoye},
  journal={arXiv preprint arXiv:2506.23749},
  year={2025}
}

@article{wang2025confucius,
  title={Confucius Code Agent: An Open-sourced AI Software Engineer at Industrial Scale},
  author={Wang, Zhaodong and Qi, Zhenting and Wong, Sherman and Hu, Nathan and Lin, Samuel and Ge, Jun and Gao, Erwin and Yang, Yining and Maurer, Ben and Chen, Wenlin and others},
  journal={arXiv preprint arXiv:2512.10398},
  year={2025}
}

@article{sapkota2025ai,
  title={Ai agents vs. agentic ai: A conceptual taxonomy, applications and challenges},
  author={Sapkota, Ranjan and Roumeliotis, Konstantinos I and Karkee, Manoj},
  journal={arXiv preprint arXiv:2505.10468},
  year={2025}
}

@article{lin2024llms,
  title={Llms as continuous learners: Improving the reproduction of defective code in software issues},
  author={Lin, Yalan and Ma, Yingwei and Cao, Rongyu and Li, Binhua and Huang, Fei and Gu, Xiaodong and Li, Yongbin},
  journal={arXiv preprint arXiv:2411.13941},
  year={2024}
}

@article{feng2025integrating,
  title={Integrating various software artifacts for better llm-based bug localization and program repair},
  author={Feng, Qiong and Ma, Xiaotian and Sheng, Jiayi and Feng, Ziyuan and Song, Wei and Liang, Peng},
  journal={ACM Transactions on Software Engineering and Methodology},
  year={2025},
}

@inproceedings{
  robeyns2025a,
  title={A Self-Improving Coding Agent},
  author={Maxime Robeyns and Martin Szummer and Laurence Aitchison},
  booktitle={Scaling Self-Improving Foundation Models without Human Supervision},
  year={2025},
}

@inproceedings{
  xu2026swingarena,
  title={{SWINGARENA}: Adversarial Programming Arena for Long-context GitHub Issue Solving},
  author={Wendong XU and Jing Xiong and Chenyang Zhao and Qiujiang Chen and Haoran Wang and Hui Shen and Zhongwei Wan and Jianbo Dai and Taiqiang Wu and He Xiao and Chaofan Tao and Zhuoqing Mao and Ying Sheng and Zhijiang Guo and Hongxia Yang and Bei Yu and Lingpeng Kong and Quanquan Gu and Ngai Wong},
  booktitle={The Fourteenth International Conference on Learning Representations},
  year={2026},
}

@inproceedings{
  zhang2025synthesizing,
  title={Synthesizing Software Engineering Data in a Test-Driven Manner},
  author={Lei Zhang and Jiaxi Yang and Min Yang and Jian Yang and Mouxiang Chen and Jiajun Zhang and Zeyu Cui and Binyuan Hui and Junyang Lin},
  booktitle={Forty-second International Conference on Machine Learning},
  year={2025},
}

@article{guo2025swe,
  title={Swe-factory: Your automated factory for issue resolution training data and evaluation benchmarks},
  author={Guo, Lianghong and Wang, Yanlin and Li, Caihua and Tao, Wei and Yang, Pengyu and Chen, Jiachi and Song, Haoyu and Tang, Duyu and Zheng, Zibin},
  journal={arXiv preprint arXiv:2506.10954},
  year={2025}
}

@article{yu2025building,
  title={Building coding agents via entropy-enhanced multi-turn preference optimization},
  author={Yu, Jiahao and Cheng, Zelei and Wu, Xian and Xing, Xinyu},
  journal={arXiv preprint arXiv:2509.12434},
  year={2025}
}

@article{xiao2026reducing,
  author  = {Yuan-An Xiao and Pengfei Gao and
             Chao Peng and Yingfei Xiong},
  title   = {Reducing Cost of LLM Agents with Trajectory Reduction},
  journal = {Proceedings of the ACM on Software Engineering},
  volume  = {3},
  number  = {FSE},
  pages   = {FSE056:1--FSE056:22},
  year    = {2026},
}

@article{yang2025lingxi,
  title={Lingxi: Repository-Level Issue Resolution Framework Enhanced by Procedural Knowledge Guided Scaling},
  author={Yang, Xu and Zhou, Jiayuan and Pacheco, Michael and Zhu, Wenhan and He, Pengfei and Wang, Shaowei and Liu, Kui and Pan, Ruiqi},
  journal={arXiv preprint arXiv:2510.11838},
  year={2025}
}

@article{cao2025siadafix,
  title={SIADAFIX: issue description response for adaptive program repair},
  author={Cao, Xin and Yu, Nan},
  journal={arXiv preprint arXiv:2510.16059},
  year={2025}
}

@inproceedings{gao2026more,
  author    = {Pengfei Gao and Chao Peng},
  title     = {More with Less: An Empirical Study of Turn-Control Strategies for Efficient Coding Agents},
  booktitle = {Proceedings of the 48th IEEE/ACM International Conference on Software Engineering},
  year      = {2026},
  pages      = {1--12},
}

@article{sonwane2025bugpilot,
  title={Bugpilot: Complex bug generation for efficient learning of swe skills},
  author={Sonwane, Atharv and White, Isadora and Lee, Hyunji and Pereira, Matheus and Caccia, Lucas and Kim, Minseon and Shi, Zhengyan and Singh, Chinmay and Sordoni, Alessandro and C{\^o}t{\'e}, Marc-Alexandre and others},
  journal={arXiv preprint arXiv:2510.19898},
  year={2025}
}

@article{zhou2025tom,
  title={Tom-swe: User mental modeling for software engineering agents},
  author={Zhou, Xuhui and Chen, Valerie and Wang, Zora Zhiruo and Neubig, Graham and Sap, Maarten and Wang, Xingyao},
  journal={arXiv preprint arXiv:2510.21903},
  year={2025}
}

@inproceedings{han-etal-2026-tdflow,
    title = "{TDF}low: Agentic Workflows for Test Driven Development",
    author = "Han, Kevin  and
      Maddikayala, Siddharth  and
      Knappe, Tim  and
      Patel, Om  and
      Liao, Austen  and
      Barati Farimani, Amir",
    booktitle = "Proceedings of the 19th Conference of the {E}uropean Chapter of the {A}ssociation for {C}omputational {L}inguistics (Volume 1: Long Papers)",
    year = "2026",
    pages = "1511--1527",
}

@inproceedings{zhang2025hierarchical,
  title={Hierarchical Reward Modeling for Fault Localization in Large Code Repositories},
  author={Zhang, Jiwei and Lian, Jianxun and Qin, Haiming and Zhou, Mingyang and Lu, KeZhong and Mao, Rui and Liao, Hao},
  booktitle={Findings of the Association for Computational Linguistics: EMNLP 2025},
  pages={17782--17796},
  year={2025}
}

@article{majgaonkar2025understanding,
  title={Understanding Code Agent Behaviour: An Empirical Study of Success and Failure Trajectories},
  author={Majgaonkar, Oorja and Fei, Zhiwei and Li, Xiang and Sarro, Federica and Ye, He},
  journal={arXiv preprint arXiv:2511.00197},
  year={2025}
}

@article{ma2025swe,
  title={SWE-fficiency: Can Language Models Optimize Real-World Repositories on Real Workloads?},
  author={Ma, Jeffrey Jian and Hashemi, Milad and Yazdanbakhsh, Amir and Swersky, Kevin and Press, Ofir and Li, Enhui and Reddi, Vijay Janapa and Ranganathan, Parthasarathy},
  journal={arXiv preprint arXiv:2511.06090},
  year={2025}
}

@article{hayashi2025self,
  title={Self-Abstraction from Grounded Experience for Plan-Guided Policy Refinement},
  author={Hayashi, Hiroaki and Pang, Bo and Zhao, Wenting and Liu, Ye and Gokul, Akash and Bansal, Srijan and Xiong, Caiming and Yavuz, Semih and Zhou, Yingbo},
  journal={arXiv preprint arXiv:2511.05931},
  year={2025}
}

@article{xia2025live,
  title={Live-SWE-agent: Can Software Engineering Agents Self-Evolve on the Fly?},
  author={Xia, Chunqiu Steven and Wang, Zhe and Yang, Yan and Wei, Yuxiang and Zhang, Lingming},
  journal={arXiv preprint arXiv:2511.13646},
  year={2025}
}

@article{li2025infcode,
  title={InfCode: Adversarial Iterative Refinement of Tests and Patches for Reliable Software Issue Resolution},
  author={Li, KeFan and Wang, Mengfei and Zhang, Hengzhi and Li, Zhichao and Yuan, Yuan and Li, Mu and Gao, Xiang and Sun, Hailong and Hu, Chunming and Lv, Weifeng},
  journal={arXiv preprint arXiv:2511.16004},
  year={2025}
}

@inproceedings{xiong2025think,
  title={Think-Search-Patch: A Retrieval-Augmented Reasoning Framework for Repository-Level Code Repair},
  author={Xiong, Bojian and Lei, Yikun and Liu, Xikai and Zhang, Shaowei and Zhu, Pengyun and Liu, Yan and Leng, Yongqi and Shi, Ling and Zhong, Meizhi and Zhang, Yurong and others},
  booktitle={Proceedings of the 2025 Conference on Empirical Methods in Natural Language Processing: Industry Track},
  pages={1555--1566},
  year={2025}
}

@inproceedings{
  cuadron2026saber,
  title={{SABER}: Small Actions, Big Errors {\textemdash} Safeguarding Mutating Steps in {LLM} Agents},
  author={Alejandro Cuadron and Pengfei Yu and Yang Liu and Arpit Gupta},
  booktitle={ICLR 2026 Workshop on Memory for LLM-Based Agentic Systems},
  year={2026},
}

@article{liu2026process,
  title={Process-centric analysis of agentic software systems},
  author={Liu, Shuyang and Chen, Yang and Krishna, Rahul and Sinha, Saurabh and Ganhotra, Jatin and Jabbarvand, Reyhaneh},
  journal={Proceedings of the ACM on Programming Languages},
  volume={10},
  number={OOPSLA1},
  pages={1961--1988},
  year={2026},
}

@inproceedings{tripathy2026swenergy,
  title={Swenergy: An empirical study on energy efficiency in agentic issue resolution frameworks with slms},
  author={Tripathy, Arihant and Harshit, Ch Pavan and Vaidhyanathan, Karthik},
  booktitle={Proceedings of the 2026 International Workshop on Agentic Engineering},
  pages={104--111},
  year={2026}
}

@article{prathifkumar2025does,
  title={Does SWE-Bench-Verified Test Agent Ability or Model Memory?},
  author={Prathifkumar, Thanosan and Mathews, Noble Saji and Nagappan, Meiyappan},
  journal={arXiv preprint arXiv:2512.10218},
  year={2025}
}

@article{zhu2025training,
  title={Training Versatile Coding Agents in Synthetic Environments},
  author={Zhu, Yiqi and Gandhi, Apurva and Neubig, Graham},
  journal={arXiv preprint arXiv:2512.12216},
  year={2025}
}

@article{wang2025swe,
  title={SWE-Bench++: A Framework for the Scalable Generation of Software Engineering Benchmarks from Open-Source Repositories},
  author={Wang, Lilin and Ramalho, Lucas and Celestino, Alan and Pham, Phuc Anthony and Liu, Yu and Sinha, Umang Kumar and Portillo, Andres and Osunwa, Onassis and Maduekwe, Gabriel},
  journal={arXiv preprint arXiv:2512.17419},
  year={2025}
}

@article{thai2025swe,
  title={SWE-EVO: Benchmarking Coding Agents in Long-Horizon Software Evolution Scenarios},
  author={Thai, Minh VT and Le, Tue and Manh, Dung Nguyen and Nhat, Huy Phan and Bui, Nghi DQ},
  journal={arXiv preprint arXiv:2512.18470},
  year={2025}
}

@article{wei2025toward,
  title={Toward training superintelligent software agents through self-play swe-rl},
  author={Wei, Yuxiang and Sun, Zhiqing and McMilin, Emily and Gehring, Jonas and Zhang, David and Synnaeve, Gabriel and Fried, Daniel and Zhang, Lingming and Wang, Sida},
  journal={arXiv preprint arXiv:2512.18552},
  year={2025}
}

@article{reddy2025swerank+,
  title={SweRank+: Multilingual, Multi-Turn Code Ranking for Software Issue Localization},
  author={Reddy, Revanth Gangi and Liu, Ye and Zhao, Wenting and Doo, JaeHyeok and Suresh, Tarun and Lee, Daniel and Xiong, Caiming and Zhou, Yingbo and Yavuz, Semih and Joty, Shafiq},
  journal={arXiv preprint arXiv:2512.20482},
  year={2025}
}

@article{zhang2025one,
  title={One Tool Is Enough: Reinforcement Learning for Repository-Level LLM Agents},
  author={Zhang, Zhaoxi and Duan, Yitong and Zhang, Yanzhi and Xu, Yiming and Wang, Zhixiang and Liang, Kun and Li, Yang and Liang, Jiahui and Xia, Deguo and Huang, Jizhou and others},
  journal={arXiv preprint arXiv:2512.20957},
  year={2025}
}

@article{liu2025context,
  title={Context as a tool: Context management for long-horizon swe-agents},
  author={Liu, Shukai and Yang, Jian and Jiang, Bo and Li, Yizhi and Guo, Jinyang and Liu, Xianglong and Dai, Bryan},
  journal={arXiv preprint arXiv:2512.22087},
  year={2025}
}

@inproceedings{
  shum2026swerm,
  title={{SWE}-{RM}: Execution-free Feedback for Software Engineering Agents},
  author={KaShun SHUM and Binyuan Hui and Jiawei Chen and Lei Zhang and X. W. and Jiaxi Yang and Yuzhen Huang and Junyang Lin and Junxian He},
  booktitle={The Fourteenth International Conference on Learning Representations},
  year={2026},
}

@article{liu2025graphlocator,
  title={GraphLocator: Graph-guided Causal Reasoning for Issue Localization},
  author={Liu, Wei and Peng, Chao and Gao, Pengfei and Liu, Aofan and Zhang, Wei and Zhao, Haiyan and Jin, Zhi},
  journal={arXiv preprint arXiv:2512.22469},
  year={2025}
}

@article{tao2026swe,
  title={Swe-lego: Pushing the limits of supervised fine-tuning for software issue resolving},
  author={Tao, Chaofan and Chen, Jierun and Jiang, Yuxin and Kou, Kaiqi and Wang, Shaowei and Wang, Ruoyu and Li, Xiaohui and Yang, Sidi and Du, Yiming and Dai, Jianbo and others},
  journal={arXiv preprint arXiv:2601.01426},
  year={2026}
}

@article{raghavendra2026agentic,
  title={Agentic Rubrics as Contextual Verifiers for SWE Agents},
  author={Raghavendra, Mohit and Gunjal, Anisha and Liu, Bing and He, Yunzhong},
  journal={arXiv preprint arXiv:2601.04171},
  year={2026}
}

@article{guo2026eet,
  title={EET: Experience-Driven Early Termination for Cost-Efficient Software Engineering Agents},
  author={Guo, Yaoqi and Xiao, Ying and Zhang, Jie M and Harman, Mark and Lou, Yiling and Liu, Yang and Chen, Zhenpeng},
  journal={arXiv preprint arXiv:2601.05777},
  year={2026}
}

@article{wang2026memgovern,
  title={MemGovern: Enhancing Code Agents through Learning from Governed Human Experiences},
  author={Wang, Qihao and Cheng, Ziming and Zhang, Shuo and Liu, Fan and Xu, Rui and Lian, Heng and Wang, Kunyi and Yu, Xiaoming and Yin, Jianghao and Hu, Sen and others},
  journal={arXiv preprint arXiv:2601.06789},
  year={2026}
}

@article{soni2026swe,
  title={SWE-Tester: Training Open-Source LLMs for Issue Reproduction in Real-World Repositories},
  author={Soni, Aditya Bharat and Ghosh, Rajat and Bhargava, Vaishnavi and Chen, Valerie and Dutta, Debojyoti},
  journal={arXiv preprint arXiv:2601.13713},
  year={2026}
}

@article{wang2026swe,
  title={SWE-Pruner: Self-Adaptive Context Pruning for Coding Agents},
  author={Wang, Yuhang and Shi, Yuling and Yang, Mo and Zhang, Rongrui and He, Shilin and Lian, Heng and Chen, Yuting and Ye, Siyu and Cai, Kai and Gu, Xiaodong},
  journal={arXiv preprint arXiv:2601.16746},
  year={2026}
}

@article{guo2026evoconfig,
  title={EvoConfig: Self-Evolving Multi-Agent Systems for Efficient Autonomous Environment Configuration},
  author={Guo, Xinshuai and Kuang, Jiayi and Pan, Linyue and Li, Yinghui and Li, Yangning and Zheng, Hai-Tao and Shen, Ying and Yin, Di and Sun, Xing},
  journal={arXiv preprint arXiv:2601.16489},
  year={2026}
}

@article{sepidband2026rgfl,
  title={RGFL: Reasoning Guided Fault Localization for Automated Program Repair Using Large Language Models},
  author={Sepidband, Melika and Taherkhani, Hamed and Pham, Hung Viet and Hemmati, Hadi},
  journal={arXiv preprint arXiv:2601.18044},
  year={2026}
}

@article{zeng2026davinci,
  title={davinci-dev: Agent-native mid-training for software engineering},
  author={Zeng, Ji and Fu, Dayuan and Mi, Tiantian and Zhuang, Yumin and Huang, Yaxing and Li, Xuefeng and Ye, Lyumanshan and Xie, Muhang and Hua, Qishuo and Huang, Zhen and others},
  journal={arXiv preprint arXiv:2601.18418},
  year={2026}
}

@article{xu2026learning,
  title={Learning Adaptive Parallel Execution for Efficient Code Localization},
  author={Xu, Ke and Xiao, Siyang and Liang, Ming and Yu, Yichen and Wang, Zhixiang and Xu, Jingxuan and Chen, Dajun and Jiang, Wei and Li, Yong},
  journal={arXiv preprint arXiv:2601.19568},
  year={2026}
}

@inproceedings{cheng2026dynamic,
  author={Cheng, Runxiang and Tufano, Michele and Cambronero, Jos{\'e} and Wei, Renyao and Shi, Sherry and Uy, Grant and Rondon, Pat and Ivan{\v{c}}i{\'c}, Franjo},
  title     = {Dynamic Cogeneration of Bug Reproduction Test in Agentic Program Repair},
  booktitle = {Proceedings of the 34th ACM Joint European Software Engineering Conference and Symposium on the Foundations of Software Engineering Companion},
  year      = {2026},
}

@article{wang2026he,
  title={HE-SNR: Uncovering Latent Logic via Entropy for Guiding Mid-Training on SWE-BENCH},
  author={Wang, Yueyang and Fu, Jiawei and Bi, Baolong and Wang, Xili and Liu, Xiaoqing},
  journal={arXiv preprint arXiv:2601.20255},
  year={2026}
}

@article{shen2026sera,
  title={SERA: Soft-Verified Efficient Repository Agents},
  author={Shen, Ethan and Tormoen, Danny and Shah, Saurabh and Farhadi, Ali and Dettmers, Tim},
  journal={arXiv preprint arXiv:2601.20789},
  year={2026}
}

@article{peng2026swe,
  title={SWE-Spot: Building Small Repo-Experts with Repository-Centric Learning},
  author={Peng, Jinjun and Saebo, Magnus and Zhong, Tianjun and Cheng, Yi-Jie and Yang, Junfeng and Ray, Baishakhi and Chen, Simin and Ding, Yangruibo},
  journal={arXiv preprint arXiv:2601.21649},
  year={2026}
}

@article{li2026outcome,
  title={Outcome-Conditioned Reasoning Distillation for Resolving Software Issues},
  author={Li, Chenglin and Xu, Yisen and Wang, Zehao and Tan, Shin Hwei and others},
  journal={arXiv preprint arXiv:2601.23257},
  year={2026}
}

@article{guo2026menvagent,
  title={MEnvAgent: Scalable Polyglot Environment Construction for Verifiable Software Engineering},
  author={Guo, Chuanzhe and Wu, Jingjing and He, Sijun and Chen, Yang and Kuang, Zhaoqi and Fan, Shilong and Chen, Bingjin and Bao, Siqi and Liu, Jing and Wu, Hua and others},
  journal={arXiv preprint arXiv:2601.22859},
  year={2026}
}

@article{benkovich2026agyn,
  title={Agyn: A Multi-Agent System for Team-Based Autonomous Software Engineering},
  author={Benkovich, Nikita and Valkov, Vitalii},
  journal={arXiv preprint arXiv:2602.01465},
  year={2026}
}

@article{SWE-Universe,
  title={SWE-Universe: Scale Real-World Verifiable Environments to Millions},
  author={Chen, Mouxiang and Zhang, Lei and Feng, Yunlong and Wang, Xuwu and Zhao, Wenting and Cao, Ruisheng and Yang, Jiaxi and Chen, Jiawei and Li, Mingze and Ma, Zeyao and others},
  journal={arXiv preprint arXiv:2602.02361},
  year={2026}
}

@article{xu2026swe,
  title={SWE-Refactor: A Repository-Level Benchmark for Real-World LLM-Based Code Refactoring},
  author={Xu, Yisen and Yang, Jinqiu and others},
  journal={arXiv preprint arXiv:2602.03712},
  year={2026}
}

@article{SWE-World,
  title={SWE-World: Building Software Engineering Agents in Docker-Free Environments},
  author={Sun, Shuang and Song, Huatong and Huang, Lisheng and Jiang, Jinhao and Le, Ran and Lv, Zhihao and Chen, Zongchao and Hu, Yiwen and Luo, Wenyang and Zhao, Wayne Xin and others},
  journal={arXiv preprint arXiv:2602.03419},
  year={2026}
}

@article{song2026swe,
  title={Swe-master: Unleashing the potential of software engineering agents via post-training},
  author={Song, Huatong and Huang, Lisheng and Sun, Shuang and Jiang, Jinhao and Le, Ran and Cheng, Daixuan and Chen, Guoxin and Hu, Yiwen and Chen, Zongchao and Jia, Yiming and others},
  journal={arXiv preprint arXiv:2602.03411},
  year={2026}
}

@article{luo2026closing,
  title={Closing the Loop: Universal Repository Representation with RPG-Encoder},
  author={Luo, Jane and Yin, Chengyu and Zhang, Xin and Li, Qingtao and Liu, Steven and Huang, Yiming and Wu, Jie and Liu, Hao and Huang, Yangyu and Kang, Yu and others},
  journal={arXiv preprint arXiv:2602.02084},
  year={2026}
}

@article{tang2026svrepair,
  title={SVRepair: Structured Visual Reasoning for Automated Program Repair},
  author={Tang, Xiaoxuan and Wang, Jincheng and Luo, Liwei and Xu, Jingxuan and Zhou, Sheng and Chen, Dajun and Jiang, Wei and Li, Yong},
  journal={arXiv preprint arXiv:2602.06090},
  year={2026}
}

@article{li2026contextbench,
  title={Contextbench: A benchmark for context retrieval in coding agents},
  author={Li, Han and Zhu, Letian and Zhang, Bohan and Feng, Rili and Wang, Jiaming and Pan, Yue and Barr, Earl T and Sarro, Federica and Chu, Zhaoyang and Ye, He},
  journal={arXiv preprint arXiv:2602.05892},
  year={2026}
}

@article{ding2026swe,
  title={SWE-Replay: Efficient Test-Time Scaling for Software Engineering Agents},
  author={Ding, Yifeng and Zhang, Lingming},
  journal={arXiv preprint arXiv:2601.22129},
  year={2026}
}

@article{zhu2026swe,
  title={Swe context bench: A benchmark for context learning in coding},
  author={Zhu, Jiayuan and Wu, Junde and Hu, Minhao and Zhu, Shengda and Pan, Jiazhen and Shen, Weixiang and Yang, Yijun and Liu, Fenglin and Hao, Jianye and Jin, Yueming and others},
  journal={arXiv preprint arXiv:2602.08316},
  year={2026}
}

@article{zhao2026immersion,
  title={Immersion in the github universe: Scaling coding agents to mastery},
  author={Zhao, Jiale and Chen, Guoxin and Meng, Fanzhe and Li, Minghao and Chen, Jie and Xu, Hui and Sun, Yongshuai and Zhao, Wayne Xin and Song, Ruihua and Zhang, Yuan and others},
  journal={arXiv preprint arXiv:2602.09892},
  year={2026}
}

@article{tian2026swe,
  title={SWE-Bench Mobile: Can Large Language Model Agents Develop Industry-Level Mobile Applications?},
  author={Tian, Muxin and Wang, Zhe and Yang, Blair and Tang, Zhenwei and Zhu, Kunlun and Dong, Honghua and Li, Hanchen and Xie, Xinni and Wang, Guangjing and You, Jiaxuan},
  journal={arXiv preprint arXiv:2602.09540},
  year={2026}
}

@article{yuan2026swe,
  title={Swe-minisandbox: Container-free reinforcement learning for building software engineering agents},
  author={Yuan, Danlong and Wu, Wei and Wang, Zhengren and Zhao, Xueliang and Zhang, Huishuai and Zhao, Dongyan},
  journal={arXiv preprint arXiv:2602.11210},
  year={2026}
}

@inproceedings{
  zhou2026featurebench,
  title={FeatureBench: Benchmarking Agentic Coding for Complex Feature Development},
  author={Qixing Zhou and JiaCheng Zhang and Haiyang Wang and Rui Hao and Jiahe Wang and Minghao Han and Yuxue Yang and Shuzhe Wu and Feiyang Pan and Lue Fan and Dandan Tu and Zhaoxiang Zhang},
  booktitle={The Fourteenth International Conference on Learning Representations},
  year={2026},
}

@article{xie2026hybrid,
  title={Hybrid-gym: Training coding agents to generalize across tasks},
  author={Xie, Yiqing and Liu, Emmy and Zhang, Gaokai and Kotalwar, Nachiket and Gandhi, Shubham and Acharya, Sathwik and Wang, Xingyao and Rose, Carolyn and Neubig, Graham and Fried, Daniel},
  journal={arXiv preprint arXiv:2602.16819},
  year={2026}
}

@article{gloaguen2026coding,
  title={Coding Agents Don't Know When to Act},
  author={Gloaguen, Thibaud and M{\"u}ndler, Niels and M{\"u}ller, Mark and Raychev, Veselin and Vechev, Martin},
  journal={arXiv preprint arXiv:2605.07769},
  year={2026}
}

@article{garg2026debug2fix,
  title={Debug2Fix: Supercharging Coding Agents with Interactive Debugging Capabilities},
  author={Garg, Spandan and Huang, Yufan},
  journal={arXiv e-prints},
  year={2026}
}

@article{kon2026swe,
  title={SWE-Prot$\backslash$'eg$\backslash$'e: Learning to Selectively Collaborate With an Expert Unlocks Small Language Models as Software Engineering Agents},
  author={Kon, Patrick Tser Jern and Pradeep, Archana and Chen, Ang and Ellis, Alexander P and Hunt, Warren and Wang, Zijian and Yang, John and Thompson, Samuel},
  journal={arXiv preprint arXiv:2602.22124},
  year={2026}
}

@article{deng2026your,
  title={Your Code Agent Can Grow Alongside You with Structured Memory},
  author={Deng, Yi-Xuan and Liu, Xiaoqin and Zhang, Yi and Yang, Guo-Wei and Yang, Shuojin},
  journal={arXiv preprint arXiv:2603.13258},
  year={2026}
}

@inproceedings{xiang2026evaluating,
  title={Evaluating and Improving Automated Repository-Level Rust Issue Resolution with LLM-based Agents},
  author={Xiang, Jiahong and He, Wenxiao and Wang, Xihua and Tian, Hongliang and Zhang, Yuqun},
  booktitle={2026 IEEE/ACM 48th International Conference on Software Engineering},
  year={2026},
  pages={13},
}

@article{badertdinov2026swe,
  title={Swe-rebench v2: Language-agnostic swe task collection at scale},
  author={Badertdinov, Ibragim and Nekrashevich, Maksim and Shevtsov, Anton and Golubev, Alexander},
  journal={arXiv preprint arXiv:2602.23866},
  year={2026}
}

@article{zhang2026sgagent,
  title={SGAgent: Suggestion-Guided LLM-Based Multi-Agent Framework for Repository-Level Software Repair},
  author={Zhang, Quanjun and Gao, Chengyu and Han, Yu and Shang, Ye and Fang, Chunrong and Chen, Zhenyu and Xiao, Liang},
  journal={ACM Transactions on Software Engineering and Methodology},
  year={2026},
}

@article{yu2026swe,
  title={SWE-ABS: Adversarial Benchmark Strengthening Exposes Inflated Success Rates on Test-based Benchmark},
  author={Yu, Boxi and Cao, Yang and Zhang, Yuzhong and Lin, Liting and Xu, Junjielong and Zhong, Zhiqing and Xu, Qinghua and Wang, Guancheng and Cao, Jialun and Cheung, Shing-Chi and others},
  journal={arXiv preprint arXiv:2603.00520},
  year={2026}
}

@article{zeng2026swe,
  title={Swe-hub: A unified production system for scalable, executable software engineering tasks},
  author={Zeng, Yucheng and Li, Shupeng and Dong, Daxiang and Xu, Ruijie and Chen, Zimo and Zheng, Liwei and Li, Yuxuan and Zhou, Zhe and Zhao, Haotian and Tian, Lun and others},
  journal={arXiv preprint arXiv:2603.00575},
  year={2026}
}

@article{he2026swe,
  title={Swe-adept: An llm-based agentic framework for deep codebase analysis and structured issue resolution},
  author={He, Kang and Roy, Kaushik},
  journal={arXiv preprint arXiv:2603.01327},
  year={2026}
}

@article{pan2026reporepair,
  title={RepoRepair: Leveraging Code Documentation for Repository-Level Automated Program Repair},
  author={Pan, Zhongqiang and Li, Chuanyi and Zhong, Wenkang and Feng, Yi and Luo, Bin and Ng, Vincent},
  journal={arXiv preprint arXiv:2603.01048},
  year={2026}
}

@article{liu2026architecture,
  title={Architecture-Aware Multi-Design Generation for Repository-Level Feature Addition},
  author={Liu, Mingwei and Chen, Zhenxi and Pei, Zheng and Wang, Zihao and Wang, Yanlin and Zheng, Zibin},
  journal={arXiv preprint arXiv:2603.01814},
  year={2026}
}

@article{SWE-CI,
  title={Swe-ci: Evaluating agent capabilities in maintaining codebases via continuous integration},
  author={Chen, Jialong and Xu, Xander and Wei, Hu and Chen, Chuan and Zhao, Bing},
  journal={arXiv preprint arXiv:2603.03823},
  year={2026}
}

@article{wang2026rubric,
  title={A rubric-supervised critic from sparse real-world outcomes},
  author={Wang, Xingyao and Chen, Valerie and Ji, Heng and Neubig, Graham},
  journal={arXiv preprint arXiv:2603.03800},
  year={2026}
}

@article{suri2026codescout,
  title={CodeScout: Contextual Problem Statement Enhancement for Software Agents},
  author={Suri, Manan and Li, Xiangci and Shojaie, Mehdi and Han, Songyang and Hsu, Chao-Chun and Garg, Shweta and Deshmukh, Aniket Anand and Kumar, Varun},
  journal={arXiv preprint arXiv:2603.05744},
  year={2026}
}

@article{fei2026echo,
  title={Echo: Graph-Enhanced Retrieval and Execution Feedback for Issue Reproduction Test Generation},
  author={Fei, Zhiwei and Pan, Yue and Sarro, Federica and Ge, Jidong and Liu, Marc and Ng, Vincent and Ye, He},
  journal={arXiv preprint arXiv:2603.07326},
  year={2026}
}

@article{wen2026swe,
  title={SWE-Fuse: Empowering Software Agents via Issue-free Trajectory Learning and Entropy-aware RLVR Training},
  author={Wen, Xin-Cheng and Chen, Binbin and Lan, Haoxuan and Yu, Hang and Di, Peng and Gao, Cuiyun},
  journal={arXiv preprint arXiv:2603.07927},
  year={2026}
}

@article{ganhotra2026resolving,
  title={Resolving Java Code Repository Issues with iSWE Agent},
  author={Ganhotra, Jatin and Serhan, Sami and Nassar, Antonio Abu and Shinnar, Avraham and Nevo, Ziv and Hirzel, Martin},
  journal={arXiv preprint arXiv:2603.11356},
  year={2026}
}

@article{fu2026davinci,
  title={daVinci-Env: Open SWE Environment Synthesis at Scale},
  author={Fu, Dayuan and Wu, Shenyu and Wu, Yunze and Peng, Zerui and Huang, Yaxing and Sun, Jie and Zeng, Ji and Jiang, Mohan and Zhang, Lin and Li, Yukun and others},
  journal={arXiv preprint arXiv:2603.13023},
  year={2026}
}

@article{huang2026beyond,
  title={Beyond Verifiable Rewards: Rubric-Based GRM for Reinforced Fine-Tuning SWE Agents},
  author={Huang, Jiawei and Yang, Qingping and Zheng, Renjie and Chen, Jiaze},
  journal={arXiv preprint arXiv:2604.16335},
  year={2026}
}

@article{sutawika2026codescout,
  title={Codescout: An effective recipe for reinforcement learning of code search agents},
  author={Sutawika, Lintang and Soni, Aditya Bharat and Gandhi, Apurva and Yassine, Taha and Vijayvargiya, Sanidhya and Li, Yuchen and Zhou, Xuhui and Zhang, Yilin and Maben, Leander Melroy and Neubig, Graham and others},
  journal={arXiv preprint arXiv:2603.17829},
  year={2026}
}

@article{ma2026failuremem,
  title={FailureMem: A Failure-Aware Multimodal Framework for Autonomous Software Repair},
  author={Ma, Ruize and Jiang, Yilei and Zhang, Shilin and Ma, Zheng and Feng, Yi and Ng, Vincent and Wang, Zhi and Yue, Xiangyu and Li, Chuanyi and Lu, Lewei},
  journal={arXiv preprint arXiv:2603.17826},
  year={2026}
}

@article{liang2026swe,
  title={SWE-Next: Scalable Real-World Software Engineering Tasks for Agents},
  author={Liang, Jiarong and Lyu, Zhiheng and Liu, Zijie and Chen, Xiangchao and Nie, Ping and Zou, Kai and Chen, Wenhu},
  journal={arXiv preprint arXiv:2603.20691},
  year={2026}
}

@article{kim2026trajeval,
  title={TRAJEVAL: Decomposing Code Agent Trajectories for Fine-Grained Diagnosis},
  author={Kim, Myeongsoo and Wang, Dingmin and Cui, Siwei and Farmahinifarahani, Farima and Garg, Shweta and Ray, Baishakhi and Zhuo, Terry Yue and Mukherjee, Rajdeep and Kumar, Varun},
  journal={arXiv preprint arXiv:2603.24631},
  year={2026}
}

@inproceedings{tian2026agentbased,
  title={Agent-Based Ensemble Reasoning for Repository-Level Issue Resolution},
  author={Tian, Zhao and Gao, Pengfei and Chen, Junjie and Peng, Chao},
  booktitle={2026 IEEE/ACM 48th International Conference on Software Engineering},
  year={2026},
  pages={13},
}

@article{zhao2026beyond,
  title={Beyond Localization: Recoverable Headroom and Residual Frontier in Repository-Level RAG-APR},
  author={Zhao, Pengtao and Yang, Boyang and Le, Bach and Liu, Feng and Tian, Haoye},
  journal={arXiv preprint arXiv:2603.29067},
  year={2026}
}

@article{jia2026compressing,
  title={Compressing code context for LLM-based issue resolution},
  author={Jia, Haoxiang and Barr, Earl T and Mechtaev, Sergey},
  journal={arXiv preprint arXiv:2603.28119},
  year={2026}
}

@article{ludwig2026swe,
  title={From SWE-ZERO to SWE-HERO: Execution-free to Execution-based Fine-tuning for Software Engineering Agents},
  author={Ludwig, Nikolai and Ahmad, Wasi Uddin and Majumdar, Somshubra and Ginsburg, Boris},
  journal={arXiv preprint arXiv:2604.01496},
  year={2026}
}

@article{shastry2026beyond,
  title={Beyond Isolated Tasks: A Framework for Evaluating Coding Agents on Sequential Software Evolution},
  author={Shastry, KN and Senrayan, Ganesh and Satapara, Shrey and Panda, Pranoy and Devaguptapu, Chaitanya},
  journal={arXiv preprint arXiv:2604.03035},
  year={2026}
}

@article{li2026beyond,
  title={Beyond Fixed Tests: Repository-Level Issue Resolution as Coevolution of Code and Behavioral Constraints},
  author={Li, Kefan and Yuan, Yuan and Wang, Mengfei and Zheng, Shihao and Wang, Wei and Yang, Ping and Li, Mu and Lv, Weifeng},
  journal={arXiv preprint arXiv:2604.04580},
  year={2026}
}

@article{team2026yet,
  title={Yet Even Less Is Even Better For Agentic, Reasoning, and Coding LLMs},
  author={Team, CodeArts Model and Ye, Yang and Tan, Jingyuan and Jiang, Tianyue and Ye, Ruizhe and He, Qiankun and Yang, Jiarui and Dong, Jian and Liang, Sicong and Yue, Chongjian and others},
  journal={arXiv preprint arXiv:2604.00824},
  year={2026}
}

@article{yu2026does,
  title={Does Pass Rate Tell the Whole Story? Evaluating Design Constraint Compliance in LLM-based Issue Resolution},
  author={Yu, Kai and Zhou, Zhenhao and Zeng, Junhao and Wang, Ying and Du, Xueying and Yuan, Zhiqiang and Liu, Junwei and Zhou, Ziyu and Wang, Yujia and Wang, Chong and others},
  journal={arXiv preprint arXiv:2604.05955},
  year={2026}
}

@article{kuang2026reagent,
  title={REAgent: Requirement-Driven LLM Agents for Software Issue Resolution},
  author={Kuang, Shiqi and Tian, Zhao and Lin, Kaiwei and Tao, Chaofan and Wang, Shaowei and Bai, Haoli and Shang, Lifeng and Chen, Junjie},
  journal={arXiv preprint arXiv:2604.06861},
  year={2026}
}

@article{li2026oracle,
  title={ORACLE-SWE: Quantifying the Contribution of Oracle Information Signals on SWE Agents},
  author={Li, Kenan and Jin, Qirui and Zhu, Liao and Huang, Xiaosong and Wu, Yijia and Zhang, Yikai and Zhang, Xin and Jin, Zijian and Huang, Yufan and Nallipogu, Elsie and others},
  journal={arXiv preprint arXiv:2604.07789},
  year={2026}
}

@article{zhang2026agent,
  title={Do agent rules shape or distort? guardrails beat guidance in coding agents},
  author={Zhang, Xing and Wang, Guanghui and Cui, Yanwei and Qiu, Wei and Li, Ziyuan and Zhu, Bing and He, Peiyang},
  journal={arXiv e-prints},
  year={2026}
}

@article{liu2026plan,
  title={From Plan to Action: How Well Do Agents Follow the Plan?},
  author={Liu, Shuyang and Dehghan, Saman and Ganhotra, Jatin and Hirzel, Martin and Jabbarvand, Reyhaneh},
  journal={arXiv e-prints},
  year={2026}
}

@article{lian2026swe,
  title={Swe-agile: A software agent framework for efficiently managing dynamic reasoning context},
  author={Lian, Shuquan and Liu, Juncheng and Chen, Yazhe and Chen, Yuhong and Li, Hui},
  journal={arXiv preprint arXiv:2604.11716},
  year={2026}
}

@article{SWE-TRACE,
  title={SWE-TRACE: Optimizing Long-Horizon SWE Agents Through Rubric Process Reward Models and Heuristic Test-Time Scaling},
  author={Han, Hao and Xie, Jin and Ma, Xuehao and Zhu, Weiquan and Zhang, Ziyao and Long, ZhiLiang and Chen, Hongkai and Ye, Qingwen},
  journal={arXiv preprint arXiv:2604.14820},
  year={2026}
}

@article{kim2026codestruct,
  title={CODESTRUCT: Code Agents over Structured Action Spaces},
  author={Kim, Myeongsoo and Hsu, Joe and Wang, Dingmin and Garg, Shweta and Kumar, Varun and Ramanathan, Murali Krishna},
  journal={arXiv preprint arXiv:2604.05407},
  year={2026}
}

@article{kim2026scaling,
  title={Scaling Test-Time Compute for Agentic Coding},
  author={Kim, Joongwon and Yang, Wannan and Niu, Kelvin and Zhang, Hongming and Zhu, Yun and Helenowski, Eryk and Silva, Ruan and Chen, Zhengxing and Iyer, Srinivasan and Zaheer, Manzil and others},
  journal={arXiv preprint arXiv:2604.16529},
  year={2026}
}

@inproceedings{meng2026llmbased,
  title={LLM-based Agents for Automated Bug Fixing: How Far Are We?},
  author={Meng, Xiangxin and Ma, Zexiong and Gao, Pengfei and Peng, Chao},
  booktitle={2026 IEEE/ACM 48th International Conference on Software Engineering},
  year={2026},
}

@article{xu2026neurosymbolic,
  title={Neurosymbolic Repo-level Code Localization},
  author={Xu, Xiufeng and Wu, Xiufeng and Zhang, Zejun and Li, Yi},
  journal={arXiv preprint arXiv:2604.16021},
  year={2026}
}

@article{wang2026icore,
  title={iCoRe: An Iterative Correlation-Aware Retriever for Bug Reproduction Test Generation},
  author={Wang, Junyi and Cao, Jialun and Liu, Zhongxin},
  journal={Proceedings of the ACM on Software Engineering},
  volume={3},
  number={FSE},
  pages={FSE186:1--FSE186:26},
  year={2026},
}

@article{SWE-Edit,
  title={SWE-Edit: Rethinking Code Editing for Efficient SWE-Agent},
  author={Zhang, Yikai and Pei, Jiaxin and Li, Kenan and Wang, Maoquan and Pan, Jin and Kang, Yu and Fu, Shengyu and Nallipogu, Elsie and Hu, Junjie and Huang, Yufan and others},
  journal={arXiv preprint arXiv:2604.26102},
  year={2026}
}

@article{seddik2026arise,
  title={ARISE: A Repository-level Graph Representation and Toolset for Agentic Fault Localization and Program Repair},
  author={Seddik, Shahd and Fard, Fatemeh},
  journal={arXiv preprint arXiv:2605.03117},
  year={2026}
}

@article{ahmed2026reproduction,
  title={Reproduction Test Generation for Java SWE Issues},
  author={Ahmed, Toufique and Ganhotra, Jatin and Shinnar, Avraham and Hirzel, Martin},
  journal={arXiv preprint arXiv:2605.04320},
  year={2026}
}

@article{li2026boostapr,
  title={BoostAPR: Boosting Automated Program Repair via Execution-Grounded Reinforcement Learning with Dual Reward Models},
  author={Li, Yuanhao and Wang, Hongbo and Shang, Xiaotang and Tang, Xunzhu and Cao, Yiming and Chen, Xuhong},
  journal={arXiv preprint arXiv:2605.09134},
  year={2026}
}

@article{wang2026hindsight,
  title={Hindsight Hint Distillation: Scaffolded Reasoning for SWE Agents from CoT-free Answers},
  author={Wang, Shengjie and Li, Guanghe and Yang, Zonghan and Gao, Yang},
  journal={arXiv preprint arXiv:2605.11556},
  year={2026}
}

@article{sahoo2026agentlens,
  title={AgentLens: Revealing The Lucky Pass Problem in SWE-Agent Evaluation},
  author={Sahoo, Priyam and Mittal, Gaurav and Li, Xiaomin and Ma, Shengjie and Steenhoek, Benjamin and Lin, Pingping and Hu, Yu},
  journal={arXiv preprint arXiv:2605.12925},
  year={2026}
}

@article{lam2026swe,
  title={SWE-Chain: Benchmarking Coding Agents on Chained Release-Level Package Upgrades},
  author={Lam, Man Ho and Wang, Chaozheng and Liu, Hange and Xiao, Jingyu and Li, Haau-sing and Huang, Jen-tse and Zhuo, Terry Yue and Lyu, Michael R},
  journal={arXiv preprint arXiv:2605.14415},
  year={2026}
}

@article{mamun2026blagent,
  title={BLAgent: Agentic RAG for File-Level Bug Localization},
  author={Mamun, Md Afif Al and Uddin, Gias},
  journal={arXiv preprint arXiv:2605.17965},
  year={2026}
}

@article{SWE-Mutation,
  title={SWE-Mutation: Can LLMs Generate Reliable Test Suites in Software Engineering?},
  author={Sun, Yuxuan and Zhao, Yuze and Wang, Yufeng and Du, Yao and Ma, Zhiyuan and Wang, Jinbo and Zhang, Mengdi and Zhang, Kai and Huang, Zhenya},
  journal={arXiv preprint arXiv:2605.22175},
  year={2026}
}

@article{li2026repomirage,
  title={RepoMirage: Probing Repository Context Reasoning in Code Agents with Perturbations},
  author={Li, Hanyu and Zhang, Yichi and Zhu, Speed and Su, Hang and Zhu, Jun and Dong, Yinpeng},
  journal={arXiv preprint arXiv:2605.26177},
  year={2026}
}

@misc{anthropic2026claudefable5,
  author       = {{Anthropic}},
  title        = {{Claude Fable 5}},
  year         = {2026},
  month        = jul,
  howpublished = {\url{https://www.anthropic.com/claude/fable}},
  note         = {Published July 1, 2026. Accessed July 2, 2026}
}

@article{seaman1999qualitative,
  title={Qualitative methods in empirical studies of software engineering},
  author={Seaman, Carolyn B.},
  journal={IEEE Transactions on software engineering},
  volume={25},
  number={4},
  pages={557--572},
  year={1999},
}

@article{HGM,
      title={Huxley-G\"odel Machine: Human-Level Coding Agent Development by an Approximation of the Optimal Self-Improving Machine}, 
      author={Wenyi Wang and Piotr Piękos and Li Nanbo and Firas Laakom and Yimeng Chen and Mateusz Ostaszewski and Mingchen Zhuge and Jürgen Schmidhuber},
      year={2025},
      journal={arXiv preprint arXiv:2510.21614},
}

@article{DGM,
  title={Darwin godel machine: Open-ended evolution of self-improving agents},
  author={Zhang, Jenny and Hu, Shengran and Lu, Cong and Lange, Robert and Clune, Jeff},
  journal={arXiv preprint arXiv:2505.22954},
  year={2025}
}

@inproceedings{tobin2017domain,
  title={Domain randomization for transferring deep neural networks from simulation to the real world},
  author={Tobin, Josh and Fong, Rachel and Ray, Alex and Schneider, Jonas and Zaremba, Wojciech and Abbeel, Pieter},
  booktitle={2017 IEEE/RSJ international conference on intelligent robots and systems},
  pages={23--30},
  year={2017},
}

@inproceedings{nanda2026wink,
  title={Wink: Recovering from misbehaviors in coding agents},
  author={Nanda, Rahul and Maddila, Chandra and Jha, Smriti and Khan, Euna Mehnaz and Paltenghi, Matteo and Chandra, Satish},
  booktitle={Proceedings of the 3rd ACM International Conference on AI-Powered Software},
  pages={208--217},
  year={2026}
}

@article{schulman2017proximal,
  title={Proximal policy optimization algorithms},
  author={Schulman, John and Wolski, Filip and Dhariwal, Prafulla and Radford, Alec and Klimov, Oleg},
  journal={arXiv preprint arXiv:1707.06347},
  year={2017}
}

@article{petersen2015guidelines,
  title={Guidelines for conducting systematic mapping studies in software engineering: An update},
  author={Petersen, Kai and Vakkalanka, Sairam and Kuzniarz, Ludwik},
  journal={Information and software technology},
  volume={64},
  pages={1--18},
  year={2015},
}

@article{liu2024lost,
  title={Lost in the middle: How language models use long contexts},
  author={Liu, Nelson F and Lin, Kevin and Hewitt, John and Paranjape, Ashwin and Bevilacqua, Michele and Petroni, Fabio and Liang, Percy},
  journal={Transactions of the association for computational linguistics},
  volume={12},
  pages={157--173},
  year={2024}
}

@article{niu2025deep,
  title={When deep learning meets information retrieval-based bug localization: A survey},
  author={Niu, Feifei and Li, Chuanyi and Liu, Kui and Xia, Xin and Lo, David},
  journal={ACM Computing Surveys},
  volume={57},
  number={11},
  pages={1--41},
  year={2025},
  publisher={ACM New York, NY}
}

\appendix
\section*{Appendix}

\section{Additional Background of Automatic Issue Resolution}\label{ap:issue_resolution_overview}

\begin{figure}[h]
    \centering
    \includegraphics[width=0.8\linewidth]{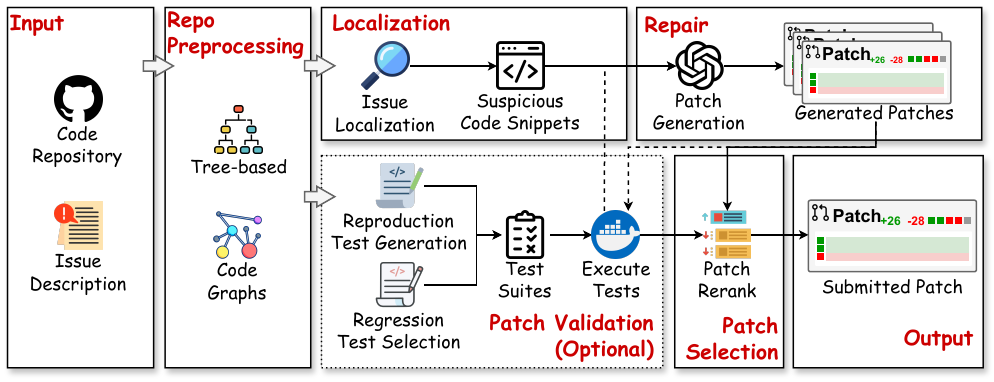}
    \caption{Overview of the basic framework of automated issue resolution.}
    \label{fig: solving_overview}
\end{figure}

Figure~\ref{fig: solving_overview} presents the framework of automated issue resolution formed by the five logical phases introduced in Section~\ref{sec:2.2}.
This section describes each phase in detail.

\textit{Repo Preprocessing.}
This phase aims to construct an accessible and comprehensible knowledge representation for the code repository.
Based on the pre-built knowledge, the LLM or agent can more conveniently and efficiently mine useful information from the repository.
For example, constructing a code knowledge graph~\cite{locagent, KGCompass} facilitates the retrieval of inheritance or invocation relationships between classes and functions by the LLM.

\textit{Localization.}
Given the issue description, this phase aims to accurately locate the most relevant code snippets to be edited that are likely responsible for the reported issue.
Localization methods may leverage information retrieval~\cite{SweRank}, static and dynamic program analysis~\cite{AutoCodeRover}, and LLM-based reasoning~\cite{agentless} to identify suspicious files, classes, or lines of code. 
Effective localization significantly narrows down the search space for patch generation and improves the overall efficiency of the repair process.

\textit{Repair.}
Given the code snippets to be modified and the relevant context, this phase aims to generate patches for issue resolution by leveraging LLMs through prompting techniques.
Specifically, the agent produces edits by replacing designated sections of code files~\cite{openhands} and utilizes git tools~\cite{SWEAgent} to generate patches. 
Typically, multiple candidate patches are generated at this stage to maximize the likelihood of producing a correct fix.

\textit{Patch Validation.}
This phase is adopted by some of the existing techniques. 
Given the issue description, it aims to generate reproduction test cases or select regression test suites from the repository.
Reproduction tests attempt to simulate the issue scenario based on the original issue report~\cite{AEGIS}, while regression tests ensure that the candidate patch does not introduce new errors elsewhere in the codebase~\cite{agentless}.
By leveraging testing techniques, this phase filters out some incorrect or harmful patches.
\revReplace{R3.19}{
Additionally, some techniques run the reproduction or regression tests against candidate patches and use the failing results to guide iterative patch regeneration~\cite{SWEAgent, openhands}, indicating that validation outcomes may also be fed back to the repair phase.    
}

\textit{Patch Selection.}
Given a set of candidate patches, this phase aims to select the most promising one for submission to resolve the issue.
For example, the agent may select a patch based on self-consistency using majority voting~\cite{agentless} or employ a strong reasoning model with LLM-as-a-judge for evaluation~\cite{MASAI}.
The selected patch will be applied to the repository as the solution to the issue.

\section{Literature Search and Selection}\label{ap:paper_collect}

\begin{figure}[h]
    \centering
    \includegraphics[width=0.7\linewidth]{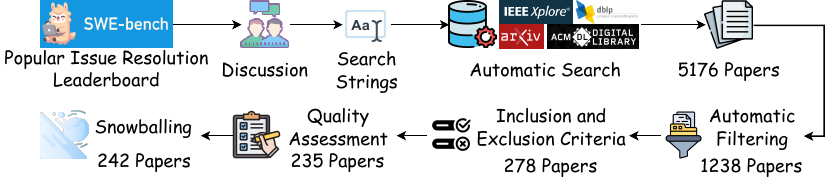}
    \caption{Overview of the process of paper collection and filtering.}
    \label{fig:paper_collection}
\end{figure}

\subsection{Search Process}
The complete search process is illustrated in Figure~\ref{fig:paper_collection}.
\subsubsection{Search Strings}

As Jimenez et al.~\cite{swebench} pioneered the issue resolution task, the leaderboard maintained by the SWE-bench~\cite{swebench,swebenchmutimodal,swebenchverified} team, to some extent, reflects the research progress in this domain. 
\revReplace{R2.1}{Therefore, we collect 19 papers corresponding to the techniques submitted to the SWE-bench leaderboard as seed papers, manually inspect them, and formulate the search strings for subsequent automated database retrieval through discussion between two of the authors.
Specifically, we first extract task-level terms from the seed papers, then expand them with synonyms and subtask terms observed during the manual inspection, and finally group them by Software Engineering (SE) task scope and Artificial Intelligence (AI) technical paradigm.}
The search keywords are summarized in Table~\ref{tab: keywords}.
\revReplace{R1.2}{Within each domain, keywords are organized into two groups. For SE, the groups follow task granularity: G1 targets end-to-end issue resolution, and G2 targets its key subtasks. For AI, the groups follow technical paradigms: G1 focuses on language models, and G2 on agents.
The SWE-bench leaderboard is used only to seed, not to directly populate the paper pool. The subsequent database search and snowballing are not limited to SWE-bench-derived techniques, and therefore also cover studies beyond the SWE-bench family.
}

\begin{table}[h]
    \centering
    \caption{Keywords related to SE/AI and issue resolution task for automated search.}
    \label{tab: keywords}
    \begin{adjustbox}{max width=\textwidth}
        \begin{tabular}{lp{10cm}p{7cm}}
            \toprule
            \textbf{Group}&\textbf{SE Domain Keywords} & \textbf{AI Domain Keywords} \\ \midrule
            G1& SWE, Issue Solving, Issue Resolution, Issue Fixing, Issue Repair, Program Repair & Large Language Model, LLM, Language Model, LM, Code Large Language Model, Code LLM\\
            \midrule
            G2& Issue Reproduction, Bug Reproduction, Reproduction Test Generation, Issue Localization, Fault Localization, Patch Generation, Patch Rerank, Patch Selection, Patch Validation, Patch Verification&Agent, Software Engineering Agent, SE Agent, SWE Agent\\
            \bottomrule
        \end{tabular}
    \end{adjustbox}
\end{table}

\subsubsection{Search Database}

We have conducted automated searches in four widely adopted databases, including IEEE Xplore, the ACM Digital Library, arXiv, and DBLP, and selected studies published on or before May 31, 2026.
\revReplace{R1.2/R2.3}{As shown in Table~\ref{tab: keywords}, we construct search strings based on the group of keywords. 
Specifically, within each group, keywords are combined using the ``OR'' operator, while the SE and AI groups are combined using the ``AND'' operator, yielding four exact queries.
We execute each of these four queries as a full-text search against IEEE Xplore, the ACM Digital Library, and arXiv. DBLP indexes bibliographic metadata rather than full text, so on DBLP the same queries are matched against publication titles.
The four query combinations return 1,590 papers for \texttt{SE-G1 AND AI-G1}, 997 papers for \texttt{SE-G1 AND AI-G2}, 1,339 papers for \texttt{SE-G2 AND AI-G1}, and 1,250 papers for \texttt{SE-G2 AND AI-G2}, totaling 5,176 raw hits.
Given the explosive growth of research on issue resolution, primarily occurring within the past year, we impose no venue restrictions and adopt full-text search wherever it is supported to maximize recall and rely on the subsequent automated filtering, inclusion/exclusion, and quality assessment stages to control for noise.
We consider only publications dated October 2023 or later, as this month marks the initial proposal of the issue resolution task.
We then merge duplicate records returned by overlapping queries and databases.}
\revReplace{R2.1/R2.2/R2.12}{In addition, we filter papers by publication venue: formally published papers are retained only if they appear in top-tier SE and AI conference proceedings and journals, while arXiv preprints are kept for subsequent screening.
The considered venues and the number of papers retained from each are TOSEM (196), ICSE (137), FSE (99), ASE (70), ISSTA (49), ICML (43), ACL (36), NeurIPS (31), ICLR (29), TSE (24), MSR (13), EMSE (11), ICSME (11), SANER (7), EMNLP (7), ICPC (5), AAAI (5), and NAACL (2), totaling 775 formally published papers alongside 463 arXiv preprints.}
After this automated filtering, we retain \revReplace{R1.2/R2.1}{1,238} papers.

\subsection{Inclusion and Exclusion Criteria}

Since the automated search only constructs the paper pool through keyword searches, this may result in the inclusion of some papers that are not related to this survey.
To ensure relevance and consistency, we define a set of inclusion and exclusion criteria based on prior guidelines~\cite{hou2024large, liu2024large} to ensure that the papers in the pool match the scope and research questions of our investigation.

\noindent\textbf{Inclusion Criteria.}
\revReplace{R1.1}{Although our survey centers on the \textit{end-to-end} issue resolution task, subtask-level work on individual stages (e.g., localization, reproduction) is also part of this problem space. Moreover, since benchmarks serve as supporting evaluation infrastructure and empirical studies serve as supporting findings, these papers should be included in our survey scope.
Thus, we design inclusion criteria from benchmark, technique, and empirical study dimensions.}
A paper will be included in this survey if it meets any of the criteria below.
\begin{itemize}[left=0pt, topsep=0em]
    \item This paper constructs a dataset or benchmark to evaluate the performance of language models on end-to-end issue resolution tasks.
    \item This paper constructs a dataset or benchmark to evaluate the performance of large language models (LLMs) on a specific stage of the issue resolution task, such as issue localization or reproduction test generation.
    \item The paper proposes or improves an approach, method, technique, or tool/framework that leverages LLMs to resolve issues or to address a specific stage of the issue resolution process.
    \item This paper proposes a specific technique related to LLMs, such as supervised fine-tuning (SFT), for building agents that accomplish issue resolution tasks.
    \item This paper conducts an empirical or experimental study on issue resolution benchmarks or techniques.
\end{itemize}

\noindent\textbf{Exclusion Criteria.}
If a paper meets any of the following criteria, it will be excluded from the scope of our study.
\begin{itemize}[left=0pt, topsep=0em]
    \item This paper does not involve any stage of the issue resolution task, such as repository-level code completion.
    \item This paper does not rely on LLMs for issue resolution.
    \item This paper primarily evaluates the general performance of LLMs, such as code generation ability or security.
    \item This paper only mentions LLMs or issue resolution in the related work or future work sections, rather than making them the core focus of the study.
    \item This paper primarily discusses directions, visions, or ethical considerations, without proposing concrete or implementable technical solutions or findings.

\end{itemize}

\revReplace{R1.2/R2.1}{Two authors independently apply the inclusion and exclusion criteria to each paper in the pool. 
A paper is retained if at least one author votes to include it. 
After this stage, 278 papers are retained for quality assessment. 
The two authors reach a Cohen's $\kappa$ of 0.8487 at this stage, indicating almost perfect agreement.}
\revReplace{R3.8}{During this stage, the two authors also manually classify each retained paper into three study types according to the inclusion criteria it satisfies: the first two criteria correspond to benchmark papers, the following two to technique papers, and the last one to empirical studies.
A paper satisfying criteria of more than one type is assigned to all applicable types. 
This classification is performed independently, reaching a Cohen's $\kappa$ of 0.9065, indicating almost perfect agreement, and disagreements are resolved through discussion with a third author. 
To facilitate the statistical analysis in Appendix~\ref{sec:collection_results}, we assign papers belonging to multiple categories one primary category through discussion.}

\subsection{Quality Assessment}
In addition, following prior survey methodologies~\cite{jiang2024survey, hou2024large}, we define a set of Quality Assessment Criteria (QAC).
\revReplace{R1.3}{Since our corpus spans three study types and a single checklist cannot fairly evaluate all of them, we adapt the QAC per category.
A set of common criteria applies to every paper, while each set of type-specific criteria applies to every paper assigned to that type.}
\revReplace{R3.2}{Both common criteria and the type-specific criteria are adapted from the QAC of prior surveys~\cite{jiang2024survey, hou2024large, niu2025deep}.}
\revReplace{R1.3}{The full set of type-adapted criteria is shown in Table~\ref{tab:qac_criteria}.}
\revReplace{R1.3/R3.8}{Each paper is scored on the common criteria together with all criteria for its assigned study types.}
Each QAC item is scored as “yes”, “partial”, or “no”, corresponding to 1, 0.5, and 0, respectively.
\revReplace{R1.3/R3.2}{Following the practice of prior surveys, we exclude a paper if its total score falls below 80\%~\cite{hou2024large, jiang2024survey} of the maximum attainable over the common criteria together with the type-specific criteria of each type assigned to it, which we refer to as the criteria applicable to that paper.}

\begin{table*}[t]
    \centering
    \small
    \caption{\revReplace{R1.3}{Type-adapted Quality Assessment Criteria.}}
    \label{tab:qac_criteria}
    \begin{adjustbox}{max width=\textwidth}
    \begin{tabular}{lp{7cm}|lp{7cm}}
        \toprule
        Type & QAC & Type & QAC \\
        \midrule
        All & Is the study related to issue resolution or its sub-tasks? & Technique & Does the study use/train LLMs to build an agentic system? \\
        All & Does the study provide clear and reproducible technical/methodological details? & Technique & Does the study clearly explain how LLMs are used? \\
        All & Does the study present a clear research motivation? & Technique & Does the study specify which LLMs are used? \\
        All & Does the study make a clear contribution to the issue resolution task? & Technique & Does the study provide a clear description of its experimental setup, including experimental environments and dataset information? \\
        All & Are the experimental results and conclusions of the study consistent with its research objectives? & Technique & Does the study include at least one baseline comparison to demonstrate its effectiveness? \\
        \midrule
        Benchmark & Does the study clearly describe its data sources and construction or curation process? & Empirical & Does the study clearly state its research questions? \\
        Benchmark & Does the study report quality-control measures, such as validation, deduplication, or contamination checks? & Empirical & Does the study adopt a clearly described experimental methodology? \\
        Benchmark & Does the study define clear evaluation metrics and protocols? & Empirical & Does the study clearly confirm its empirical findings? \\
        \bottomrule
    \end{tabular} 
    \end{adjustbox}
\end{table*}

\revReplace{R1.2/R1.3}{The same two authors independently applied the type-adapted QAC to all 278 papers retained after the inclusion and exclusion stage and then made binary retain/exclude decisions. They reached a Cohen's $\kappa$ of 0.9290 on these filtering decisions, indicating almost perfect agreement.
When their independent decisions disagreed, a third author adjudicated the case to reach a final decision.}

\subsection{Snowballing}

After quality assessment, we obtain a preliminary paper set.
To ensure coverage and reduce the risk of missing relevant studies, we apply a snowballing strategy~\cite{liu2024large} to expand the set by identifying transitively related work.
\revReplace{R2.4}{We conduct both backward and forward snowballing between June 1 and June 10, 2026.\footnote{This period corresponds to the most recent update of our paper collection before finalizing this survey.}}
Backward snowballing inspects reference lists of collected papers, while forward snowballing uses Google Scholar to identify citing studies.
\revReplace{R1.2/R1.3}{This process yields 19 candidate papers for further screening. After applying the same inclusion, exclusion, and type-adapted quality assessment criteria, 7 papers are retained and added to the 235 papers obtained after quality assessment, resulting in the final set of 242 papers.}

\subsection{Collection Results and Statistics}\label{sec:collection_results}
As shown in Figure~\ref{fig:paper_collection}, \revReplace{R2.1}{we first discuss and determine the search keywords based on the seed papers collected from the SWE-bench leaderboard~\cite{swebench}.}
\revReplace{R1.2}{Using these keywords, we construct four exact queries that return 5,176 raw hits. After duplicate removal and automated filtering, 1,238 papers are retained.}
After automated filtering, applying the inclusion and exclusion criteria (278 papers), applying the quality assessment criteria (235 papers), and performing the snowballing strategy, we finally identify a collection of 242 papers focused on issue resolution.
\revReplace{R3.2}{The papers resulting from each filtering step are released on the artifacts page accompanying this survey.}

\begin{figure*}[htb]
    \centering 
    \begin{minipage}{0.57\textwidth}
        \centering 
        \includegraphics[height=4.5cm]{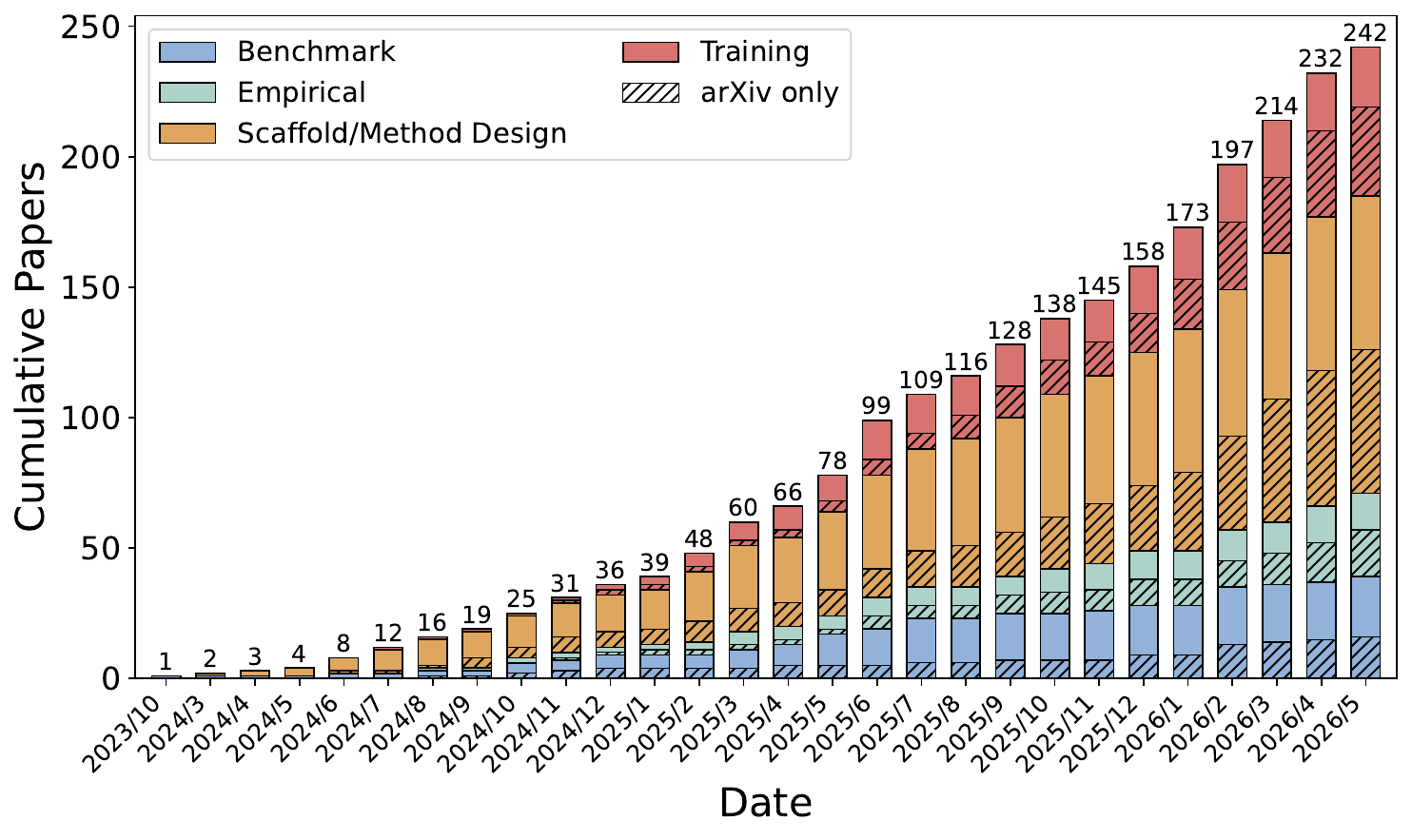}
        \caption{Trend in the number of papers by month.} 
        \label{fig:paper_nums}
    \end{minipage}
    \hfill 
    \begin{minipage}{0.42\textwidth} 
        \centering 
        \includegraphics[height=4.5cm]{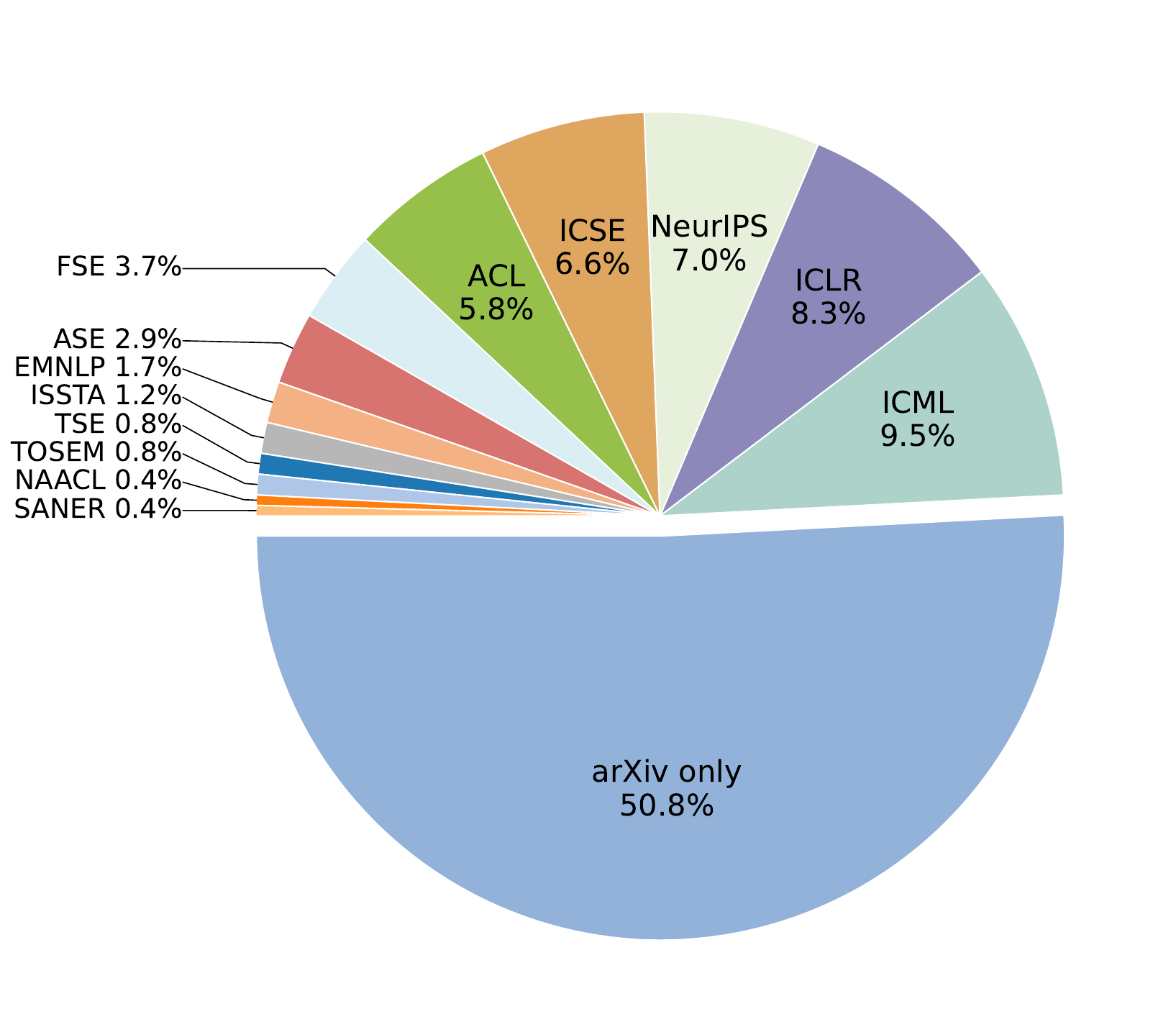}
        \caption{Distribution of papers across publication venues.} 
        \label{fig:paper_distribution}
    \end{minipage}
\end{figure*}

Figure~\ref{fig:paper_nums} shows the cumulative number of papers published over time as of \revReplace{R1.2}{May 2026}.
\revReplace{R3.8}{In this figure, each paper is counted under its primary category, and technique papers are further split into scaffold/method design and training studies following the two dimensions of the technique taxonomy in Section~\ref{sec:technique}.}
We observe a significant gap in this field from October 2023 to May 2024.
The number of studies in this area has grown explosively since May 2024, which further highlights the necessity of this survey.
Figure~\ref{fig:paper_distribution} presents the publication status of the papers included in this survey. 
The issue resolution task currently attracts primary attention from the artificial intelligence and software engineering research communities, with \revReplace{R1.2}{32.6\%} and \revReplace{R1.2}{16.5\%} of the papers published in top-tier venues related to LLM and SE, respectively.
\revReplace{R3.1}{Duplicate records are identified and merged by title during data collection. When both an arXiv preprint and a formally published version of the same paper are found, the paper is attributed to its publication venue rather than to arXiv.
We further cross-check the arXiv papers against the latest publicly available acceptance lists of the considered conferences, and reassign 25 papers that have been accepted but not yet formally published to their corresponding venues.}
\revReplace{R1.4}{Notably, 50.8\% of the papers have not yet undergone peer review and are available on arXiv.
To gauge whether this affects the reliability of our conclusions, we repeat the taxonomy construction procedure of Section~\ref{sec:taxonomy_construction} using only the peer-reviewed subset of the corpus.
The resulting top-level taxonomy is unchanged: the same first-level dimensions emerge for benchmarks, techniques, and empirical studies, suggesting that non-peer-reviewed papers have limited influence on our survey's organization.
Moreover, our inclusion, exclusion, and type-adapted quality assessment criteria apply uniformly to all papers regardless of venue, which can mitigate the effects.
}

\section{Benchmark Statistics}

\subsection{End-to-End Benchmark Statistics}\label{ap: bench_stat}
Table~\ref{tab:benchmarks} lists the statistics of end-to-end issue resolution benchmarks.

\begin{table}
\centering
\caption{\revReplace{R3.3}{The statistics and analysis of end-to-end issue resolution benchmarks. System Kind abbreviations: Lib. = library or package; Frm. = framework; Tool = developer, build, testing, or DevOps tool; App. = end-user application; Svc. = service or platform; Sys. = systems or infrastructure software; Mixed OSS = heterogeneous open-source software projects. System Domain terms and abbreviations: AI = Artificial Intelligence; ML = Machine Learning; DS = Data Science; MLOps = ML Operations; CLI = Command-Line Interface; Dev. Tooling = software-development support (e.g., testing, code quality, documentation, and build); Diverse = heterogeneous repositories without a dominant application domain. Language abbreviations: Py = Python; J = Java; JS = JavaScript; TS = TypeScript.}}
\begin{adjustbox}{width=\textwidth}
\begin{tabular}{cccc>{\centering\arraybackslash}m{2.4cm}>{\centering\arraybackslash}m{2.9cm}>{\centering\arraybackslash}m{4.5cm}>{\centering\arraybackslash}m{2.2cm}cc}
\toprule
Benchmark & Release Time & \#Instances & \#Repositories & Language & System Kind & System Domain & Issue Type & Curation & Source\\
\midrule
SWE-bench~\cite{swebench} & 2023-10 & 2,294 & 12 & Py & Lib./Frm./Tool & DS/ML, Scientific Computing, Web, Dev. Tooling & Bug, Feature & Manual & GitHub\\
SWE-bench Lite~\cite{swebench} & 2023-10 & 300 & 12 & Py & Lib./Frm./Tool & DS/ML, Scientific Computing, Web, Dev. Tooling & Bug & Manual & SWE-bench\\
SWE-bench Lite-S~\cite{agentless} & 2024-07 & 252 & 12 & Py & Lib./Frm./Tool & DS/ML, Scientific Computing, Web, Dev. Tooling & Bug & Manual & SWE-bench\\
SWE-bench Verified~\cite{swebenchverified} & 2024-08 & 500 & 12 & Py & Lib./Frm./Tool & DS/ML, Scientific Computing, Web, Dev. Tooling & Bug & Manual & SWE-bench\\
SWE-bench-java~\cite{swebench_java} & 2024-08 & 91 & 6 & J & Lib./Frm./Tool & Data Serialization, Web, DevOps & Bug & Manual & GitHub\\
SWE-bench Multimodal~\cite{swebenchmutimodal} & 2024-10 & 517 & 12 & JS & Lib./Frm. & Web, Visualization & Bug & Manual & GitHub\\
SWE-Bench+~\cite{swebench_plus} & 2024-10 & 548 & 12 & Py & Lib./Frm./Tool & DS/ML, Scientific Computing, Web, Dev. Tooling & Bug & Manual & SWE-bench\\
Visual SWE-bench~\cite{CodeV} & 2024-12 & 133 & 11 & Py & Lib./Frm. & DS/ML, Visualization & Bug & Manual & GitHub\\
SWEE-Bench~\cite{SWA} & 2025-03 & 885 & 366 & Py & Lib. & Diverse (PyPI packages) & Bug & Automatic & GitHub\\
SWA-Bench~\cite{SWA} & 2025-03 & 535 & 44 & Py & App. & DevOps, Media, Networking, CLI & Bug & Automatic & GitHub\\
FEA-Bench~\cite{FEABench} & 2025-03 & 1,401 & 83 & Py & Lib./Frm./Tool & DS/ML, Scientific Computing, Web, Database, Dev. Tooling & Feature & Manual & GitHub\\
Multi-SWE-bench~\cite{mutiswebench} & 2025-04 & 1,632 & 39 & J, JS, TS, Go, C, Cpp, Rust & Lib./Frm./Tool/Sys. & Web, Data Serialization, Dev. Tooling & Bug, Feature, Performance & Manual & GitHub\\
LiveSWEBench~\cite{liveswebench2025} & 2025-04 & 126 & 5 & Py, JS, TS, J, Cpp & Lib./Frm./Tool/App. & AI/ML, Web, Education, Testing & Bug & Manual & GitHub\\
SWE-PolyBench~\cite{swepoly} & 2025-04 & 2,110 & 21 & Py, J, JS, TS & Lib./Frm./Tool/App. & Web, AI/ML, Cloud/Middleware, Dev. Tooling & Bug, Feature, Refactoring & Automatic & GitHub\\
OmniGIRL~\cite{omnigirl} & 2025-05 & 959 & 15 & Py, J, JS, TS & Lib./Frm./Tool & Code Quality, Web, Time Utilities, Networking, Statistics, Testing, Cryptography & Bug & Manual & GitHub\\
SWE-bench Multilingual~\cite{swebenchmutilingual} & 2025-05 & 300 & 42 & J, JS, TS, Go, C, Cpp, PHP, Ruby, Rust & Lib./Frm./Tool/Sys. & Web, Data Storage, Dev. Tooling & Bug, Feature & Manual & GitHub\\
SwingArena~\cite{xu2026swingarena} & 2025-05 & 400 & N/A & Rust, Py, Go, Cpp & Mixed OSS & Diverse & Bug & Manual & GitHub\\
SWE-rebench~\cite{SWE-rebench} & 2025-05 & 21,000+ & 3,468 & Py & Mixed OSS & Diverse & Bug & Automatic & GitHub\\
SWE-bench-Live~\cite{swebench-live} & 2025-05 & 1,319 & 93 & Py & Mixed OSS & AI/ML, DevOps, Web, Scientific Computing, Database & Bug & Automatic & GitHub\\
GSO~\cite{gso} & 2025-05 & 102 & 10 & Py, C, Cpp, Cython, Rust & Lib./Frm./Sys. & Scientific Computing, Data Analysis, Image Processing, Web, Data Validation, ML & Performance & Manual & GitHub\\
SWE-MERA~\cite{swe-mera} & 2025-07 & 300 & 200 & Py & Mixed OSS & Diverse & Bug & Automatic & GitHub\\
SWE-Perf~\cite{swe-perf} & 2025-07 & 140 & 9 & Py & Lib./Frm./Tool & DS/ML, Scientific Computing & Performance & Automatic & SWE-bench\\
NoCode-bench~\cite{nocode-bench} & 2025-07 & 634 & 10 & Py & Lib./Frm./Tool & DS/ML, Scientific Computing, Web, Dev. Tooling & Feature & Manual & GitHub\\
SWE-Bench Pro~\cite{SWEbenchPro} & 2025-09 & 1,865 & 41 & Py, Go, JS, TS & App./Svc./Tool & Consumer Apps, B2B/Enterprise, Dev. Tooling & Bug, Feature & Manual & GitHub/Proprietary\\
SWE-fficiency~\cite{ma2025swe} & 2025-11 & 498 & 9 & Py & Lib./Frm. & DS/ML, High-Performance Computing, Scientific Computing & Performance & Manual & GitHub\\
SWE-Bench++~\cite{wang2025swe} & 2025-12 & 11,133 & 3,971 & Py, J, JS, TS, Go, Rust, C, Cpp, Ruby, PHP, C\# & Mixed OSS & Diverse & Bug, Feature & Automatic & GitHub\\
SWE-EVO~\cite{thai2025swe} & 2025-12 & 48 & 7 & Py & Lib./Frm./Tool & DS/ML, Package Management, HTTP, Data Processing & Bug, Feature & Manual & GitHub\\
SWE-Refactor~\cite{xu2026swe} & 2026-02 & 1,099 & 18 & J & Lib./Frm./Tool & Utilities, Testing, Code Quality, Web Middleware & Refactoring & Automatic & GitHub\\
FeatureBench~\cite{zhou2026featurebench} & 2026-02 & 200 & 24 & Py & Lib./Frm./Tool & DS/ML, MLOps, Web, Scientific Computing, Dev. Tooling & Feature & Automatic & GitHub\\
Rust-SWE-bench~\cite{xiang2026evaluating} & 2026-02 & 500 & 34 & Rust & Lib./Frm./Tool/Sys. & Systems Infrastructure, Web, CLI Utilities & Bug, Feature, Refactoring & Manual & GitHub\\
SWE-Bench Mobile~\cite{tian2026swe} & 2026-02 & 50 & 1 & Swift, Objective-C & App. & Mobile (iOS) & Feature & Manual & Proprietary\\
SWE-CI~\cite{SWE-CI} & 2026-03 & 100 & 68 & Py & Mixed OSS & Diverse & Bug, Feature, Maintenance & Automatic & GitHub\\
SWE-STEPS~\cite{shastry2026beyond} & 2026-04 & 168 & 6 & Py & Lib./Frm./Tool & Build Automation, AI, Scientific Computing, Database, Testing & Bug, Feature, Maintenance & Automatic & GitHub\\
SWE-Chain~\cite{lam2026swe} & 2026-05 & 155 & 9 & Py & Lib./Frm./Tool & Web, Testing, DS, HTTP, Package Management & Bug, Feature & Manual & PyPI\\
\bottomrule
\end{tabular}
\end{adjustbox}
\label{tab:benchmarks}
\end{table}

\subsection{Subtask Benchmark Statistics}\label{ap: bench_stat_sub}
Table~\ref{tab:sub_task_bench} lists the statistics of subtask benchmarks, including issue localization and reproduction test generation.

\begin{table}
\centering
\caption{\revReplace{R3.3}{The statistics and analysis of issue resolution subtask benchmarks.}}
\begin{adjustbox}{width=\textwidth}
\begin{tabular}{cccccccc}
\toprule
Benchmark&  Release Time&  \#Instances&   \#Repositories&  Domain&  Issue Type&  Curation&Source\\
\midrule
\multicolumn{8}{c}{\cellcolor{tabletitle} \textbf{Localization}} \\
\midrule
LocBench~\cite{locagent}& 2025-03& 660&  N/A& DS/ML, Web, Scientific Computing, Tools& \makecell{Bug, Feature,\\ Performance, Security}& Manual&GitHub\\
MULocBench~\cite{mulocbench}& 2025-09& 1,100&  46& DS/ML, Web, Tools, Systems& Bug, Feature& Manual&GitHub\\
ContextBench~\cite{li2026contextbench}& 2026-02& 1,136& 66& DS/ML, Web, Tools, Systems& Bug, Feature& Manual&GitHub\\
SWE Context Bench~\cite{zhu2026swe}& 2026-02& 399& 12& DS/ML, Web, Scientific Computing, Tools& Bug& Manual&SWE-bench\\
\midrule
\multicolumn{8}{c}{\cellcolor{tabletitle} \textbf{Reproduction}} \\
\midrule
SWT-Bench~\cite{SWT_bench}& 2024-06& 1,978&  12& DS/ML, Web, Scientific Computing, Tools& Bug& Automatic&SWE-bench\\
TestGenEval~\cite{testgeneval}& 2024-10& 1,210&  11& DS/ML, Web, Scientific Computing, Tools& Test& Automatic&SWE-bench\\
TDD-Bench~\cite{tddbench}& 2024-11& 449&  12& DS/ML, Web, Scientific Computing, Tools& Bug& Manual&SWE-bench\\
\bottomrule
\end{tabular}
\end{adjustbox}
\label{tab:sub_task_bench}
\end{table}

\subsection{Evaluation Metrics}\label{ap:metrics}

\begin{table}[ht]
\centering
\caption{\revReplace{R3.3}{Categorization of evaluation metrics for issue resolution. The Usage column lists representative benchmarks and techniques that adopt each metric.}}
\label{tab:metrics}
\begin{adjustbox}{max width=\textwidth}
\begin{tabular}{lllp{10cm}}
\toprule[1pt]
\textbf{Category} & \textbf{Metric} & \textbf{Scope} & \textbf{Usage} \\
\midrule
\multirow{6}{*}{Execution-based}
& Applied\% & End-to-End & \cite{swebench, MAGIS, SWT_bench, MASAI, repounderstander, RepoGraph, AEGIS, FEABench, SWA, issue2test, KGCompass, swepoly, SWE-rebench, swebench-live, swe-perf, nocode-bench, SWEbenchPro, thai2025swe, tian2026swe, zeng2026swe} \\
& Resolved\% / Pass@k & End-to-End & \cite{swebench, SWEAgent, agentless, swebench_java, swebenchmutimodal, FEABench, mutiswebench, swebench-live, SWEbenchPro, thai2025swe, tian2026swe, zeng2026swe, AutoCodeRover, CodeXGraph, testgeneval, SWE-Gym, CodeMonkeys, DARS, Satori-SWE, golubev2025training, soni2026swe, kim2026scaling} \\
& Reproduction Success Rate & Reproduction & \cite{SWT_bench, lin2024llms, AEGIS, tddbench, Otter, issue2test, nocode-bench, experepair, AssertFlip, e-otter, BLAST, soni2026swe, SWE-World, zhao2026immersion, fei2026echo, xiang2026evaluating, meng2026llmbased, wang2026icore, ahmed2026reproduction} \\
& Delta Change Coverage & Reproduction & \cite{SWT_bench, AEGIS, tddbench, Otter, issue2test, AssertFlip, e-otter, soni2026swe, fei2026echo, li2026beyond, wang2026icore} \\
& Speedup & End-to-End & \cite{gso, swe-perf, ma2025swe} \\
& Opt@k & End-to-End & \cite{gso, swe-perf, ma2025swe} \\
\midrule
\multirow{6}{*}{Match-based}
& File Matched Rate & Localization & \cite{swebench, swepoly, agentless, OrcaLoca} \\
& Function Matched Rate & Localization & \cite{swebench, swepoly, agentless, OrcaLoca} \\
& Top-k & Localization & \cite{CodeXGraph, locagent, mulocbench, li2026contextbench, zhu2026swe, CoSIL, SweRank, coret, SACL, RepoSearcher, wang2025improving, zhang2025hierarchical, xiong2025think, reddy2025swerank+, liu2025graphlocator, luo2026closing, mamun2026blagent, sepidband2026rgfl, xu2026neurosymbolic, wang2026icore} \\
& MAP & Localization & \cite{BLAZE, BugCerberus, locagent, mulocbench, li2026contextbench, zhu2026swe, CoSIL, SACL, RepoSearcher, zhang2025hierarchical, liu2025graphlocator, wang2026icore} \\
& MRR & Localization & \cite{BLAZE, BugCerberus, CoRNStack, locagent, mulocbench, li2026contextbench, zhu2026swe, CoSIL, coret, SACL, RepoSearcher, zhang2025hierarchical, xiong2025think, liu2025graphlocator, sepidband2026rgfl, wang2026icore, mamun2026blagent} \\
& Precision, Recall, and F1 & Localization & \cite{MAGIS, swebench, SpecRover, marscode, FEABench, locagent, swepoly, Nemotron-Cortexa, SACL, meta-rag, RepoSearcher, xiong2025think, mulocbench, li2026contextbench, liu2025graphlocator, zhang2025one, zhang2025hierarchical, xu2026learning, sutawika2026codescout, xu2026neurosymbolic, seddik2026arise} \\
\bottomrule[1pt]
\end{tabular}
\end{adjustbox}
\end{table}

\revReplace{R3.3}{Table~\ref{tab:metrics} maps each execution-based and match-based metric to the representative benchmarks and techniques adopting it.
Statistics-based metrics are not listed because they are generic measures applicable to any LLM-based agentic system, whereas we focus on metrics specific to issue resolution.}

\subsubsection{Execution-based Metrics}

In general, benchmarks related to code generation often rely on test suites for evaluation. 
As a challenging code generation task, issue resolution depends on two types of test suites for end-to-end evaluation. 
The first type consists of test cases whose outcomes change from ``FAIL'' to ``PASS'' after applying the golden patch, i.e., the tests that verify whether the issue is resolved (denoted as $F\to P$).
The second type consists of test cases whose outcomes remain ``PASS'' both before and after applying the golden patch, i.e., regression tests (denoted as $P\to P$).
Based on the execution results of these two types of test suites, existing work mainly uses the following metrics.

\begin{itemize}[left=0pt, topsep=0em]

\item \textbf{Applied\%:} The proportion of candidate patches that can be successfully applied to the repository and initiate the execution of the test suites.

\item \textbf{Resolved\%:} The proportion of candidate patches that pass all $F \to P$ and $P \to  P$ tests.

\item \textbf{Pass@k (a.k.a. Resolved@k):} The proportion of cases in which at least one of the $k$ patches generated by the agentic system passes all $F \to P$ and $P \to P$ tests.

\item \textbf{Reproduction Success Rate:} In the issue reproduction stage, the generated reproduction test qualifies as an $F \to P$ test and, after applying the generated test patch, passes all $P \to P$ tests.

\item \textbf{Delta Change Coverage:} \revReplace{R2.6}{It evaluates how well a generated reproduction test covers the code changes introduced by the ground-truth patch~\cite{wang2026icore}. Formally, for instance $i$ with modified patch-line set $M_i$ and covered patch-line set $C_i(\hat{t}_i)$ under generated test $\hat{t}_i$, $\Delta C_i=|M_i \cap C_i(\hat{t}_i)|/|M_i|$.}

\item \textbf{Speedup:} \revReplace{R3.3}{In performance optimization tasks, the ratio of the execution time of the target tests before applying the candidate patch to that after applying it, measuring the runtime improvement delivered by the patch~\cite{gso, swe-perf}.}

\item \textbf{Opt@k:} \revReplace{R3.3}{The proportion of instances in which at least one of the $k$ candidate patches passes all correctness tests and achieves at least a predefined fraction of the speedup delivered by the ground-truth patch~\cite{gso}.}

\end{itemize}

\subsubsection{Match-based Metrics}
Match-based metrics are mainly used to evaluate agentic system performance in the localization stage.
\revReplace{R3.13}{These metrics compare predicted locations against reference locations derived from the developer patch. Since the developer patch is not the only possible fix, the reference locations may miss code involved in alternative correct fixes, so matching results measure agreement with the developer fix rather than complete localization correctness.}
However, due to the autonomy of agent decision-making, not all agentic systems explicitly output intermediate results of the localization stage. 
Therefore, we discuss the following metrics from two perspectives: the generated patches (File/Function Matched Rate) and the intermediate localization results (Top-k, MAP, MRR, Precision, Recall, and F1).

\begin{itemize}[left=0pt, topsep=0em]

\item \textbf{File Matched Rate:}
The proportion of instances where the set of files modified by the candidate patch intersects with the \revReplace{R3.13}{reference} file set. Formally, for instance $i$ with predicted file set $\hat{F}_i$ and \revReplace{R3.13}{reference} set $F_i$, it counts as correct if $\hat{F}_i \cap F_i \neq \varnothing$.

\item \textbf{Function Matched Rate:} The proportion of instances where the set of modified functions or methods in the candidate patch intersects with the \revReplace{R3.13}{reference} function set, that is $\hat{G}_i \cap G_i \neq \varnothing$ for instance $i$.

\item \textbf{Top-k (a.k.a. Hit@k):} Given a ranked list of candidate locations produced during localization, the proportion of instances where at least one \revReplace{R3.13}{reference} location appears within the top $k$ positions. \revReplace{R3.13}{When there are multiple reference locations, success is recorded if any of them is within rank $k$.}

\item \textbf{Mean Average Precision (MAP):} 
It assesses ranking quality by averaging the precision values computed at the ranks of all correctly identified buggy elements within the recommendation list.
Formally, MAP is defined as follows:
$
MAP = \frac{1}{n} \sum_{j=1}^{n} \text{AvgP}_j, \
\text{AvgP}_j = \frac{1}{|K_j|} \sum_{k \in K_j} \text{Prec}@k,\
\text{Prec}@k = \frac{1}{k} \sum_{i=1}^{k} \text{IsRelevant}(i)
$

where, $|K_j|$ is the number of buggy elements for the $j$-th bug, and $\text{IsRelevant}(i)$ returns 1 if the $i$-th element is buggy, and 0 otherwise. $K_j$ denotes the ranks of all buggy elements associated with the $j$-th bug.

\item \textbf{Mean Reciprocal Rank (MRR):}
It assesses ranking quality by computing the reciprocal of the position of the earliest correctly identified buggy element in the recommendation list.
Formally, MRR is defined as:
$
    MRR = \frac{1}{n}\sum_{j=1}^{n}\frac{1}{\text{rank}_j}
$
where \(\text{rank}_j\) is the ranking position of the first buggy method that was modified to resolve the \(j\)-th bug in the recommendation list.

\item \textbf{Precision, Recall, and F1:} \revReplace{R2.6/R3.13}{They evaluate the overlap between predicted and reference buggy locations from complementary perspectives~\cite{liu2025graphlocator}. Formally, for instance $i$ with predicted location set $\hat{L}_i$ and reference location set $L_i$, $\text{Precision}=|\hat{L}_i \cap L_i|/|\hat{L}_i|$, $\text{Recall}=|\hat{L}_i \cap L_i|/|L_i|$, and $F1=2\cdot\text{Precision}\cdot\text{Recall}/(\text{Precision}+\text{Recall})$.}

\end{itemize}

\subsubsection{Statistics-based Metrics}

Agentic systems typically rely on LLMs to perform multi-turn reasoning.
Therefore, besides effectiveness, it is also necessary to evaluate system efficiency. 
This is measured by tracking the average number of consumed input and output tokens (\#Token), dollar cost (\$Cost), and time consumed (Time) per issue during system execution.

\section{Techniques}
\subsection{End-to-End Scaffolds Overview.}\label{ap:e2e_scaffold_overview}

Table~\ref{tab:e2e_scaffolds} presents and analyzes the chronological development of existing end-to-end scaffolds and the performance-cost trade-off. 

\begin{table}[h]
\centering
\caption{End-to-end scaffolds overview. MA = Multi-Agent; MCTS = Monte Carlo tree search; Valid. = Validation; Reprod. = Reproduction; Regress. = Regression; Sel. = Selection. Cost is the average USD expense per issue. Resolved\% and Cost are colored by the benchmark on which they are reported: \resVerified{SWE-bench Verified}, \resLite{SWE-bench Lite}, and \resOther{other benchmarks}.}
\begin{adjustbox}{width=\textwidth}
\setlength{\tabcolsep}{3pt}
\begin{tabular}{c c c c c c c c c c c c}
\toprule
\multirow{2}{*}{\textbf{Method}} & \multirow{2}{*}{\textbf{\makecell{Release\\Time}}} & \multirow{2}{*}{\textbf{\makecell{MA}}} & \textbf{Repo Preprocessing} & \multirow{2}{*}{\textbf{Localization}} & \multicolumn{2}{c}{\textbf{Patch Valid.}} & \textbf{Patch Sel.} & \multirow{2}{*}{\textbf{\makecell{Extra\\Characteristic}}} & \multirow{2}{*}{\textbf{Backbone}} & \multirow{2}{*}{\textbf{\makecell{Resolved\%}}} & \multirow{2}{*}{\textbf{Cost}}\\
\cmidrule(lr){4-4} \cmidrule(lr){6-7} \cmidrule(lr){8-8}
&  &  & Representation &  & Reprod. & Regress. & Rerank & & & & \\
\midrule
\multicolumn{12}{c}{\cellcolor{tabletitle} \textbf{Agent-based Methods}} \\
\midrule
MAGIS~\cite{MAGIS} & 2024-03 & \cmark & \xmark & BM25 & \xmark & \xmark &\xmark& Code Review & GPT-4 & - & - \\
AutoCodeRover~\cite{AutoCodeRover} & 2024-04 & \cmark & \xmark & Navigation/Spectrum & \xmark & \xmark &\xmark & - & GPT-4 & \resLite{19.0} & \resLite{\$0.43} \\
SWE-agent~\cite{SWEAgent} & 2024-05 & \xmark & \xmark & Navigation & \cmark & \xmark &\xmark & - & GPT-4 Turbo & \resLite{18.0} & \resLite{\$1.67} \\
CodeR~\cite{CodeR} & 2024-06 & \cmark & \xmark & BM25/Spectrum & \cmark & \xmark &\xmark & Code Task Graph & GPT-4 Turbo & \resLite{28.3} & \resLite{\$3.09} \\
MASAI~\cite{MASAI} & 2024-06 & \cmark & \xmark & Navigation & \cmark & \xmark &\cmark & Test Template Generation & GPT-4o & \resLite{28.3} & \resLite{\$1.96} \\
Alibaba LingmaAgent~\cite{repounderstander} & 2024-06 & \cmark & Code Graph~\cite{repounderstander}& BM25/Graph & \xmark & \xmark &\xmark & MCTS & Claude 3.5 Sonnet & \resLite{38.3} & \resLite{\$2.18} \\
OpenHands CodeAct~\cite{openhands} & 2024-07 & \xmark & \xmark & Navigation & \cmark & \cmark &\xmark & General SE Agent & Claude 3.5 Sonnet & \resLite{26.0} & \resLite{\$1.10} \\
CodeXGraph~\cite{CodeXGraph} & 2024-08 & \xmark & Code Graph~\cite{CodeXGraph}& Graph & \xmark & \xmark &\xmark & - & GPT-4o & \resLite{23.0} & - \\
SpecRover~\cite{SpecRover} & 2024-08 & \cmark & \xmark & Navigation & \cmark & \cmark &\cmark & Intent Infer & Claude 3.5 Sonnet & \resLite{31.0} & \resLite{\$0.65} \\
MarsCode Agent~\cite{marscode} & 2024-09 & \cmark & Code Graph~\cite{marscode}& Navigation & \cmark & \xmark &\cmark & Language Server & GPT-4o & \resLite{34.0} & - \\
HyperAgent~\cite{hyperagent} & 2024-09 & \cmark & \xmark & Navigation & \xmark & \xmark &\xmark & - & GPT-4o & \resVerified{31.4} & \resVerified{\$2.01} \\
SuperCoder~\cite{supercoder} & 2024-09 & \cmark & \xmark & Embedding & \xmark & \xmark &\xmark & - & GPT-4/Claude 3.5 Sonnet & \resLite{34.0} & - \\
SWE-Search~\cite{SWE-Search} & 2024-10 & \xmark & \xmark & Navigation & \xmark & \xmark &\xmark & MCTS & GPT-4o & \resLite{31.0} & - \\
Infant Agent~\cite{infant} & 2024-11 & \cmark & \xmark & Navigation & \xmark & \xmark &\xmark & - & GPT-4o & \resLite{30.0} & - \\
Learn-by-interact~\cite{Learn-by-interact} & 2025-01 & \xmark & \xmark & Navigation & \xmark & \xmark &\xmark & Synthesized Data & Claude 3.5 Sonnet & \resVerified{60.0} & - \\
DARS~\cite{DARS} & 2025-03 & \xmark & Code Graph~\cite{RepoGraph}& Navigation & \cmark & \xmark &\cmark & - & Claude 3.5 Sonnet & \resLite{47.0} & \resLite{\$12.24} \\
InfantAgent-Next~\cite{InfantAgent-Next} & 2025-05 & \cmark & \xmark & Navigation & \xmark & \xmark &\xmark & Multimodal Support & GPT-4o + DeepSeek-V3 & \resLite{31.7} & - \\
OpenHands-Versa~\cite{OpenHands-Versa} & 2025-06 & \xmark & \xmark & Navigation & \cmark & \cmark &\xmark & Multimodal General Agent & Claude 4 Sonnet & \resOther{34.4} & \resOther{\$1.79} \\
EXPEREPAIR~\cite{experepair} & 2025-06 & \cmark & Tree-based~\cite{agentless} & Navigation & \cmark & \xmark &\cmark & Dual Memory/Experience & Claude 4 Sonnet + o4-mini & \resVerified{74.6} & \resVerified{\$1.91} \\
Agent KB~\cite{AgentKB} & 2025-07 & \xmark & \xmark & Navigation & \cmark & \cmark &\xmark & Experience & Claude 3.7 Sonnet & \resLite{48.3} & - \\
Prometheus~\cite{Prometheus} & 2025-07 & \cmark & Code Graph~\cite{Prometheus}& Navigation/Graph & \cmark & \xmark &\xmark & - & DeepSeek-V3 & \resLite{28.7} & \resLite{\$0.23} \\
SWE-Exp~\cite{SWE-EXP} & 2025-07 & \cmark & \xmark & Navigation & \xmark & \xmark &\xmark & Experience & DeepSeek-V3 & \resVerified{42.0} & \resVerified{\$0.13} \\
SWE-Debate~\cite{SWE-Debate} & 2025-07 & \cmark & Code Graph~\cite{SWE-Debate}& Navigation/Graph & \xmark & \xmark &\xmark & Debate & DeepSeek-V3 & \resVerified{41.4} & - \\
TRAE~\cite{trae2025,tian2026agentbased} & 2025-07 & \cmark & \xmark & Navigation & \cmark & \cmark &\cmark & - & Claude 3.7 Sonnet & \resVerified{66.4} & - \\
SE-Agent~\cite{SE-Agent} & 2025-08 & \xmark & \xmark & Navigation & \xmark & \xmark &\xmark & Self-Evolution & Claude 3.7 Sonnet & \resVerified{61.2} & - \\
Lita~\cite{lita} & 2025-09 & \xmark & \xmark & Navigation & \cmark & \xmark &\xmark & Minimalist Design & Claude 4 Opus & \resVerified{62.6} & - \\
Lingxi~\cite{yang2025lingxi} & 2025-10 & \cmark & \xmark & Navigation & \cmark & \cmark &\xmark & Procedural Knowledge & Claude 4 Sonnet & \resVerified{74.6} & - \\
InfCode~\cite{li2025infcode} & 2025-11 & \cmark & \xmark & Navigation & \cmark & \xmark &\cmark & Adversarial Refine & DeepSeek-V3 & \resLite{40.3} & \resLite{\$0.26} \\
Confucius Code Agent~\cite{wang2025confucius} & 2025-12 & \cmark & \xmark & Navigation & \cmark & \xmark &\xmark & Hierarchical Memory & Claude 4 Sonnet & \resVerified{74.6} & - \\
Agyn~\cite{benkovich2026agyn} & 2026-02 & \cmark & \xmark & Navigation & \xmark & \cmark &\xmark & Role-Based Team & GPT-5 + GPT-5-Codex & \resOther{72.2} & - \\
Debug2Fix~\cite{garg2026debug2fix} & 2026-02 & \cmark & \xmark & Navigation & \xmark & \xmark &\xmark & Interactive Debugging & Claude 4.5 Sonnet & \resOther{40.4} & - \\
SGAgent~\cite{zhang2026sgagent} & 2026-02 & \cmark & Code Graph~\cite{zhang2026sgagent} & Graph & \cmark & \cmark &\cmark & Locate-Suggest-Fix & Claude 3.5 Sonnet & \resLite{51.3} & \resLite{\$1.48} \\
SWE-Adept~\cite{he2026swe} & 2026-03 & \cmark & Tree-based~\cite{he2026swe} & Navigation & \cmark & \cmark &\cmark & Hypothesis-Driven & Claude 4.5 Sonnet & \resLite{71.3} & - \\
iSWE Agent~\cite{ganhotra2026resolving} & 2026-03 & \cmark & \xmark & Navigation & \xmark & \xmark &\xmark & Java-Specialized & Claude 4.5 Sonnet & \resOther{33.6} & \resOther{\$1.86} \\
REAgent~\cite{kuang2026reagent} & 2026-04 & \cmark & \xmark & Navigation & \cmark & \cmark &\cmark & Requirements-Driven & DeepSeek-V3.2 & \resVerified{46.0} & - \\
Agent-CoEvo~\cite{li2026beyond} & 2026-04 & \cmark & \xmark & Navigation & \cmark & \xmark &\cmark & Coevolutionary Search & DeepSeek-V3 & \resLite{41.3} & \resLite{\$1.11} \\

\midrule
\multicolumn{12}{c}{\cellcolor{tabletitle} \textbf{Pipeline-based Methods}} \\
\midrule
Agentless~\cite{agentless} & 2024-07 & \xmark & Tree-based~\cite{agentless} & Navigation/Embedding & \cmark & \cmark &\cmark & - & GPT-4o & \resLite{32.0} & \resLite{\$0.70} \\
RepoGraph~\cite{RepoGraph} & 2024-10 & \xmark & \makecell{Code Graph~\cite{RepoGraph}\\Tree-based~\cite{agentless}} & Navigation & \xmark & \xmark &\cmark & Code Relation Graph & GPT-4o & \resLite{29.7} & \resLite{\$0.39} \\
SWESynInfer~\cite{SWE-GPT} & 2024-11 & \xmark & \xmark & Navigation & \xmark & \xmark &\xmark & - & Lingma SWE-GPT 72B~\cite{SWE-GPT} & \resVerified{30.2} & - \\
SWE-Fixer~\cite{SWE-Fixer} & 2025-01 & \xmark & Tree-based~\cite{agentless} & BM25/Navigation & \xmark & \xmark &\xmark & - & SWE-Fixer-72B~\cite{SWE-Fixer} & \resVerified{32.8} & - \\
Agentless-Mini~\cite{SWE-RL} & 2025-02 & \xmark & Tree-based~\cite{agentless} & Navigation & \cmark & \cmark &\cmark & - & Llama3-SWE-RL-70B~\cite{SWE-RL} & \resVerified{41.0} & - \\
CodeV~\cite{CodeV} & 2025-02 & \xmark & Tree-based~\cite{agentless} & Navigation & \xmark & \xmark &\cmark & Multimodal Support & Qwen2-VL + Qwen2.5-Coder-32B & \resOther{12.8} & - \\
CodeMonkeys~\cite{CodeMonkeys} & 2025-02 & \xmark & \xmark & Navigation & \cmark & \xmark &\cmark & Multi-Model & Claude 3.5 Sonnet & \resVerified{57.4} & \resVerified{\$4.58} \\
PatchPilot~\cite{PatchPilot} & 2025-02 & \xmark & Tree-based~\cite{agentless} & Navigation/Embedding & \cmark & \cmark &\cmark & Iterative Refinement & Claude 3.5 Sonnet & \resVerified{53.6} & \resVerified{\$0.99} \\
KGCompass~\cite{KGCompass} & 2025-03 & \xmark & \xmark & Graph & \cmark & \xmark &\cmark & PR Knowledge Graph & Claude 4 Sonnet & \resLite{58.3} & \resLite{\$0.20} \\
Jiang et al.~\cite{LCLM} & 2025-05& \xmark & \xmark & - & \xmark & \xmark &\xmark & Direct Solve & Gemini 2.5 Pro & \resVerified{50.8} & - \\
CGM~\cite{CGM} & 2025-05& \xmark & Code Graph~\cite{CGM} & Embedding & \xmark & \xmark &\xmark & GraphRAG & Qwen2.5-72B-Instruct~\cite{CGM} & \resVerified{50.4} & - \\
GUIRepair~\cite{GUIRepair} & 2025-06 & \xmark & Tree-based~\cite{agentless} & Navigation/Embedding & \cmark & \xmark &\cmark & Render Check & o4-mini & \resOther{33.9} & \resOther{\$0.36} \\
SemAgent~\cite{SEMAgent} & 2025-06 & \xmark & \xmark & Spectrum & \cmark & \xmark &\cmark & Semantic Understanding & Claude 3.7 Sonnet & \resLite{44.7} & \resLite{\$6.90} \\
Nemotron-Cortexa~\cite{Nemotron-Cortexa} & 2025-06 & \xmark & Tree-based~\cite{agentless} & Navigation/Embedding & \cmark & \xmark &\cmark & Code Embedding Model & Claude 3.5 Sonnet & \resVerified{52.6} & \resVerified{\$0.51} \\
SynFix~\cite{SynFix} & 2025-07 & \xmark & Code Graph~\cite{SynFix} & Graph/Embedding & \cmark & \cmark &\xmark & Code Relation Graph & GPT-4o & \resLite{52.3} & \resLite{\$0.56} \\
SIADAFIX~\cite{cao2025siadafix} & 2025-10 & \cmark & \xmark & Navigation & \cmark & \cmark &\cmark & Adaptive Workflow & Claude 4 Sonnet & \resLite{60.7} & - \\
TDFlow~\cite{han-etal-2026-tdflow} & 2025-10 & \xmark & \xmark & Navigation & \cmark & \cmark &\cmark & Test-Driven Workflow & GPT-5 + Claude 4 Sonnet & \resVerified{68.0} & \resVerified{\$4.12} \\
Think-Search-Patch~\cite{xiong2025think} & 2025-11 & \xmark & \xmark & Embedding & \xmark & \xmark &\xmark & Two-Stage Training & Qwen2.5-Coder-14B~\cite{xiong2025think} & \resLite{8.3} & - \\
SVRepair~\cite{tang2026svrepair} & 2026-02 & \xmark & \xmark & Navigation & \xmark & \cmark &\xmark & Multimodal Visual & SVR-8B~\cite{tang2026svrepair} + o3 & \resOther{36.5} & - \\
RepoRepair~\cite{pan2026reporepair} & 2026-03 & \xmark & \xmark & Embedding & \xmark & \cmark &\cmark & Documentation-Enhanced & Claude 4 Sonnet + DeepSeek-V3 & \resLite{45.7} & \resLite{\$0.44} \\
RAIM~\cite{liu2026architecture} & 2026-03 & \xmark & \xmark & Navigation & \xmark & \cmark &\cmark & Feature Addition & Gemini 2.5 Pro & \resOther{39.5} & - \\
ARISE~\cite{seddik2026arise} & 2026-05 & \xmark & Code Graph~\cite{seddik2026arise} & Graph & \xmark & \xmark &\xmark & Graph Toolset & Qwen3.6-35B-A3B & \resLite{34.0} & - \\

\bottomrule
\end{tabular}
\end{adjustbox}
\label{tab:e2e_scaffolds}
\end{table}

\subsection{Repository Representation Methods}\label{ap:repo_representation}

This section details representative methods of the two repository representation types summarized in Section~\ref{sec:e2e_scaffold}.A.

\begin{figure}[h]
    \centering
    \includegraphics[width=0.5\linewidth]{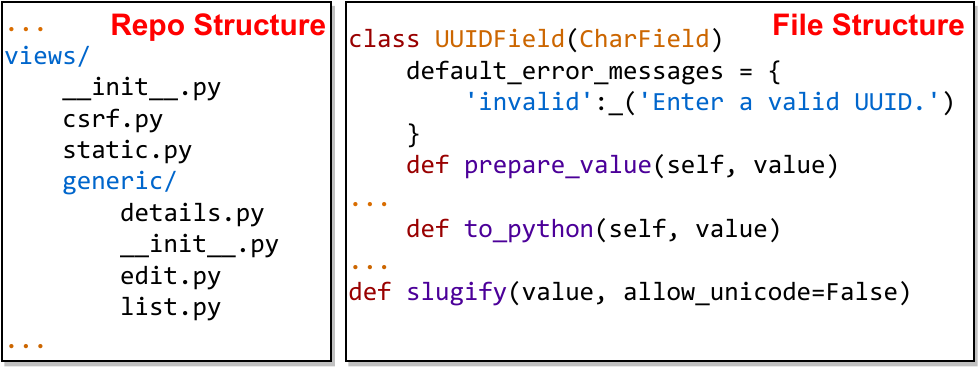}
    \caption{Tree-based representation of repository and code files.}
    \label{fig:repo_file_structure}
\end{figure}

\ulit{A.1 Tree-based Representation.}
Tree-based methods use static parsers to organize repository files, classes, and functions into lightweight hierarchies that fit within the LLM context window.
For example, Agentless~\cite{agentless} statically parses the repository into a repository-level tree of files and a file-level tree of classes and functions, illustrated in Figure~\ref{fig:repo_file_structure}, so that the LLM sees the repository skeleton without reading any function body.
Because they are simple and inexpensive to construct, similar representations are widely used by pipeline-based scaffolds and related systems~\cite{RepoGraph, CodeV, PatchPilot, SWE-Fixer, GUIRepair, Nemotron-Cortexa, experepair}.

\ulit{A.2 Code Graph.}
Code-graph methods represent deeper repository semantics through relations such as function calls, inheritance, imports, references, and data flows.
Existing systems differ mainly in graph granularity and access mechanism: they construct function-level call graphs~\cite{repounderstander, SynFix, CGM}, line-level definition-reference graphs~\cite{RepoGraph, DARS}, or richer dependency graphs over code and documentation~\cite{marscode, Prometheus, CodeXGraph, SWE-Debate}.
Some supply graph-derived context directly to the LLM, whereas CodeXGraph~\cite{CodeXGraph} and Prometheus~\cite{Prometheus} expose graph-database query interfaces.
For example, CodeXGraph~\cite{CodeXGraph} persists the extracted nodes and edges in a graph database and lets the LLM retrieve context by generating Cypher queries over it.

\subsection{Localization Methods in End-to-End Scaffolds}\label{ap:localization}

This section details representative methods of the five localization types summarized in Section~\ref{sec:e2e_scaffold}.B.

\ulit{B.1 BM25.}
BM25~\cite{bm25} performs lexical retrieval and is well suited to long issue descriptions and code documents~\cite{swebench}.
Existing scaffolds use it either for initial suspicious-file retrieval~\cite{MAGIS, SWE-Fixer} or as a signal combined with search and reasoning, such as MCTS in Alibaba LingmaAgent~\cite{repounderstander} and spectrum scores in CodeR~\cite{CodeR}.
For example, SWE-Fixer~\cite{SWE-Fixer} scores every repository file against the issue description as a query and forwards only the top-ranked files to its repair model.

\ulit{B.2 Spectrum-based Methods.}
Spectrum-based fault localization (SBFL) contrasts coverage from passing and failing tests and ranks program elements with suspiciousness metrics such as Tarantula~\cite{jones2002visualization} and Ochiai~\cite{abreu2007accuracy}.
Issue-resolution systems use these rankings directly as fault-revealing context~\cite{AutoCodeRover}, combine them with lexical retrieval~\cite{CodeR}, or derive semantic cues from bug reports~\cite{SEMAgent}.
For example, AutoCodeRover~\cite{AutoCodeRover} runs the test suite with coverage, ranks methods by suspiciousness, and passes the top-ranked ones to the agent as fault-revealing context.
Because reproduction tests are often unavailable or difficult to generate~\cite{SWT_bench}, the adoption of SBFL remains limited.

\ulit{B.3 Embedding-based Methods.}
Embedding-based methods project issue descriptions and code segments into a shared semantic space and retrieve code by vector similarity~\cite{agentless,RepoGraph}.
This supports coarse-grained localization in scaffolds such as Agentless, SuperCoder, PatchPilot, GUIRepair, and SynFix~\cite{agentless, supercoder, PatchPilot, GUIRepair, SynFix}.
For example, Agentless~\cite{agentless} splits repository files into code segments and ranks them by the cosine similarity between their embeddings and that of the issue description.
Recent variants enrich vector retrieval with code graphs~\cite{CGM} or fine-tune the embedding model for repository-level file retrieval~\cite{Nemotron-Cortexa}.

\ulit{B.4 Navigation-based Methods.}
Navigation-based methods let the LLM inspect a repository through either specialized search operations or a general shell interface.
Agent-based scaffolds dynamically select fine-grained tools~\cite{AutoCodeRover, CodeR, SpecRover} or general bash commands~\cite{SWEAgent, MASAI, openhands, marscode, hyperagent, SWE-Search, infant, Learn-by-interact, InfantAgent-Next, DARS, OpenHands-Versa, experepair, AgentKB, Prometheus, SWE-EXP, SWE-Debate, trae2025, SE-Agent}, whereas pipeline-based scaffolds constrain navigation to predefined file-to-function hierarchies~\cite{agentless, RepoGraph, SWE-GPT, PatchPilot, SWE-Fixer, CodeV, Nemotron-Cortexa, GUIRepair, SWE-RL}.
For example, SWE-agent~\cite{SWEAgent} repeats an observe--act loop over its agent--computer interface, deciding each search or edit operation from the returned repository state, whereas Agentless~\cite{agentless} follows a fixed cascade that narrows from the structure tree to files, then to classes and functions, and finally to edit locations.
Other designs either aggregate LLM-ranked files~\cite{CodeMonkeys} or use long-context models to inspect an entire repository~\cite{LCLM}.

\ulit{B.5 Graph-based Methods.}
Graph-based localization retrieves issue-relevant context from the repository graphs introduced in Section~\ref{sec:e2e_scaffold}.A.2.
Most methods reason over graph neighborhoods or paths: Alibaba LingmaAgent constrains exploration with MCTS~\cite{repounderstander}, KGCompass combines graph traversal with LLM-proposed candidates~\cite{KGCompass}, SynFix combines embedding retrieval with one-hop expansion~\cite{SynFix}, and SWE-Debate compares multiple fault-propagation paths~\cite{SWE-Debate}.
For example, SWE-Debate~\cite{SWE-Debate} enumerates candidate fault-propagation paths from the entities named in the issue and lets multiple agents debate them, adopting the surviving path as the localization result.
A second design exposes graph-database queries, allowing the LLM to retrieve context through generated Cypher commands~\cite{CodeXGraph, Prometheus}.
These methods capture structural dependencies beyond lexical similarity but depend on the quality and accessibility of the underlying graph.

\subsection{Training Datasets Statistics}\label{ap:train_data}
Table~\ref{tab:training_data} lists the statistics of existing training datasets for training the issue resolution model.

\begin{table}[h]
\centering
\caption{Comparison of existing training datasets for software engineering agents. For Env. Size, ``-'' indicates the dataset is not executable. ``Not Released'' indicates the dataset authors do not release the official snapshot. ``Not Stat.'' indicates the dataset authors do not report the size of the snapshot.}
\begin{adjustbox}{width=0.65\textwidth}
\begin{tabular}{cccccc}
\toprule
\textbf{Dataset}&  \textbf{Release Time}&  \textbf{\#Instances}&   \textbf{Executable?} &\textbf{Env. Size}&  \textbf{Type}\\
\midrule
SWE-bench(Train)~\cite{swebench}&  2023-10&  19,008& \xmark &-&  Real\\
R2E~\cite{R2E}&  2024-07&  246& \cmark &270GBs&  Synthetic\\
SWE-Gym~\cite{SWE-Gym}&  2024-12&  2,438& \cmark &6TBs&  Real\\
SWE-bench-extra~\cite{swe-bench-extra}&  2024-12&  6,411& \cmark &Not Released&  Real\\
SWE-Fixer~\cite{SWE-Fixer}&  2025-01&  110k& \xmark &-&  Real\\
R2E-Gym(Subset)~\cite{r2e_gym}&  2025-04&  4,578& \cmark &4TBs&  Synthetic\\
SWE-Synth~\cite{SWE-Synth}&  2025-04&  9,459& \cmark &Not Released&  Synthetic\\
SWE-smith~\cite{SWE-smith}&  2025-04&  50,137& \cmark &295GBs&  Synthetic\\
Multi-SWE-RL~\cite{mutiswebench}&  2025-04&  4,723& \cmark &Not Stat.&  Real\\
SWE-rebench~\cite{SWE-rebench}& 2025-05& 20k& \cmark & Not Stat.&Real\\
SWE-Dev(a)~\cite{swe-dev_sjtu}&  2025-06&  14.5k& \cmark &100GBs&  Real\\
SWE-Dev(b)~\cite{swe-dev_thu}&  2025-06&  26k& \cmark &Not Released&  Synthetic\\
Skywork-SWE\tablefootnote{Skywork-SWE dataset has not been open-sourced yet.}~\cite{Skywork-SWE}&  2025-06&  10k& \cmark &Not Released&  Real\\
SWE-Flow~\cite{zhang2025synthesizing}&  2025-06&  16k& \cmark &Not Stat.&  Synthetic\\
SWE-Mirror~\cite{swe-mirror}&  2025-09&  60k& \cmark &100GBs&  Synthetic\\
SWE-Universe~\cite{SWE-Universe}&  2026-02&  807,693& \cmark &Not Released&  Real\\
Scale-SWE-Data~\cite{zhao2026immersion}&  2026-02&  100k& \cmark &Not Stat.&  Real\\
Hybrid-Gym~\cite{xie2026hybrid}&  2026-02&  4,470& \cmark &Not Stat.&  Synthetic\\
SWE-rebench V2~\cite{badertdinov2026swe}&  2026-02&  32,079& \cmark &26.36TBs&  Real\\
SWE-Next~\cite{liang2026swe}&  2026-03&  2,308& \cmark &639GBs&  Real\\
STITCH~\cite{team2026yet}&  2026-04&  2,067& \cmark &Not Stat.&  Real\\

\bottomrule
\end{tabular}

\end{adjustbox}
\label{tab:training_data}
\end{table}

\subsection{Training Methods Statistics}\label{ap:train_method}

Table~\ref{tab:training_method} summarizes representative models, their training paradigms, and their performance on SWE-bench Verified.

\begin{table}
\centering
\caption{Training methods and performance statistics of domain-specific models for the issue resolution task. Resolved\% represents the reported score of each model on SWE-bench Verified when used with its corresponding scaffold. ``Process'' indicates process-oriented methods; ``Distillation'' indicates teacher-model distillation methods; ``PRM'' indicates process reward model; ``ORM'' indicates outcome reward model.}
\begin{adjustbox}{width=\textwidth}
\begin{tabular}{cccccccccc}
\toprule
\multirow{2}{*}{\textbf{Method/Model}} &  
\multirow{2}{*}{\textbf{Release Time}} &  
\multirow{2}{*}{\textbf{Params}} & 
\multicolumn{2}{c}{\textbf{SFT?}} & 
\multicolumn{2}{c}{\textbf{RL?}} & 
\multirow{2}{*}{\textbf{Scaffold}} & 
\multirow{2}{*}{\textbf{Base Model}} & 
\multirow{2}{*}{\textbf{Resolved\%}} \\
\cmidrule(lr){4-5} \cmidrule(lr){6-7}
 & & & \textbf{Process} & \textbf{Distillation} & \textbf{PRM} & \textbf{ORM} & & & \\
\midrule
SWE-GPT~\cite{SWE-GPT}& 2024-11&  72B& \cmark &\xmark  &\xmark   &\xmark   & SWESynInfer~\cite{SWE-GPT}&Qwen2.5-Instruct&30.2\\
SWE-Gym~\cite{SWE-Gym}& 2024-12&  32B& \xmark  &\cmark &\xmark   &\xmark   & OpenHands~\cite{openhands}&Qwen2.5-Coder-Instruct&32.0\\
ReSAT~\cite{ReSAT}& 2024-12&  7B& \cmark &\xmark  &\xmark   &\xmark   & Agentless~\cite{agentless}&CodeQwen-1.5-Chat&7.2\\
SWE-Fixer~\cite{SWE-Fixer}& 2025-01&  72B& \cmark &\xmark  &\xmark   &\xmark   & SWE-Fixer~\cite{SWE-Fixer}&Qwen2.5-Base&32.8\\
SWE-RL~\cite{SWE-RL}&  2025-02& 70B & \xmark &\xmark &\xmark   &\cmark   & Agentless-Mini~\cite{SWE-RL}&Llama3.3-Instruct&41.0\\
SoRFT~\cite{SoRFT}& 2025-02&  32B& \cmark &\cmark &\cmark   &\xmark   & Agentless~\cite{agentless}&Qwen2.5-Coder-Instruct&30.8\\
SWE-Reasoner~\cite{swe-resoner}& 2025-03&  32B& \cmark &\cmark &\xmark &\cmark & SWESynInfer+~\cite{swe-resoner}&Qwen2.5-Coder-Instruct&46.0\\
SEAlign~\cite{SEAlign}&  2025-03&  14B& \xmark&\xmark&\xmark   &\cmark&  OpenHands~\cite{openhands}&Qwen2.5-Coder-Instruct&21.8\\
R2E-Gym~\cite{r2e_gym}&  2025-04& 32B &\xmark&\cmark &\xmark   &\xmark   & OpenHands~\cite{openhands} &Qwen2.5-Coder-Instruct&34.4\\
SWE-agent-LM~\cite{SWE-smith}&  2025-04& 32B & \xmark&\cmark &\xmark   &\xmark   & SWE-agent~\cite{SWEAgent} &Qwen2.5-Coder-Instruct&40.2\\
Co-PatcheR~\cite{Co-PatcheR} & 2025-05& 3*14B& \cmark &\cmark & \xmark   &\xmark & PatchPilot~\cite{PatchPilot}& Qwen2.5-Coder-Instruct&46.0\\
Satori-SWE~\cite{Satori-SWE} & 2025-05& 32B& \cmark &\cmark & \xmark&\cmark   & Satori-SWE~\cite{Satori-SWE}& Qwen2.5-Coder-Instruct&41.6\\
Agent-RLVR~\cite{Agent-RLVR} & 2025-06& 72B& \xmark &\xmark & \xmark   &\cmark& Agentless~\cite{agentless}& Qwen2.5-Instruct&27.8\\
MCTS-Refine~\cite{MCTS-Refined} & 2025-06& 72B& \cmark &\xmark& \xmark   &\xmark& Agentless~\cite{agentless}& Qwen2.5-Instruct&35.0\\
SWE-Dev-LM~\cite{swe-dev_thu}&  2025-06& 32B &\xmark  &\cmark&\xmark   &\cmark& OpenHands~\cite{openhands} &Qwen-2.5-Coder-Instruct&36.6\\
Skywork-SWE~\cite{Skywork-SWE}&  2025-06& 32B &\xmark  &\cmark&\xmark   &\xmark&  OpenHands~\cite{openhands}&Qwen-2.5-Coder-Instruct&38.0\\
DeepSWE~\cite{DeepSWE}&  2025-07& 32B &\xmark  &\xmark&\xmark   &\cmark& DeepSWE~\cite{DeepSWE} & Qwen3&59.0\\
RepoForge~\cite{RepoForge} & 2025-08&8B & \xmark &\cmark& \xmark   &\cmark& OpenHands~\cite{openhands}& Qwen3(Non-Thinking)&16.4\\
Golubev et al.~\cite{golubev2025training} &2025-08 &72B & \xmark  &\xmark& \xmark    &\cmark& Golubev et al.~\cite{golubev2025training} &Qwen2.5-Instruct &39.0\\
Devstral-Small~\cite{devstal_small}&  2025-08& 24B &\xmark  &\cmark&\xmark   &\cmark& OpenHands~\cite{openhands} &Mistral Small 3&46.8\\
SWE-Swiss~\cite{SWESwiss}&  2025-08& 32B &\cmark  &\cmark&\xmark   &\cmark& SWE-Swiss~\cite{SWESwiss} & Qwen2.5-Instruct&60.2\\
SWE-Mirror-LM~\cite{swe-mirror}&  2025-09& 32B &\xmark  &\cmark &\xmark   &\xmark& OpenHands~\cite{openhands} &Qwen-2.5-Coder-Instruct&52.2\\
Kimi-Dev~\cite{kimi-dev}&  2025-09& 72B &\cmark  &\cmark&\xmark   &\cmark&  Kimi-Dev~\cite{kimi-dev}&Qwen2.5-Base&60.4\\
CWM~\cite{CWM}&  2025-09& 32B &\xmark  &\cmark &\xmark   &\cmark& CWM~\cite{CWM} & -&65.8\\
EntroPO~\cite{yu2025building}&  2025-09& 30B &\xmark  &\cmark &\xmark   &\cmark& R2E-Gym~\cite{r2e_gym} & Qwen3-Coder&52.2\\
FrogBoss~\cite{sonwane2025bugpilot}& 2025-10& 32B & \xmark &\cmark &\xmark &\xmark & R2E-Gym~\cite{r2e_gym} &Qwen3-32B&54.6\\
SWE-Play~\cite{zhu2025training}& 2025-12& 32B & \xmark &\cmark &\xmark &\xmark & OpenHands~\cite{openhands} &Qwen2.5-Coder-32B-Instruct&31.2\\
SSR~\cite{wei2025toward}& 2025-12& 32B & \xmark &\xmark &\xmark &\cmark & CWM~\cite{CWM} &CWM-sft~\cite{CWM}&51.4\\
SWE-Compressor~\cite{liu2025context}& 2025-12& 32B & \xmark &\cmark &\xmark &\xmark & OpenHands~\cite{openhands} &Qwen2.5-Coder-32B&57.6\\
SWE-RM~\cite{shum2026swerm}& 2025-12& 30B & \xmark &\xmark &\xmark &\cmark & OpenHands~\cite{openhands} &Qwen3-30B-A3B&54.8\\
SWE-Lego~\cite{tao2026swe}& 2026-01& 32B & \xmark &\cmark &\xmark &\xmark & OpenHands~\cite{openhands} &Qwen3-32B&52.6\\
daVinci-Dev~\cite{zeng2026davinci}& 2026-01& 72B & \xmark &\cmark &\xmark &\xmark & SWE-agent~\cite{SWEAgent} &Qwen2.5-72B-Base&58.5\\
SERA~\cite{shen2026sera}& 2026-01& 32B & \xmark &\cmark &\xmark &\xmark & SWE-agent~\cite{SWEAgent} &Qwen3-32B&54.2\\
SWE-Spot~\cite{peng2026swe}& 2026-01& 4B & \xmark &\cmark &\xmark &\xmark & mini-SWE-agent~\cite{SWEAgent} &Qwen3-4B-Instruct&17.1\\
SWE-World~\cite{SWE-World}& 2026-02& 32B & \xmark &\cmark &\xmark &\cmark & R2E-Gym~\cite{r2e_gym} &Qwen2.5-Coder-32B-Instruct&55.0\\
SWE-Master~\cite{song2026swe}& 2026-02& 32B & \xmark &\cmark &\xmark &\cmark & R2E-Gym~\cite{r2e_gym} &Qwen2.5-Coder-32B-Instruct&61.4\\
SWE-Protégé~\cite{kon2026swe}& 2026-02& 7B & \xmark &\cmark &\cmark &\cmark & SWE-agent~\cite{SWEAgent} &Qwen2.5-Coder-7B-Instruct&42.4\\
SWE-Fuse~\cite{wen2026swe}& 2026-03& 32B & \xmark &\cmark &\xmark &\cmark & mini-SWE-agent-plus~\cite{SWEAgent} &Qwen3-32B&60.2\\
SWE-HERO~\cite{ludwig2026swe}& 2026-04& 32B & \xmark &\cmark &\xmark &\xmark & OpenHands~\cite{openhands} &Qwen2.5-Coder-32B-Instruct&62.2\\
SWE-AGILE~\cite{lian2026swe}& 2026-04& 8B & \xmark &\cmark &\xmark &\cmark & R2E-Gym~\cite{r2e_gym} &Qwen3-8B&24.1\\
SWE-TRACE~\cite{SWE-TRACE}& 2026-04& 30B & \xmark &\cmark &\cmark &\cmark & mini-SWE-agent~\cite{SWEAgent} &Qwen3-30B-A3B&63.5\\
BoostAPR~\cite{li2026boostapr}& 2026-05& 32B & \xmark &\cmark &\xmark &\cmark & OpenHands~\cite{openhands} &Qwen2.5-Coder-32B-Instruct&40.7\\
HHD~\cite{wang2026hindsight}& 2026-05& 72B & \xmark &\cmark &\xmark &\xmark & OpenHands~\cite{openhands} &Qwen2.5-72B-Base&51.2\\

\bottomrule
\end{tabular}

\end{adjustbox}
\label{tab:training_method}
\end{table}

\end{document}